\documentclass{article}
\usepackage{graphicx}
\usepackage{authblk}
\usepackage{amsmath}
\usepackage{amssymb}
\usepackage{bm}
\usepackage[square]{natbib}
\usepackage[table]{xcolor}
\usepackage[skip=10pt plus1pt, indent=0pt]{parskip}
\usepackage[verbose=true,letterpaper]{geometry}
\usepackage{hyperref}
\usepackage[utf8]{inputenc}
\usepackage[T1]{fontenc}
\usepackage{booktabs} 
\usepackage{tabularx}
\usepackage{amsfonts}
\usepackage{lmodern}
\usepackage{microtype}
\usepackage{fancyhdr}
\usepackage{subcaption}
\usepackage{bbm}
\usepackage{float}
\usepackage{chngcntr}
\usepackage[most]{tcolorbox}

\AtBeginDocument{
  \newgeometry{
    textheight=9in,
    textwidth=6.5in,
    top=1in,
    headheight=14pt,
    headsep=25pt,
    footskip=30pt
  }
}
\newdimen\figrasterwd
\figrasterwd\textwidth

\newcommand{\shorttitle}{permApprox: permutation $p$\,-value approximation}
\fancyheadoffset{0pt}
\title{permApprox: a general framework for accurate permutation $p$\,-value approximation}

\author[1]{Stefanie Peschel\thanks{Correspondence: stefanie.peschel@mail.de}}
\author[2,3]{Anne-Laure Boulesteix}
\author[4,5,6]{Erika von Mutius}
\author[1,3,7,8]{Christian L. Müller}
\affil[1]{Department of Statistics, Ludwig-Maximilians-Universit\"at M\"unchen, Munich, Germany}
\affil[2]{Institute for Medical Biometry and Epidemiology, Ludwig-Maximilians-Universit\"at M\"unchen, Munich, Germany}
\affil[3]{Munich Center for Machine Learning, Munich, Germany}
\affil[4]{Institute of Asthma and Allergy Prevention, Helmholtz Munich, Neuherberg, Germany}
\affil[5]{Department of Pediatrics, Dr. von Hauner Children's Hospital, LMU University Hospital, Ludwig-Maximilians-Universit\"at M\"unchen, Munich, Germany}
\affil[6]{Comprehensive Pneumology Center Munich, Member of the German Center for Lung Research (DZL), Munich, Germany}
\affil[7]{Institute of Computational Biology, Helmholtz Munich, Neuherberg, Germany}
\affil[8]{Center for Computational Mathematics, Flatiron Institute, New York, USA}
\date{}

\begin{document}
\maketitle

\begin{abstract}
Permutation procedures are common practice in hypothesis testing when distributional assumptions about the test statistic are not met or unknown. With only few permutations, empirical $p$-values lie on a coarse grid and may even be zero when the observed test statistic exceeds all permuted values. Such zero $p$-values are statistically invalid and hinder multiple testing correction. Parametric tail modeling with the Generalized Pareto Distribution (GPD) has been proposed to address this issue, but existing implementations can again yield zero $p$-values when the estimated shape parameter is negative and the fitted distribution has a finite upper bound.

We introduce a method for accurate and zero-free $p$-value approximation in permutation testing, embedded in the \texttt{permApprox} workflow and \texttt{R} package. Building on GPD tail modeling, the method enforces a support constraint during parameter estimation to ensure valid extrapolation beyond the observed statistic, thereby strictly avoiding zero $p$-values. The workflow further integrates robust parameter estimation, data-driven threshold selection, and principled handling of hybrid $p$-values that are discrete in the bulk and continuous in the extreme tail.

Extensive simulations using two-sample $t$-tests and Wilcoxon rank-sum tests show that \texttt{permApprox} produces accurate, robust, and zero-free $p$-value approximations across a wide range of sample and effect sizes. Applications to single-cell RNA-seq and microbiome data demonstrate its practical utility: \texttt{permApprox} yields smooth and interpretable $p$-value distributions even with few permutations. By resolving the zero-$p$-value problem while preserving accuracy and computational efficiency, \texttt{permApprox} enables reliable permutation-based inference in high-dimensional and computationally intensive settings.
\end{abstract}


%%%%%%%%%%%%%%%%%%%%%%%%%%%%%%%%%%%%%%%%%%
\section{Introduction} \label{sec:introduction}
%%%%%%%%%%%%%%%%%%%%%%%%%%%%%%%%%%%%%%%%%%

Permutation tests are a flexible and widely used alternative to parametric hypothesis tests when the null distribution of a test statistic is unknown or difficult to derive \citep{fisher1935design, pitman1937significance, good2005permutation}. They rely on the exchangeability of labels under the null hypothesis: by repeatedly shuffling the data and recalculating the test statistic, an empirical null distribution is obtained from which a $p$-value can be computed. Despite their generality, permutation procedures become computationally demanding when accurate estimation of small $p$-values is required \citep{segal2018fast,he2019permutation}.
A practical challenge arises when the observed test statistic exceeds all values in the permutation distribution, yielding an empirical $p$-value of zero. Although this result indicates strong evidence against the null hypothesis, an exact zero permutation $p$-value is invalid \citep{phipson2010permutation}. This is especially problematic since it remains zero after adjusting for multiplicity.

To avoid zero $p$-values, it has been strongly recommended to include the observed test statistic in the permutation distribution \citep{north2002note,phipson2010permutation}, which guarantees a strictly positive $p$-value.
However, this convention introduces a different limitation: the smallest attainable $p$-value becomes $1/(B+1)$, e.g.,~$\approx0.001$ for $B=1000$ permutations. Such coarse resolution can be insufficient in large-scale multiple testing, where even moderately small $p$-values may not remain significant after correction. Increasing the number of permutations can improve this resolution, but is often computationally prohibitive in practice, particularly in omics or imaging applications involving complex models or high-dimensional data \citep{john2022efficient}.

To overcome the limitations of finite permutation resolution, \citet{knijnenburg2009fewer} introduced a parametric tail approximation based on the Generalized Pareto Distribution (GPD), which enables accurate extrapolation of small $p$-values from a limited number of permutations. Their method fits a GPD to the largest subset of permuted test statistics and uses the resulting model to extrapolate beyond the observed permutation range, thereby reducing the number of permutations needed while retaining accuracy in the small-$p$ regime. Building on this idea, \citet{winkler2016faster} conducted a systematic comparison of several strategies for accelerating permutation inference in the general linear model, including (i) early stopping with negative binomial modeling, (ii) Gamma distribution fitting to the full permutation distribution, and (iii) GPD fitting restricted to the tail. They recommended the GPD-based tail approximation as a general-purpose choice due to its flexibility and weak distributional assumptions. The authors also emphasized that including or excluding the observed test statistic in the permutation set can systematically shift the resulting $p$-values, leading either to conservativeness or to ``invalid'' $p$-values.

A persistent limitation of existing GPD-based approaches lies in handling cases where the estimated shape parameter $\xi$ is negative. 
In this situation, the fitted GPD has a finite upper endpoint, and if the observed test statistic lies beyond this bound, the extrapolated $p$-value again collapses to zero. 
While \citet{knijnenburg2009fewer} suggested raising all test statistics to a power (e.g., cubing) to avoid negative shape estimates, this ad hoc transformation changes the scale of the statistic and can distort inference. 
\citet{winkler2016faster} acknowledged the issue but did not propose a remedy. To date, no general methodological solution has been established. Consequently, the issue of zeros in GPD-approximated permutation $p$-values remains largely unresolved.  

In this work, we propose a new method for accurate $p$-value approximation in permutation testing based on support-constrained GPD tail modeling, embedded within a comprehensive and robust analysis workflow implemented in the open-source \texttt{R} package \texttt{permApprox}.
The proposed method builds on the classical GPD tail approximation but introduces a \emph{support constraint} during parameter estimation to ensure that the fitted distribution extends beyond the observed statistic, thereby strictly avoiding zero $p$-values.
Beyond resolving this core methodological problem, the \texttt{permApprox} workflow structures the key practical decisions involved in GPD-based permutation inference, including (i) choosing one of the established GPD parameter estimation techniques, (ii) selecting a GPD tail threshold and number of starting exceedances, and (iii) handling the resulting hybrid $p$-values in downstream inference procedures such as multiple testing correction.

In the following, we present the \texttt{permApprox} workflow and explain how the proposed constrained GPD method and its supporting components together enable accurate, zero-free, and computationally efficient $p$-value approximation.
We compare the accuracy and robustness of the proposed approach in extensive simulation studies based on classical two-sample tests, including the Student’s $t$-test and the Wilcoxon rank-sum test, covering both parametric and nonparametric settings. 
We further showcase the practical relevance of \texttt{permApprox} in two real-world biological applications that are well suited for permutation-based inference due to their high dimensionality and complex, non-Gaussian data structures. Specifically, we analyze two types of tasks: differential abundance testing in microbiome data and differential distribution testing in single-cell RNA-seq data. These examples demonstrate that the proposed method behaves robustly across heterogeneous biological settings, consistently eliminating zero $p$-values and yielding smooth, stable tail probabilities even with a limited number of permutations.
A summary of the notation used throughout the paper is provided in Appendix \ref{app:sec:methods:notation}.

%%%%%%%%%%%%%%%%%%%%%%%%%%%%%%%%%%%%%%%%%%
\section{Permutation \texorpdfstring{$p$}{p}-value approximation} 
\label{sec:methods}
%%%%%%%%%%%%%%%%%%%%%%%%%%%%%%%%%%%%%%%%%%

%====================================
\subsection{Permutation testing and empirical \texorpdfstring{$p$}{p}-values} \label{sec:methods:permutation_testing}
%====================================

Permutation tests are a widely used non-parametric approach to hypothesis testing, particularly valuable when the sampling distribution of a test statistic is unknown or difficult to derive analytically. 
They rely on the assumption that, under the null hypothesis, the labels are exchangeable, so that permutations of the data generate realizations from the null distribution of the test statistic.
Given a test statistic $T^{(j)}$ for test $j \in \{1, \dots, m\}$, and its observed value $T_{\text{obs}}^{(j)}$, 
a permutation test constructs a null distribution $\mathcal{T}^{(j)} = \{ T_b^{*(j)} \}_{b=1}^B$ by recalculating the test statistic under $B$ random permutations of the data. 

Throughout this work, we consider right-tailed tests, i.e.\ small $p$-values correspond to large test statistics. In settings where a left-tailed or two-sided test is required, the observed and permuted statistics can be transformed (e.g.\ sign flipping or absolute value) so that tail modeling is always performed on a right tail.
To quantify the strength of evidence against the null hypothesis, the empirical $p$-value of test $j$ is then defined as
\begin{equation}
\label{eq:empirical_p-value}
p_{\mathrm{emp}}^{(j)} 
= \frac{1 + \sum_{b=1}^B \mathbb{I}\!\left(T_b^{*(j)} 
\geq T_{\mathrm{obs}}^{(j)}\right)}{1 + B},
\end{equation}
where $\mathbb{I}(\cdot)$ denotes the indicator function.

The addition of 1 to both the numerator and denominator ensures that the
empirical permutation $p$-value is strictly positive, even when the observed
test statistic exceeds all values in the permutation distribution. This
adjustment has been strongly recommended \citep{north2002note,
phipson2010permutation}, because a permutation $p$-value can never be exactly
zero in principle: the \emph{exact} permutation $p$-value is defined over all
possible permutations of the data and therefore necessarily includes the
observed statistic itself \citep{phipson2010permutation}. A zero empirical
$p$-value is thus purely an artefact of using a limited number of permutations
and can cause difficulties in practice, especially in multiple testing
settings where zero values remain unadjusted.

While this correction guarantees strictly positive $p$-values, it does not
change the fundamental limitation that empirical permutation $p$-values lie on
a discrete grid. With $B$ permutations, the attainable values are multiples of
$1/(B+1)$. When accurate estimation of very small $p$-values is required and the number of attainable permutations is limited, the resulting coarse grid can severely restrict inference in the tail. This limitation motivates the use of parametric tail approximations that extrapolate beyond the empirical permutation range. In the following subsection, we describe the approach for permutation $p$-value approximation based on the Generalized Pareto Distribution (GPD) that underlies the \texttt{permApprox} framework.

%====================================
\subsection{GPD-based \texorpdfstring{$p$}{p}-value approximation}
\label{sec:methods:gpd_tail_approximation}

To approximate small $p$-values in the tail of the permutation distribution, we adopt the GPD tail modeling framework originally proposed by \citet{knijnenburg2009fewer}. 
This approach is grounded in extreme value theory, which states that, for a sufficiently high threshold $u^{(j)}$, the conditional distribution of excesses over $u^{(j)}$ (``peaks-over-threshold``) can be approximated by a Generalized Pareto Distribution \citep{pickands1975statistical, coles2001introduction}.

For each test $j$, let $T_{\mathrm{obs}}^{(j)}$ denote the observed test statistic and $\{T_b^{*(j)}\}_{b=1}^B$ the corresponding permutation statistics.
We select a threshold $u^{(j)}$ that defines the tail region of the permutation distribution.

We refer to permutation values satisfying $T_b^{*(j)} > u^{(j)}$ as \emph{exceedances}. 
Conditional on exceeding the threshold, we define the associated \emph{excesses} as the amount by which the test statistic exceeds the threshold,
\begin{equation}
Y_b^{*(j)} = T_b^{*(j)} - u^{(j)} \;\big|\; T_b^{*(j)} > u^{(j)}, \quad b = 1, \dots, B.
\end{equation}
The collection of permutation excesses is denoted by
\begin{equation}
\mathcal{Y}^{(j)} = \{ Y_b^{*(j)} \}.
\end{equation}
The observed excess is defined analogously as
\begin{equation}
Y_{\mathrm{obs}}^{(j)} = T_{\mathrm{obs}}^{(j)} - u^{(j)} \;\big|\; T_{\mathrm{obs}}^{(j)} > u^{(j)}.
\end{equation}

To extrapolate tail probabilities beyond the empirical permutation range, a GPD is fitted to the set of permutation excesses $\mathcal{Y}^{(j)}$.
The GPD is parameterized by a scale parameter $\sigma > 0$ and a shape parameter $\xi \in \mathbb{R}$, with cumulative distribution function
\begin{equation}
\label{eq:GPD_CDF}
F(y; \sigma, \xi) = 
\begin{cases}
1 - \left(1 + \frac{\xi y}{\sigma} \right)^{-1/\xi}, & \text{if } \xi \neq 0, \\[6pt]
1 - \exp\left(-\frac{y}{\sigma}\right), & \text{if } \xi = 0,
\end{cases}
\qquad y > 0.
\end{equation}
When $\xi < 0$, the distribution has a finite upper support boundary at $s = -\sigma/\xi$.

Let $k^{(j)} = |\mathcal{Y}^{(j)}|$ denote the number of permutation excesses above the threshold $u^{(j)}$.
The empirical probability of exceeding the threshold is estimated as
$
p_{\mathrm{exc}}^{(j)} = \frac{k^{(j)}}{B}.
$
The permutation $p$-value is approximated by combining this exceedance probability with the GPD tail probability of the observed excess,
\begin{equation}
\label{eq:GPD_p-value}
p_{\mathrm{GPD}}^{(j)}
=
p_{\mathrm{exc}}^{(j)}
\times
\bar{F}_{\mathrm{GPD}}\!\left(
Y_{\mathrm{obs}}^{(j)};
\hat{\sigma}^{(j)}, \hat{\xi}^{(j)}
\right),
\end{equation}
where $\bar{F}_{\mathrm{GPD}} = 1 - F_{\mathrm{GPD}}$ denotes the upper tail probability of the fitted GPD.

\paragraph{Parametrization note.}
The GPD appears in multiple parametrizations across the literature. We adopt the standard two-parameter convention from extreme value theory with scale $\sigma>0$ and shape $\xi\in\mathbb{R}$, which implies bounded support when $\xi<0$ \citep{coles2001introduction}. Alternative parametrizations, such as $(a,k)=(\sigma,-\xi)$, are also common, e.g., \citep{castillo2005extreme}. The two formulations are directly related through this transformation.

%====================================
\subsection{Proposed constrained GPD fitting}
\label{sec:methods:constraint}

\begin{figure}[t]
  \centering
  \includegraphics[width=0.95\textwidth]{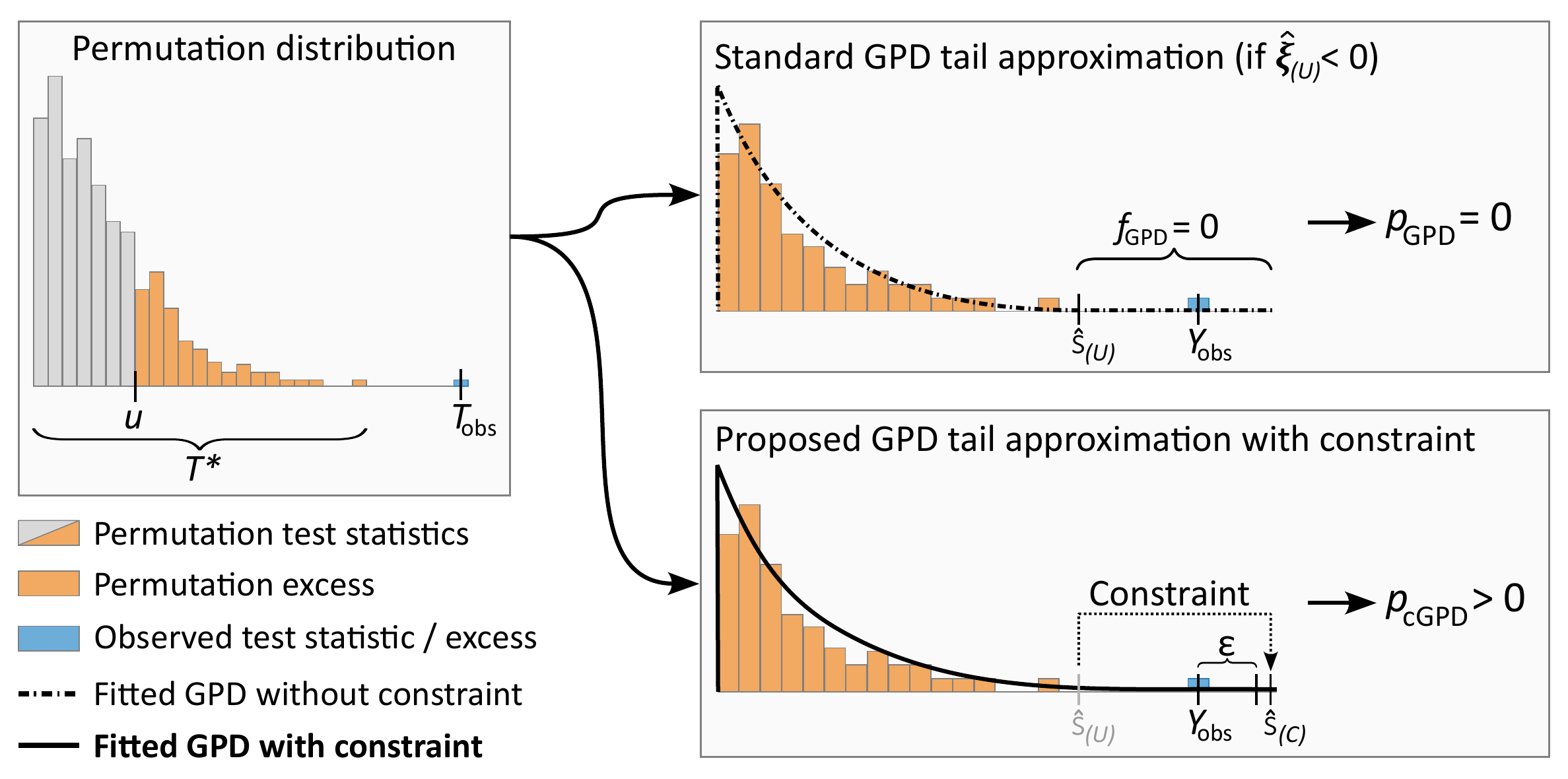}
  \caption{
  \textbf{GPD-based $p$-value approximation.}
  \textbf{Left panel:} Distribution of the permuted test statistic $T^*$ with
  threshold $u$ defining the permutation excesses (orange), and observed test
  statistic $T_{\mathrm{obs}}$.
  \textbf{Upper right:} Standard (unconstrained) GPD tail approximation when the estimated shape
  parameter is negative ($\hat{\xi}_{(U)}<0$), resulting in a finite upper support
  boundary $\hat{s}_{(U)} = -\frac{\hat{\sigma}_{(U)}}{\hat{\xi}_{(U)}}$. If the observed excess $Y_{\mathrm{obs}}$ lies beyond this boundary, the tail approximation assigns zero probability
  mass, yielding a zero GPD-based $p$-value.
  \textbf{Lower right:} Proposed constrained GPD fit, enforcing that the upper
  support boundary $\hat{s}_{(C)}$ lies strictly above the evaluation point, with offset $\varepsilon$. This guarantees a strictly positive tail probability at $Y_{\mathrm{obs}}$.
  }
  \label{fig:constrained_gpd}
\end{figure}

The GPD-based tail approximation provides a principled way to refine small permutation $p$-values beyond the empirical permutation grid. A difficulty arises, however, when the estimated GPD shape parameter is negative. In this case, the fitted GPD has a finite upper support boundary, which can result in zero-valued tail probabilities at extreme observed test statistics. 
Figure~\ref{fig:constrained_gpd} illustrates this issue and our proposed remedy.
For clarity, we omit the test index $j$ throughout this section and write all
quantities for a fixed test.

Let $(\hat{\sigma}_{(U)}, \hat{\xi}_{(U)})$ denote the parameter estimates
obtained from an unconstrained GPD fit to the permutation excesses. When
$\hat{\xi}_{(U)} < 0$, the fitted GPD has a finite upper support boundary given by
%\begin{equation}
$\hat{s}_{(U)} = -\frac{\hat{\sigma}_{(U)}}{\hat{\xi}_{(U)}}$.
%\end{equation}
If the observed excess $Y_{\mathrm{obs}}$ lies at or beyond this boundary, the
tail approximation assigns zero probability mass to the evaluation point,
yielding a zero GPD-based $p$-value. This behavior is visualized in the upper
right panel of Figure~\ref{fig:constrained_gpd}.

To prevent $p$-values from being exactly zero, we propose to enforce a support constraint during GPD parameter estimation. Specifically, we require that the upper support boundary of the fitted GPD lies strictly above the evaluation point,
\begin{equation}
\fbox{$\displaystyle
\hat{s}_{(C)} > Y_{\mathrm{obs}} + \varepsilon
$}
\end{equation}

where $\varepsilon > 0$ is a small, data-dependent safety margin. Enforcing this
constraint yields constrained parameter estimates
$(\hat{\sigma}_{(C)}, \hat{\xi}_{(C)})$ with corresponding support boundary
$\hat{s}_{(C)} = -\hat{\sigma}_{(C)} / \hat{\xi}_{(C)}$. The resulting constrained
GPD fit ensures a strictly positive tail probability at the observed statistic,
as illustrated in the lower right panel of Figure~\ref{fig:constrained_gpd}.
The constrained estimates are substituted into the tail-approximation formula in
\eqref{eq:GPD_p-value}, yielding the constrained GPD-based $p$-value
$p_{\mathrm{cGPD}}$.

%====================================
\subsection{Data-adaptive selection of the safety margin \texorpdfstring{$\varepsilon$}{epsilon}}
\label{sec:methods:epsilon}
%====================================

The safety margin $\varepsilon$ determines how far the support constraint extends beyond the observed excess and is therefore central to numerically stable tail evaluation under the constrained GPD fit. If $\varepsilon$ is too small, the evaluation point may lie extremely close to the fitted upper endpoint, which can cause numerical instability when computing $\bar F_{\mathrm{GPD}}$. If $\varepsilon$ is too large, the constraint pushes the endpoint far into the upper tail and yields overly conservative extrapolation.

In principle, $\varepsilon$ could be chosen as a simple function of the observed test statistic, for example as a fixed proportion of $T_{\mathrm{obs}}$. However, exploratory analyses revealed that such ad hoc rules are insufficient: numerical instability may occur not only for extreme test statistics, but already at moderate distances from the tail threshold.
A comparison of different $\varepsilon$ rules is provided in Appendix~\ref{app:sec:methods:eps_comparison}.
This motivated the development of a data-adaptive construction that (i) is invariant to the scale of the test statistic, (ii) saturates for extreme values, and (iii) provides additional stabilization in the region where tail extrapolation is most sensitive.

We therefore propose the \emph{Standardized Lifted Log-Saturation} (SLLS) rule. It is constructed on a standardized $Z$-scale and then mapped back to the scale of the test statistic. For test $j$, let			 
\begin{equation} 
Z_{\mathrm{obs}}^{(j)}=
\frac{\lvert T_{\mathrm{obs}}^{(j)}-\mu_j\rvert}{\sigma_j}
\end{equation}
denote the standardized observed statistic, where $\mu_j$ and $\sigma_j$ are the mean and standard deviation of the corresponding permutation distribution. We then define a standardized cap $Z_{\mathrm{cap}}>0$ to set a common reference range. In multiple testing settings, a natural choice is $Z_{\mathrm{cap}}=\max_{j} Z_{\mathrm{obs}}^{(j)}$. For single tests, few tests, or robustness against extreme observed statistics, we recommend a permutation-based cap, e.g., a high quantile of the standardized permutation distribution.
This yields the scale-free position
\begin{equation}
l_j=\frac{Z_{\mathrm{obs}}^{(j)}}{Z_{\mathrm{cap}}}\in[0,1].
\end{equation}

The SLLS rule defines a \emph{standardized} safety margin $\varepsilon_j^\ast$ as a smooth function of the relative position $l_j$, with sample size-dependent curvature and plateau parameters:

\begin{equation}
\varepsilon_j^{\ast}
=
\max\!\left\{
\varepsilon^{\ast}_{\max}(n)\,
\frac{\log\!\big(1+\kappa(n)\,l_j\big)}{\log\!\big(1+\kappa(n)\big)}
\;+\;
\rho_{\mathrm{lift}}\,
\psi(l_j)
\;,\;
\varepsilon_{\min}
\right\},
\label{eq:epsilon_standardized}
\end{equation}
where the construction combines a logarithmic saturation component with an additional localized lift.

The logarithmic term induces a monotone increase in $\varepsilon_j^\ast$ as $l_j$ increases, with smooth saturation at the plateau $\varepsilon^{\ast}_{\max}(n)$ as $l_j \to 1$. This reflects the empirical requirement that the safety margin must grow with an increasing observed statistic, while remaining bounded to avoid unnecessary conservativeness for very large values. The curvature of the transition is controlled by $\kappa(n)$. Larger values lead to a steep initial increase of $\varepsilon_j^\ast$ for small $l_j$, with saturation reached early, whereas smaller values produce a more gradual, approximately linear growth up to the cap.

While this saturated logarithmic component stabilizes tail evaluation for extreme statistics, simulation studies showed that it can yield slightly underestimated $p$-values for test statistics in an intermediate range, where tail extrapolation is already active but saturation has not yet been reached (see also Figure~\ref{app:fig:eps_comparison_ttest}, panels~(c) and~(d)). To correct this behavior without altering the asymptotic plateau, an additional localized lift is added via the function $\psi(l_j)$. 
The lift is defined using the compactly supported Wendland $C^2$ kernel from the $d=3$ Wendland family \citep{wendland1995piecewise},
\begin{equation}
\psi(l) = (1-l)^4(1+4l), \qquad l\in[0,1],
\label{eq:wendland}
\end{equation}
which is smooth, nonnegative, and vanishes together with its first two derivatives at $l=1$. These properties ensure that the lift increases the safety margin primarily for small to moderate values of $l_j$, while fading out smoothly near the cap and leaving the saturation behavior for extreme statistics unchanged.

To adapt both curvature and plateau height to sample size, the parameters are scaled as
\begin{equation}
\kappa(n)=\kappa_{\mathrm{factor}}\frac{n_{\mathrm{ref}}}{n},
\qquad
\varepsilon^\ast_{\max}(n)=\tau\,\frac{n_{\mathrm{ref}}}{n},
\label{eq:epsilon_scaling}
\end{equation}
where $n$ denotes the per-group sample size and $n_{\mathrm{ref}}$ is a fixed reference sample size. When group sizes differ, we set $n=\min(n_1,n_2)$ to obtain a conservative scaling. The reference value $n_{\mathrm{ref}}$ is fixed to $n_{\mathrm{ref}}=500$ and defines a pivot at which the strength of the regularization transitions from more conservative behavior for smaller samples to weaker regularization for larger samples. 
The constants $\kappa_{\mathrm{factor}}$ and $\tau$ control the curvature and plateau height of the saturation component, respectively, while $\rho_{\mathrm{lift}}$ determines the magnitude of the lift at $l_j=0$. A small lower bound $\varepsilon_{\min}>0$ prevents the margin from vanishing. We propose $\kappa_{\mathrm{factor}}=1000$ and $\tau=0.25$, which yielded stable behavior across a diverse range of settings.

Finally, the test-specific safety margin on the original scale of the test statistic is obtained by rescaling with the permutation standard deviation,
\begin{equation}
\varepsilon_j=\sigma_j\,\varepsilon_j^\ast.
\label{eq:epsilon_final}
\end{equation}

The empirical motivation and stepwise derivation of the SLLS rule, including the role of logarithmic growth, saturation, and the Wendland-based lift, are documented in detail in Appendix~\ref{app:sec:methods:eps_derivation}. In rare cases, numerical underflow may persist for extreme test statistics. To avoid zero $p$-values in these cases, $\varepsilon$ can be refined via the procedure described in Appendix~\ref{app:sec:methods:epsilon_refinement}. 

%%%%%%%%%%%%%%%%%%%%%%%%%%%%%%%%%%%%%%%%%%
\section{The \texttt{permApprox} workflow}
%%%%%%%%%%%%%%%%%%%%%%%%%%%%%%%%%%%%%%%%%%

\subsection{Overview and scope of the workflow}
\label{sec:workflow:overview}
%====================================

The proposed workflow provides a practical framework for approximating small permutation $p$-values using constrained GPD tail modeling. It is implemented in the open-source \texttt{permApprox} \texttt{R} package. While the statistical core of the approach is the support-constrained GPD fit described in Section~\ref{sec:methods}, accurate permutation-based inference requires a sequence of additional decisions that extend beyond the GPD fitting step itself: (i) which empirical permutation $p$-values are eligible for tail approximation, (ii) how the tail region of the permutation distribution is defined, (iii) which methods are appropriate for GPD parameter fitting, and (iv) how the support constraint is incorporated. The \texttt{permApprox} workflow integrates these decision points into a coherent and reproducible analysis pipeline.

\begin{figure}[t]
  \centering
  \includegraphics[width=\textwidth]{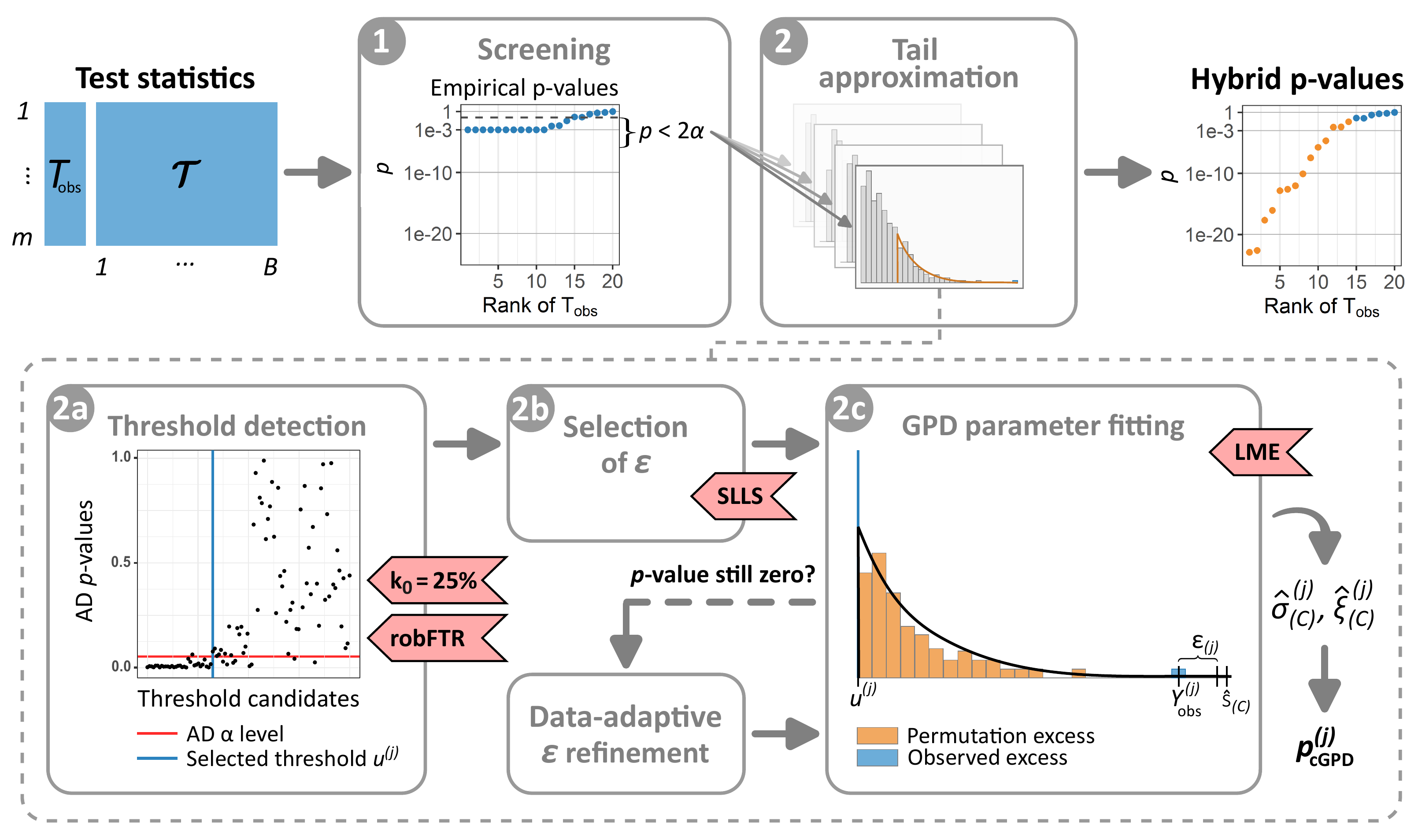}
  \caption{Graphical representation of the \texttt{permApprox} workflow.
  \textbf{Input:} A vector with observed test statistics $T_{\mathrm{obs}}$ and a matrix with permuted test statistics $\mathcal{T} = {T_b^{*(j)}}$ for $B$ permutations and $m$ tests.
  \textbf{Step 1:} Empirical permutation $p$-values are computed and screened for small values relative to the chosen significance level $\alpha$ (i.e., $p_{\mathrm{emp}} < 2\alpha$), with selected tests flagged for GPD-based tail approximation.
  \textbf{Step 2:} For a selected test $j$, the tail approximation stage consists of three main steps:
  \textbf{Step 2a:} A threshold $u^{(j)}$ is chosen via the Anderson Darling (AD) based test to define the tail region of the permutation distribution.
  \textbf{Step 2b:} The safety margin $\varepsilon^{(j)}$ is selected (with optional refinement in rare cases of machine-underflow). 
  \textbf{Step 2c:} A GPD is fitted to the permutation excesses under a support constraint ensuring valid evaluation beyond the observed excess.
  The fitted constrained GPD yields refined tail $p$-values $p_{\mathrm{cGPD}}^{(j)}$, which replace the corresponding empirical values.
  \textbf{Output:} Vector of hybrid $p$-values that are discrete in the bulk of the distribution and continuous in the extreme tail. The \textbf{red flags} mark our suggested methods as explained in Section~\ref{sec:methods:default_config}
  }
  \label{fig:main_workflow}
\end{figure}

Figure~\ref{fig:main_workflow} provides a schematic overview of the main steps of the workflow. The following subsections describe how these components are combined into a practical analysis pipeline and outline the methodological options available at each stage. While a recommended default configuration is summarized in Section~\ref{sec:methods:default_config}, the workflow is implemented as a flexible toolbox that allows users to substitute alternative methods at each stage.

%=========================================
\subsection{Screening for tail approximation}
\label{sec:workflow:screening}
%=========================================

The first decision in the \texttt{permApprox} workflow concerns whether an empirical permutation $p$-value should be refined using parametric tail modeling. Empirical permutation $p$-values are exact under the null hypothesis, but their granularity is limited by the number of permutations $B$, which primarily affects inference in the extreme tail. Tail approximation is therefore applied only to sufficiently small empirical $p$-values. Restricting the model to the extreme tail saves computation time and avoids incorporating regions of the permutation distribution that are not well described by a GPD. 

Accordingly, the workflow applies a screening step (Step~1 in Figure~\ref{fig:main_workflow}) based on the empirical permutation $p$-values. For each test $j$, an empirical $p$-value $p_{\mathrm{emp}}^{(j)}$ is computed as formalized in \eqref{eq:empirical_p-value}. Only tests with $p_{\mathrm{emp}}^{(j)} < p_{\mathrm{thr}}$ are selected for GPD-based tail approximation, where $p_{\mathrm{thr}}$ is a user-specified threshold. We follow \citet{winkler2016faster}, who suggest using twice the chosen significance level: $p_{\mathrm{thr}} = 2\,\alpha$.
For tests selected for tail approximation, the GPD fitted under the support constraint is used to compute a refined tail $p$-value $p_{\mathrm{cGPD}}^{(j)}$ (Step~2 in Figure~\ref{fig:main_workflow}), which replaces the corresponding empirical permutation value. Tests that are not selected retain their empirical $p$-values. This leads naturally to a hybrid construction that combines empirical and GPD-based components.

Specifically, for each test $j$, we define the hybrid $p$-value
\begin{equation}
\label{eq:hybrid_pvalue}
p^{(j)} :=
\begin{cases}
p_{\mathrm{emp}}^{(j)}, & \text{if } p_{\mathrm{emp}}^{(j)} \ge p_{\mathrm{thr}}, \\[6pt]
p_{\mathrm{cGPD}}^{(j)}, & \text{if } p_{\mathrm{emp}}^{(j)} < p_{\mathrm{thr}}.
\end{cases}
\end{equation}

The resulting vector $\{p^{(j)}\}_{j=1}^m$ is discrete in the bulk of the distribution, where empirical permutation $p$-values are sufficiently resolved, and continuous in the extreme tail, where parametric refinement is most beneficial. These hybrid $p$-values form the primary output of the \texttt{permApprox} workflow.

%====================================
\subsection{GPD threshold selection strategies}
\label{sec:methods:threshold}
%====================================

Fitting a GPD to the tail of a permutation distribution requires specifying a threshold $u$ above which the exceedances are assumed to follow a GPD. Threshold selection (Step~2a in Figure~\ref{fig:main_workflow}) is a central problem in extreme value theory and involves a well-known bias-variance trade-off: setting $u$ too low may include parts of the bulk that is not well described by a GPD, while setting it too high leave too few exceedances for reliable parameter estimation \citep{scarrott2012review, langousis2016threshold}.

In the \texttt{permApprox} framework, threshold selection naturally separates into two components. First, an initial tail region is defined by specifying a starting threshold $u_0$ or, equivalently, an initial number of exceedances $k_0$. This step determines the range of thresholds over which the GPD assumption is evaluated and ensures that a sufficient number of exceedances is available for stable fitting. The impact of different choices for the starting number of exceedances $k_0$, including fixed values and relative proportions of the permutation sample, is systematically investigated in Appendix~\ref{app:sec:default:exceed0}.
In a second step, the final threshold is selected within this candidate range using a data-driven criterion that assesses the adequacy of the GPD approximation.

We consider and compare five threshold selection strategies in this work. All rely on fitting the GPD at a range of candidate thresholds and evaluating goodness-of-fit (GoF) statistics or shape parameter stability. The GoF tests are based on the Anderson-Darling (AD) test statistic tailored for GPDs \citep{choulakian2001goodness}. A detailed comparison of the threshold detection strategies based on simulation studies is provided in Appendix~\ref{app:sec:default:thresh_methods}.
For valid tail extrapolation, the threshold search is restricted to values below the observed test statistic, with the additional requirement that a minimal number of exceedances remains above the threshold. If no candidate threshold satisfies the selection criterion, the empirical permutation $p$-value is retained as a fallback.

The methods are as follows:

\begin{itemize}
  \item \textbf{Failure-to-reject (FTR):} 
  A commonly used heuristic in extreme value theory in which candidate thresholds are evaluated in increasing order along the permutation tail. FTR selects the lowest threshold at which the null hypothesis of the AD test is not rejected, under the assumption that a valid GPD fit begins once the GoF test no longer indicates lack of fit \citep{choulakian2001goodness}.

  \item \textbf{Robust FTR (robFTR):} 
  A stricter variant of FTR in which a threshold is accepted only if the AD test is not rejected at three successive candidate thresholds. This guards against premature acceptance caused by isolated fluctuations in the AD $p$-values and yields more stable threshold selection.

  \item \textbf{ForwardStop:} Inspired by the method of \citet{gsell2016sequential} and adapted for threshold selection by \citet{bader2018automated}, this approach interprets the sequence of ASD $p$-values as an ordered hypothesis testing problem. We compute the ForwardStop statistic \( \bar{Y}_k = \frac{1}{k} \sum_{i=1}^k -\log(1 - p_i) \), and select the smallest index $k$ for which \( \bar{Y}_k > \alpha \). This provides a formal stopping rule while controlling the false discovery rate.

  \item \textbf{GoF $p$-value change point detection:} To detect structural changes in the GoF behavior, we apply change point analysis \citep{killick2012optimal} to the sequence of AD $p$-values. We prepend a set of synthetic small $p$-values (to simulate poor fits) and use the \texttt{changepoint} \texttt{R} package \citep{killick2024changepoint} to detect the point at which GoF improves. The first accepted threshold beyond this change point is selected.

  \item \textbf{Shape-variation:} An established heuristic in extreme value analysis is to select thresholds where the estimated shape parameter $\hat{\xi}$ stabilizes across increasing thresholds \citep{scarrott2012review, northrop2014improved}. This motivates our shape-variation approach, which selects the threshold that minimizes the rolling variance of $\hat{\xi}$ across accepted thresholds. By formalizing the stability principle, the method provides a simple, data-driven alternative to visual inspection.
\end{itemize}

%====================================
\subsection{GPD parameter estimation methods}  
\label{sec:methods:gpd_estimators}

Once the threshold has been selected and the tail region defined, a test-specific safety margin $\varepsilon$ is determined via the SLLS rule to ensure valid tail evaluation under the support constraint, as described in Section~\ref{sec:methods:epsilon} (Step~2b in Figure~\ref{fig:main_workflow}). In cases of machine-underflow, the $\varepsilon$ refinement described in Appendix~\ref{app:sec:methods:epsilon_refinement} can be used to avoid zero $p$-values. With the exceedances and $\varepsilon$ fixed, the final step of the tail approximation stage is to fit a GPD to the resulting excesses (Step~2c in Figure~\ref{fig:main_workflow}).

We evaluate several established techniques for GPD parameter estimation, each of which takes the excesses $\mathcal{Y}$ as input and returns parameter estimates $(\hat{\sigma}, \hat{\xi})$. Our goal is to obtain stable and accurate estimates suitable for tail extrapolation, particularly in the presence of the support constraint introduced in Section~\ref{sec:methods:constraint}.
All methods except the method of moments (MOM) have been adapted to incorporate this constraint. Depending on the estimation approach, the constraint is enforced by restricting the parameter space, modifying the optimization criterion, or adjusting internal candidate values.

Below, we summarize the GPD parameter estimation methods considered in this work, together with the abbreviations used throughout the paper. The performance of these methods is
evaluated in a dedicated simulation study in Appendix~\ref{app:sec:default:estimators}.

\newpage
\begin{itemize}
  \item \textbf{Method of Moments [MOM]:} Introduced by \citet{hosking1987parameter}, this approach estimates GPD parameters by equating theoretical and sample moments. It is simple and computationally efficient but does not involve optimization and cannot incorporate support constraints. We include it only for comparison in the unconstrained setting.

  \item \textbf{Maximum Likelihood Estimation [MLE2D]:} The classical two-parameter MLE \citep{pickands1975statistical} maximizes the log-likelihood over both shape and scale. It is widely used in extreme value theory and serves as a standard reference estimator.

  \item \textbf{One-Parameter MLE [MLE1D]:} \citet{castillo2015one} propose a reparametrization based on the upper support boundary $s = -\sigma/\xi$. The method reduces estimation to a one-dimensional optimization problem, offering computational advantages and greater numerical stability in some scenarios.

  \item \textbf{Likelihood Moment Estimation [LME]:} This hybrid method by \citet{zhang2007maximum} combines likelihood and moment principles and was specifically developed to improve robustness in small samples and to stabilize estimation near the boundary.

  \item \textbf{Zhang and Stephens Estimator [ZSE]:} An extension of LME that incorporates grid-based prior-like stabilization \citep{zhang2009likelihood}, with the goal of improving accuracy and reliability, especially in moderate sample sizes.

  \item \textbf{Two-Step Nonlinear Least Squares Estimator [NLS2]:} Proposed by \citet{song2012threshold}, this estimator fits a second-order tail approximation using nonlinear least squares. It was designed to improve performance in small-sample settings and to better handle bounded support.

  \item \textbf{Weighted Nonlinear Log-Least Squares Estimator [WNLLSM]:} A two-step estimator introduced by \citet{zhao2019wnllsm} that emphasizes accuracy in the tail region by applying weights in a log-transformed least squares framework.
\end{itemize}

%=========================================
\subsection{Default \texttt{permApprox} configuration and implementation}
\label{sec:methods:default_config}
%=========================================

The final configuration of the \texttt{permApprox} workflow represents the outcome of several dedicated simulation studies (see Appendix~\ref{app:sec:default}), each addressing one specific component of the procedure. 
These studies were conducted to identify the most accurate, robust, and computationally efficient settings for practical use. Our proposed configuration is as follows:

\begin{itemize}

    \item \textbf{Initial number of exceedances:} The threshold search starts with the top 25\% of permutation values ($k_0 = 0.25B$), but includes at least 250 exceedances to ensure sufficient data for stable fitting. This configuration yielded the lowest bias and variance across simulated scenarios (Supplementary Study~\ref{app:sec:default:exceed0}).

    \item \textbf{GPD threshold detection:} The robFTR method in combination with the Anderson-Darling goodness-of-fit test is employed for threshold selection, as it achieved the best trade-off between robustness to outliers and computational efficiency. A detailed comparison of threshold detection strategies is provided in Supplementary Study~\ref{app:sec:default:thresh_methods}. The empirical $p$-value is used as fall-back, when the test is rejected for all candidate thresholds. 
      
    \item \textbf{Estimator:} We use the likelihood moment estimator (LME) as the default method for GPD parameter fitting. While any of the estimators evaluated in this work could, in principle, be applied, LME provides a favorable balance between estimation accuracy, robustness under support constraints, and computational efficiency in our simulation studies (Section~\ref{app:sec:default:estimators}).
    
    \item \textbf{Constraint:} The GPD is fitted under the boundary constraint described in Section~\ref{sec:methods:constraint}, using the standardized lifted log-saturation (SLLS) rule for determining $\varepsilon$. The data-adaptive epsilon refinement procedure (Section~\ref{app:sec:methods:epsilon_refinement}) is applied only in the real-world applications, but not in the simulation studies.

\end{itemize}

The proposed workflow is implemented in the open-source \texttt{R} package \texttt{permApprox}, which can be accessed at \url{https://github.com/stefpeschel/permApprox}.
The above configuration defines the package’s default settings, and was applied in the simulation study and real-world application in Sections~\ref{sec:simulations} and \ref{sec:application}. 
Additionally, the package provides all of the alternative $p$-value approximation, GPD estimation, and threshold detection methods evaluated in this work. This allows users to reproduce, compare, and extend our analyses.

%=========================================
\subsection{Remarks on single and multiple testing}
\label{sec:methods:multiple_testing}
%=========================================

The \texttt{permApprox} framework can be applied both in single-test settings and in large-scale inference problems involving many parallel hypotheses. In the single-test setting, the output is a single hybrid $p$-value, whereas in the multiple testing setting it is a vector ${p^{(j)}}_{j=1}^m$ of hybrid $p$-values as defined in \eqref{eq:hybrid_pvalue}.

In multiple testing scenarios, the discreteness of empirical permutation $p$-values has motivated the development of several false discovery rate (FDR) procedures tailored specifically to discrete $p$-values \citep{gilbert2005modified,jiang2017discrete}. While classical procedures such as the Benjamini-Hochberg (BH) method  \citep{benjamini1995controlling} remain valid in this setting, they can exhibit conservative behavior because discrete null $p$-values tend to be stochastically larger than uniformly distributed ones \citep{benjamini2001control}.

The hybrid $p$-values produced by \texttt{permApprox} retain this discrete structure for moderate and large values, while they provide a smooth, continuous approximation in the extreme tail, where fine resolution is most critical. As a result, the hybrid $p$-values can often be treated analogously to continuous $p$-values in downstream inference, i.e., standard multiple testing procedures for continuous tests may be applied directly.

An important distinction between single and multiple testing settings arises in the selection of the cap $Z_{\mathrm{cap}}$ used in the data-adaptive $\varepsilon$ rule (Section~\ref{sec:methods:epsilon}). In multiple testing scenarios, the cap is defined by the maximum observed test statistic across all hypotheses, yielding a conservative and robust constraint. For single tests or small numbers of hypotheses, a permutation-based cap is used instead, providing greater stability in the presence of extreme observations.

%%%%%%%%%%%%%%%%%%%%%%%%%%%%%%%%%%%%%%%%%%
\section{Simulation studies}
\label{sec:simulations}
%%%%%%%%%%%%%%%%%%%%%%%%%%%%%%%%%%%%%%%%%%

In this section, we evaluate the performance of \texttt{permApprox} in comparison to several existing approaches for approximating permutation $p$-values. 
We consider both a two-sample \textit{t}-test on Gaussian data and a Wilcoxon rank-sum test on exponential data, thus covering parametric and nonparametric test settings. 
The goal is to assess the accuracy of each method’s $p$-value approximation in finite-sample scenarios, particularly in situations where the true $p$-values are extremely small but the number of permutations is limited.

%=========================================
\subsection{Compared methods}
\label{sec:sim:methods}
%=========================================

The following $p$-value approximation methods are compared. 
For all approaches, the fit is applied only to tests with empirical $p$-values below 0.1, ensuring that the tail approximation is used only when necessary. Empirical $p$-values are also used as fall-back for the GPD-based methods, if the goodness-of-fit test is rejected for all candidate thresholds. 

\begin{itemize}
    \item \textbf{Gamma:} 
    A Gamma distribution is fitted to the entire permutation distribution ($T_\text{obs}$ not included), following the approach of \citet{winkler2016faster}. The parameters of the Gamma distribution are estimated via MLE. The $p$-value is obtained as the upper-tail probability of the fitted Gamma distribution evaluated at the observed test statistic. 

    \item \textbf{GPD (Knijnenburg):} 
    The GPD tail approximation method introduced by \citet{knijnenburg2009fewer}.
    The original paper does not specify whether one- or two-parameter maximum likelihood estimation is used. We use MLE1D as it is more stable in practice. Following \citet{knijnenburg2009fewer}, FTR with AD test is used for threshold detection, starting with the top 250 exceedances.

    \item \textbf{GPD (Knijnenburg, pow3):} 
    Same as above, but all test statistics are raised to the third power prior to fitting, as proposed by \citet{knijnenburg2009fewer} to address issues arising from negative shape parameters and bounded GPD support. 

    \item \textbf{GPD (Winkler, +obs):}
    \citet{winkler2016faster} compare GPD tail fits both \emph{with} and \emph{without} including the observed test statistic in the permutation distribution. We only evaluate the ``+obs'' variant here as candidate to strictly avoid zero $p$-values. The authors use MOM for parameter estimation, which performs very poorly in some scenarios in our simulations (see Appendix~\ref{app:sec:default:estimators}). We use the standard one-parameter maximum likelihood estimate (MLE1D) instead. For threshold detection, they also use FTR with AD test, but start with the upper quartile of the permutation distribution.

    \item \textbf{GPD (permApprox):} 
    The proposed constrained GPD tail-approximation workflow with default configuration as described in Section~\ref{sec:methods:default_config}.
\end{itemize}

%=========================================
\subsection{Two-sample \textit{t}-test with Gaussian data}
\label{sec:sim:ttest}
%=========================================

To evaluate the accuracy and stability of permutation-based $p$-value approximations under controlled conditions, we first consider a two-sample \textit{t}-test applied to Gaussian data. 
This scenario serves as a benchmark, since the exact theoretical $p$-values are known and the \textit{t}-test is optimally suited to the underlying data-generating model. 
It therefore allows a direct assessment of how well each approximation method reproduces the true $p$-values when only a limited number of permutations is available.

We generated two groups of normally distributed data with equal sample sizes $n \in \{25, 50, 100, 250, 500, 1000\}$ and standard deviation $\sigma = 1$. 
For each $n$, we considered effect sizes $d \in \{0.5, 1, 1.5, 2\}$, corresponding to the standardized mean difference between the two groups. 
For each configuration, $B \in \{500, 1000, 5000, 10000\}$ random label permutations were performed to generate the null distribution of the test statistic. 
A total of 1000 independent repetitions were conducted per setting to assess variability. All approximation methods described in Section~\ref{sec:sim:methods} were applied to the simulated data. 

\begin{figure}[ht]
  \centering
  \includegraphics[width=\textwidth]{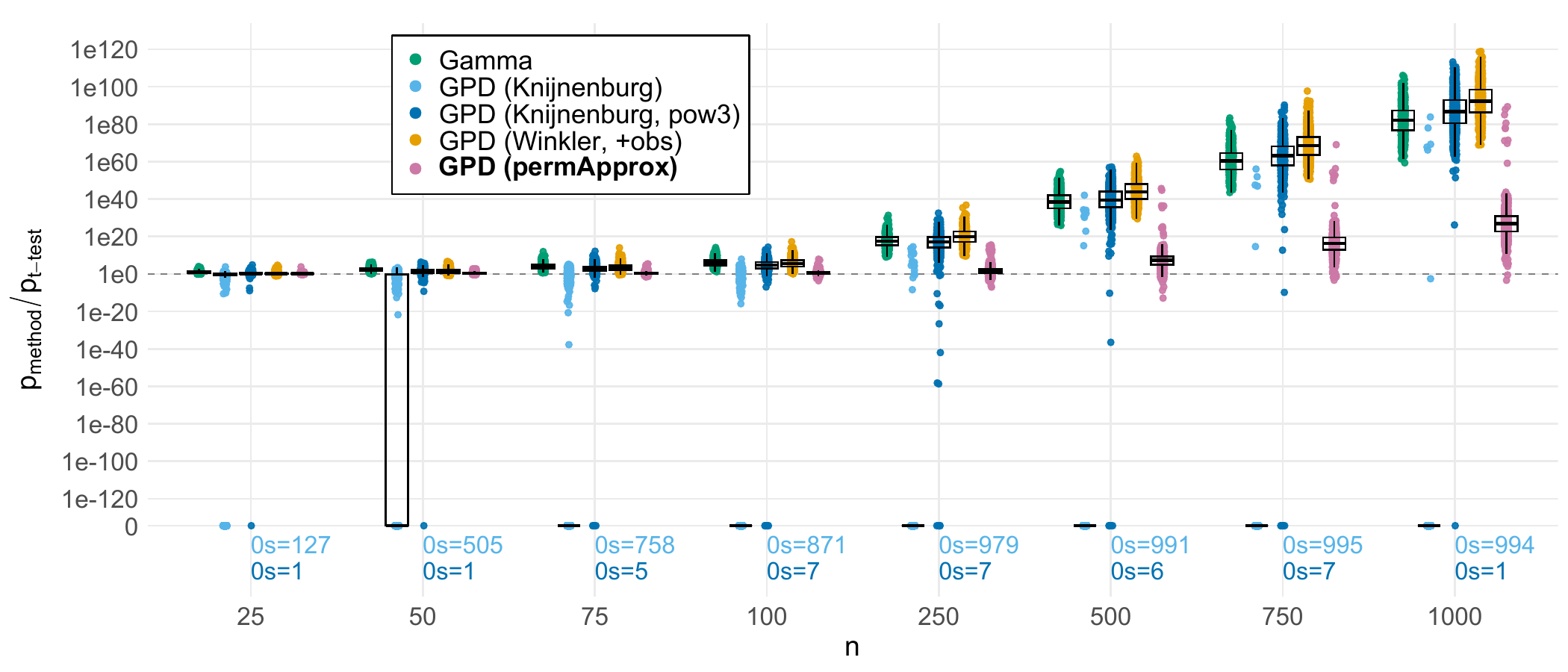}
  \caption{\textbf{Two-sample t-test with Gaussian data:} Ratios of approximated to ground-truth $p$-values: $p_{\text{method}} / p_{t\text{-test}}$, stratified by sample size $n$. The horizontal dashed line at 1 indicates perfect agreement with the \textit{t}-test reference. Effect size and number of permutations are fixed to $d=1$ and $B=1000$, respectively. Points indicate individual replicates (1000 per setting). Counts of exact zeros are shown below each group as ``\texttt{0s}=x'' in the corresponding color. For plotting only, zero $p$-values are mapped to a small constant floor so they appear at ``0'' on the $y$-axis. The y-axis is on a log$_{10}$ scale, while tick labels show the original ratios.}
  \label{fig:ttest_ratios_by_n}
\end{figure}

Figure~\ref{fig:ttest_ratios_by_n} displays the ratios of approximated permutation $p$-values to the corresponding theoretical \textit{t}-test $p$-values for $d=1$ and $B=1000$ across sample sizes. Ratios close to 1 indicate good agreement with the reference, whereas ratios above 1 correspond to conservative behavior. For small sample sizes, all methods yield ratios centered close to 1. As $n$ increases, however, pronounced differences between the approximation methods emerge.

The Gamma approximation exhibits steadily increasing ratios, reflecting its conservative nature due to modeling the full permutation distribution rather than only the tail. Including the observed statistic in the permutation distribution (``GPD (Winkler, +obs)'') further amplifies this effect, leading to even larger ratios at moderate and large sample sizes. The classical GPD approach of \citet{knijnenburg2009fewer} produces smaller, less conservative $p$-values, but quickly generates zeros even for moderate sample sizes. Already at $n=50$, nearly half of the approximated $p$-values are zero, and for large $n$, almost all vanish to zero due to negative shape estimates (the estimated shape values are provided in Supplementary Figure~\ref{app:fig:ttest_shapes_by_n}).  Raising all test statistics to the third power, as suggested by the authors, largely prevents zero $p$-values but leads, again, to overly conservative approximations. 	   
In contrast, the proposed \texttt{permApprox} workflow strictly avoids zero $p$-values by construction. While its ratios increase moderately with sample size, they remain substantially closer to 1 than those of the competing methods across all settings, indicating comparatively stable behavior relative to the \textit{t}-test reference.

The corresponding $p$-values for this setting are shown in Supplementary Figure~\ref{app:fig:ttest_pvals_by_n}. Additional Supplementary Figures~\ref{app:fig:ttest_pvals_by_d_and_B} and \ref{app:fig:ttest_ratios_by_d_and_B} provide $p$-values and ratios stratified by effect size $d$ and number of permutations $B$. The results are consistent across different effect sizes, but slightly change with the number of permutations: With more permutations, the number of zero $p$-values decreases for the classical Knijnenburg approach, but increases when test statistics are raised to the third power. Including the observed test statistic in the null distribution becomes less conservative with increasing $B$. The Gamma approximation is almost unaffected by~$B$.

%=========================================
\subsection{Wilcoxon rank-sum test with exponential data}
\label{sec:sim:wilcoxon}
%=========================================

To assess the robustness of the proposed $p$-value approximation method beyond the Gaussian setting, we repeated the simulation study of Section~\ref{sec:sim:ttest} with the nonparametric Wilcoxon rank-sum test (using the Mann-Whitney $U$ statistic). This setting departs from the assumptions of the \textit{t}-test and involves asymmetric, heavy-tailed data. In this case, inference relies on the permutation null distribution of the Wilcoxon rank-sum statistic, which does not admit a simple parametric form.
We generated two-sample data from exponential distributions with means $1$ and $1+d$, where $d \in \{0, 0.5, 1, 1.5, 2\}$ controls the effect size. 
All five approximation methods considered in the Student's \textit{t}-test setting are applied again here, namely the Gamma approximation, the classical GPD method of \citet{knijnenburg2009fewer}, its power-transformed variant (``pow3''), the ``\,+obs'' variant following \citet{winkler2016faster}, and our proposed \texttt{permApprox} framework.

\begin{figure}[!ht]
  \centering
  % Top panel: single repetition
  \begin{subfigure}[t]{\textwidth}
    \centering
    \subcaption{Single-scenario simulation with 1000 tests (200 per effect size), $n=250$, and $B=1000$.}
    \includegraphics[width=\textwidth]{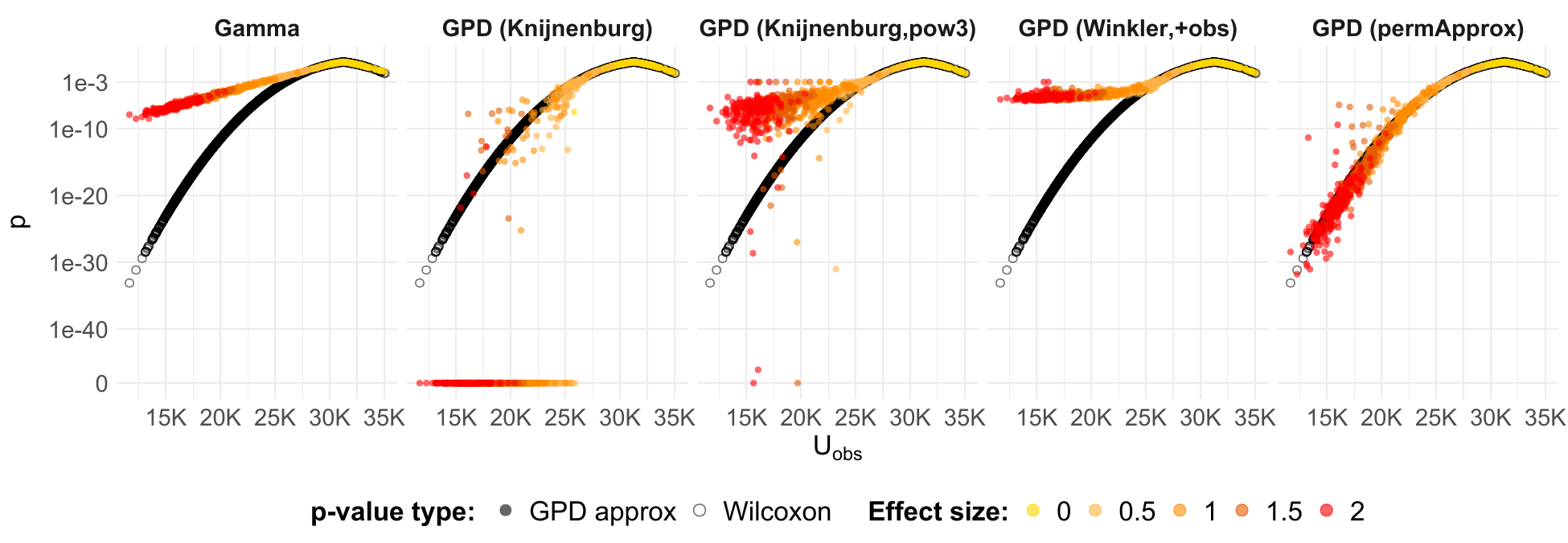}
    \label{fig:wilcox_single}
  \end{subfigure}
  
  % Bottom panel: multiple repetitions
  \begin{subfigure}[t]{0.95\textwidth}
    \centering
    \subcaption{Simulation with 1000 replicates (1 test per replicate) for $d=1$ and $B=1000$.}
    \includegraphics[width=\textwidth]{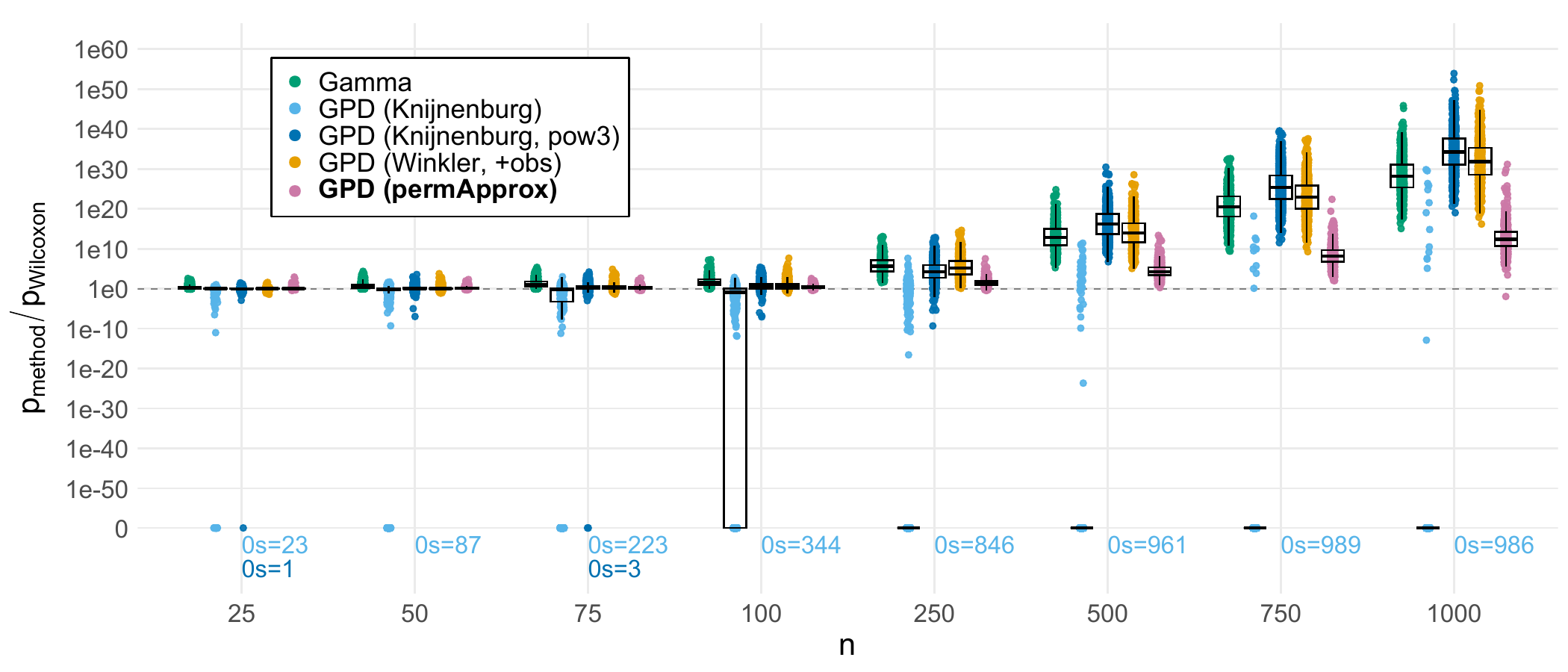}
    \label{fig:wilcox_multi}
  \end{subfigure}
  
  \caption{\textbf{Wilcoxon rank-sum test with exponential data:} Comparison of $p$-value approximations to the Wilcoxon reference.
  \textbf{(a)} Illustrative simulation scenario with 1000 tests across five effect sizes ($d \in \{0, 0.5, 1, 1.5, 2\}$) at $n=250$ samples per group. The x-axis shows the observed Mann-Whitney $U$ statistic, and points represent individual tests. The y-axis displays $p$-values on a log$_{10}$ scale, while tick labels show the original $p$-values.
  \textbf{(b)} Simulation study with 1000 independent replicates per sample size ($d=1$, $B=1000$). The x-axis indicates the per-group sample size, and the y-axis shows ratios of approximated to Wilcoxon $p$-values ($p_{\text{method}} / p_{\text{Wilcoxon}}$) on a log$_{10}$ scale (tick labels show the original ratios). The horizontal dashed line at 1 indicates agreement with the Wilcoxon reference. Points represent individual replicates (1000 per setting).
  In panel~(b), counts of exact zero $p$-values produced by each method are shown below each group as ``\texttt{0s}=x'' in the corresponding color. For visualization only, zero $p$-values are mapped to a small positive constant so they appear at the lower plotting boundary.
  }
  \label{fig:wilcox_two_panel}
\end{figure}

Figure~\ref{fig:wilcox_two_panel} shows the results of two different simulation studies.
Panel~(a) shows an illustrative simulation scenario with 1000 tests (200 per effect size), generated for a single fixed data configuration using $n=250$ samples per group and $B=1000$ permutations.
Rather than aggregating results over repeated Monte Carlo replicates, this panel is intended to provide a direct, point-wise comparison of the five approximation methods against the Wilcoxon reference $p$-values across a range of effect sizes.
As in the Gaussian setting, the classical GPD approach produces numerous zero $p$-values for stronger effects due to GPD support violations.
The power-transformed variant largely avoids zero estimates but exhibits pronounced conservativeness.
Both the ``Winkler,\,+obs'' approach and the Gamma approximation avoid zeros by construction, but yield even more conservative $p$-values.
In contrast, \texttt{permApprox} produces smooth approximations that closely track the Wilcoxon $p$-values across all effect sizes, without introducing zeros.

Panel~(b) presents a comprehensive simulation study with 1000 independent replicates per setting. In each replicate, a single test was conducted, with effect size $d=1$, $B=1000$ permutations, and varying per-group sample sizes from $n=25$ to $n=1000$. Each box summarizes the distribution of ratios of the approximated permutation $p$-values to the corresponding Wilcoxon $p$-values across replicates. 
Ratios close to 1 indicate good agreement with the Wilcoxon reference, while ratios above 1 reflect conservative behavior.
Consistent with the \textit{t}-test results, the classical GPD method produces an increasing number of zero $p$-value estimates as $n$ increases, leading to ratios that collapse to zero for a growing fraction of replicates. 
The remaining approximation methods yield ratios that systematically exceed 1, indicating conservative behavior. 
In contrast, \texttt{permApprox} yields ratios that remain comparatively close to 1 across sample sizes, while strictly avoiding zero $p$-values, suggesting stable behavior across varying test statistic scales and data distributions.
Complementary results, including the corresponding $p$-values for this setting, as well as $p$-values and ratios stratified by $d$ and $B$, are provided in the Supplementary Figures~\ref{app:fig:wilcox_pvals_by_n} to \ref{app:fig:wilcox_ratios_by_d_and_B}.

%%%%%%%%%%%%%%%%%%%%%%%%%%%%%%%%%%%%%%%%%%
\section{Application} \label{sec:application}
%%%%%%%%%%%%%%%%%%%%%%%%%%%%%%%%%%%%%%%%%%

\subsection{Differential abundance analysis in microbiome data}
\label{sec:application:dacomp}

Microbiome sequencing data are well suited to demonstrate the practical relevance of \texttt{permApprox}. These data are compositional and typically sparse, with a high proportion of zeros, making appropriate preprocessing and normalization steps necessary. As a consequence, the test statistics involved in common tasks such as differential abundance analysis rarely follow standard parametric distributions. Permutation tests are therefore widely used to obtain valid null distributions \cite{brill2021dacomp,sommer2022randomization,yang2022comprehensive}. 
However, in practice these analyses typically rely on empirical permutation $p$-values, which are directly determined by the number of permutations performed. \texttt{permApprox} can be seamlessly integrated into such workflows to refine these empirical estimates and provide more informative tail probabilities.

To illustrate this, we analyzed fecal microbiome profiles aggregated at the bacterial genus level, obtained from 16S rRNA amplicon sequencing in the PASTURE birth cohort \citep{mutius2006pasture}. The PASTURE study is a multi-centre prospective cohort investigating how farming-related exposures and early-life diet shape infant immune development. We focused on samples collected at age two months and compared infants who were exclusively breastfed for at least two months (``EBF'' group; $n=485$) with those who were never exclusively breastfed (``non-EBF'' group; $n=212$). Infant feeding mode is one of the strongest determinants of early-life gut microbiome composition, and exclusive breastfeeding has consistently been associated with characteristic microbial profiles and lower gut microbial diversity during early infancy \citep{ho2018meta, ma2020comparison, odiase2023gut}. Consistent with these findings, our own diversity analysis using three common diversity indices (Supplementary Figure~\ref{app:fig:alpha_diversity}) showed lower $\alpha$-diversity in the EBF group.

We performed differential abundance testing with the \texttt{dacomp} \texttt{R} package \citep{brill2021dacomp} to identify genera that differ between EBF and non-EBF infants at age two months. \texttt{dacomp}, which implements the framework of \citet{brill2022testing}, performs differential abundance testing by first selecting a set of reference taxa assumed not to change between groups. Each test taxon is then normalized by the aggregated reference set, and a Wilcoxon-type statistic is computed. To assess significance, \texttt{dacomp} generates a permutation distribution of this statistic by repeatedly shuffling the group labels while keeping the microbiome profiles fixed, and computes empirical permutation $p$-values as defined in equation~\eqref{eq:empirical_p-value}. Further methodological details are given in Appendix~\ref{app:sec:DA_microbiome_dacomp}.

To illustrate a limited-permutation setting, we ran \texttt{dacomp} with $B=1000$ permutations and refined the resulting empirical permutation $p$-values using \texttt{permApprox} (default configuration; Section~\ref{sec:methods:default_config}). To assess how well the refined tail probabilities approximate the behavior expected under substantially larger permutation budgets, we additionally ran \texttt{dacomp} with $B=10^6$ and $B=10^7$ permutations, which serve as high-accuracy empirical references. Throughout, we used \texttt{dacomp}’s option to test all taxa, whereby each reference taxon is temporarily excluded from the reference set to obtain a test statistic and corresponding permutation $p$-value.

\begin{figure}[h]
  \centering
  \includegraphics[width=\textwidth]{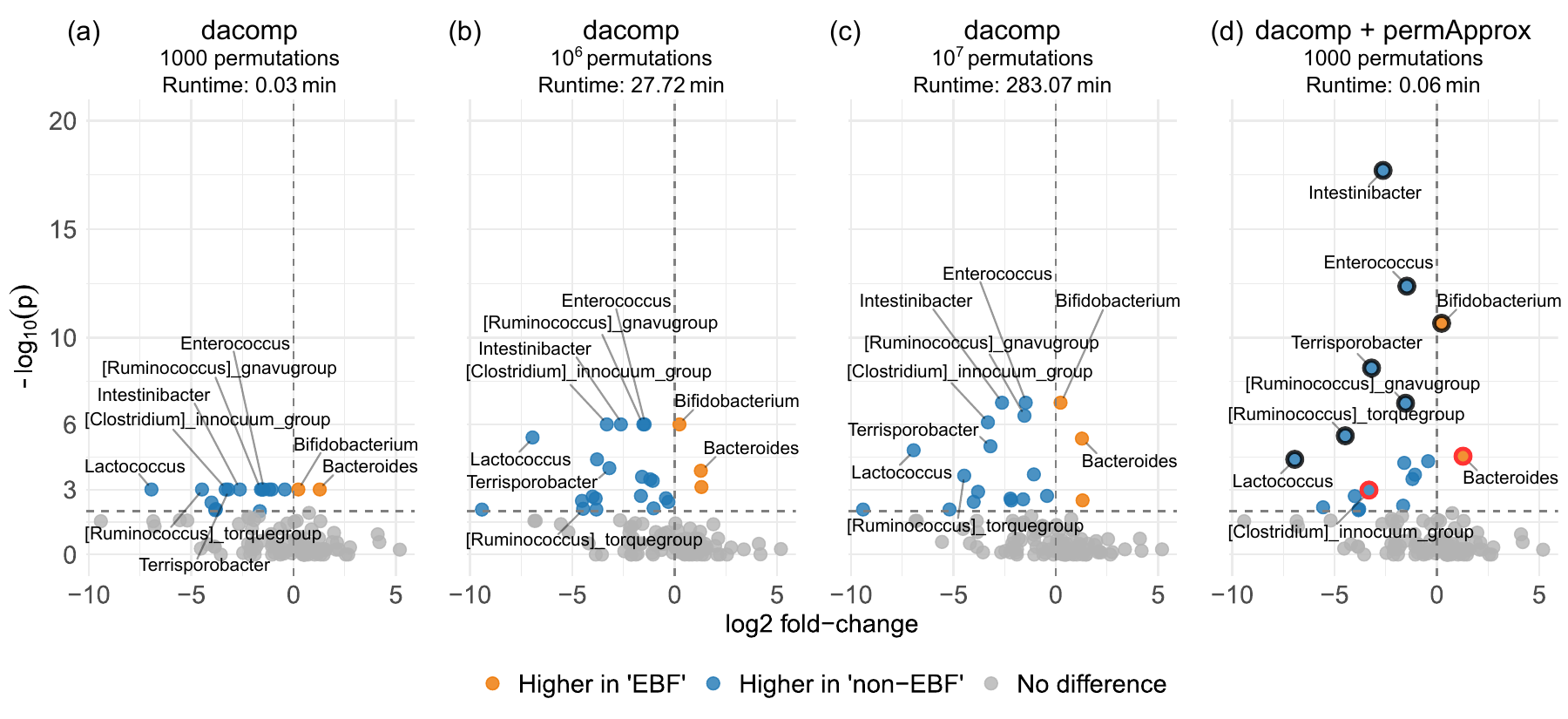}
  \caption{
  \textbf{Differential abundance analysis in the PASTURE cohort: effect of permutation budget and comparison to \texttt{permApprox}.}
  Volcano plots show log$_2$ fold-changes (EBF vs.\ non-EBF) versus raw (unadjusted) permutation $p$-values for genus-level differential abundance testing with \texttt{dacomp}.
  Orange points indicate genera enriched in the EBF group and blue points indicate genera enriched in the non-EBF group. The horizontal dashed line marks the significance level $\alpha=0.01$, and points shown in gray are not significant at this level.
  \textbf{(a)} \texttt{dacomp} with $B=10^3$ permutations: empirical permutation $p$-values are bounded below by approximately $1/1000$, resulting in a pronounced floor in the tail.
  \textbf{(b)} \texttt{dacomp} with $B=10^6$ permutations and
  \textbf{(c)} \texttt{dacomp} with $B=10^7$ permutations: increasing the permutation budget progressively refines tail resolution.
  \textbf{(d)} \texttt{dacomp} with $B=10^3$ permutations followed by \texttt{permApprox} (default configuration), yielding refined tail $p$-values that largely reproduce the trends observed at $B=10^6$--$10^7$ at a fraction of the computational cost.
  For readability, only taxa that attain a raw $p$-value below $10^{-5}$ in at least one of the three right-most panels ((b)--(d)) are labeled.
  In panel~(d), taxa whose \texttt{permApprox} $p$-values follow the trend observed from $B=10^6$ to $B=10^7$ are indicated by a black outline, whereas taxa that deviate from this trend are highlighted by a red outline.
  Panel headers report runtimes for the respective analyses.
  }
  \label{fig:dacomp_volcano}
\end{figure}

Figure~\ref{fig:dacomp_volcano} shows the resulting raw (unadjusted) permutation $p$-values. We deliberately omit multiple testing correction here to isolate the effect of the permutation budget on the $p$-values themselves. With $B=1000$ permutations (panel~(a)), the \texttt{dacomp} $p$-values are constrained to the grid of multiples of $1/(B+1)$ and therefore bounded below by $\approx1/1000$, producing a pronounced floor that hides differences among the most significant taxa. Increasing the permutation budget to $B=10^6$ and $B=10^7$ (panels~(b) and (c)) progressively resolves the tail, yielding smoother $p$-value gradients and a more informative ranking, but at a substantial computational cost. 
On a Linux-based compute node equipped with an Intel Xeon E7-4850~v4 CPU (2.10\,GHz) and 170\,GB RAM, \texttt{dacomp} run times were 27.72\,min for $B=10^6$ and 283.07\,min for $B=10^7$ permutations, respectively, compared to a few seconds for $B=1000$ with \texttt{dacomp} + \texttt{permApprox}. All computations were performed on a single CPU core. A comprehensive runtime comparison is provided in Table~\ref{app:tab:dacomp_runtime} and Figure~\ref{app:fig:dacomp_runtime}.

To assess whether \texttt{permApprox} can recover these high-resolution tail patterns from only $B=1000$ permutations, we treat the $B=10^6$ and $B=10^7$ results as increasingly precise empirical references and compare the \emph{trends} across taxa rather than expecting point-wise equality.
For readability of the plots, we restrict attention to taxa that attain a raw $p$-value below $10^{-5}$ in at least one of the three right-most panels ((b)--(d)). Only these taxa are labeled in the plots. The \texttt{permApprox} results (panel~(d)) closely track the tail behavior seen at $B=10^6$--$10^7$ for the majority of these genera. To illustrate this, taxa whose \texttt{permApprox} $p$-values follow the trend from $B=10^6$ to $B=10^7$ are marked with a black outline in the \texttt{permApprox} panel, whereas taxa that deviate from this trend are highlighted with a red outline.

Only two genera stand out as noticeable deviations: \emph{Bacteroides} and the \emph{[Clostridium] innocuum group}. For the latter, the permutation histogram for $B=1000$ permutations (Figure~\ref{app:fig:dacomp_perm_histograms}) shows a comparatively coarse permutation distribution, which makes smooth parametric tail extrapolation inherently difficult and may differ from the behavior observed at much larger $B$ values. For \emph{Bacteroides}, we did not identify a similarly clear distributional explanation. However, the discrepancy is small in magnitude and may plausibly reflect random variability in the extreme tail rather than a systematic limitation of the approximation.

While Figure~\ref{fig:dacomp_volcano} focuses on tail behavior and relative trends in the raw permutation $p$-values, we additionally summarize the resulting significance decisions across all tested genera in the Appendix.
Supplementary Tables~\ref{app:tab:dacomp_unadjusted_pvals} and \ref{app:tab:dacomp_sig_agreement_raw} report the unadjusted $p$-values and the corresponding agreement of significance decisions between methods, whereas Tables~\ref{app:tab:dacomp_adjusted_pvals} and \ref{app:tab:dacomp_sig_agreement_adj} present the same summaries based on Benjamini-Hochberg adjusted $p$-values.
Using \texttt{dacomp} with $B=10^7$ permutations as a high-precision empirical reference, the adjusted results show that the constrained \texttt{permApprox} approach applied to only $B=1000$ permutations yields highly concordant significance decisions, identifying 10 of the 11 genera declared significant by \texttt{dacomp} at $\alpha = 0.01$ across a total of 118 tests.

We also applied the unconstrained GPD approximation to examine the effect of the support constraint in this real-world application. For \emph{Terrisporobacter}, the unconstrained fit yielded a zero $p$-value, caused by a negative GPD shape parameter and the observed test statistic lying above the resulting upper support boundary. The constrained version of \texttt{permApprox} avoids this issue by ensuring that the fitted support extends beyond the observation.
Apart from preventing this single zero $p$-value, the constrained and unconstrained fits produced almost identical results for all other taxa, as reflected in Tables~\ref{app:tab:dacomp_unadjusted_pvals} and \ref{app:tab:dacomp_adjusted_pvals} in the Appendix. Additional results, including prevalence-variance diagnostics for \texttt{dacomp} reference selection, stacked bar plots of genus- and phylum-level composition, and distributions of \texttt{dacomp}-rarefied counts for the highly significant genera, are shown in Supplementary Figures~\ref{app:fig:dacomp_prev_vs_score} to \ref{app:fig:dacomp_rarefied_counts}.

Taken together, these results highlight a broader practical implication.
Although the primary motivation for \texttt{permApprox} is to prevent zero $p$-values in GPD-based tail approximation, only a single such case arises in this application, and the emphasis here is therefore on the quality of the resulting tail approximation under limited permutation budgets.
Even when increasing computational resources have reduced the practical burden of large-scale permutation testing, the results above indicate that there is often little benefit in expending such resources when comparable inference can be obtained at negligible additional cost using \texttt{permApprox}.
Moreover, the present example involves only 118 hypothesis tests and is therefore modest in size compared to many contemporary biological applications, which routinely involve thousands of features, further amplifying the computational advantages of approximation-based approaches.

%=========================================
\subsection{Differential distribution analysis in single-cell RNA-seq data}
\label{sec:application:scRNA}
%=========================================

Differential distribution (DD) analysis provides a flexible framework for detecting differences between two groups that go beyond shifts in the mean. \citet{schefzik2021fast} proposed a DD approach based on the 2-Wasserstein distance $d^2_\text{W}$, which quantifies discrepancies between entire distributions in terms of location, size, shape, and zero proportion. Their method is implemented in the \texttt{waddR} \texttt{R} package. Because the null distribution of the Wasserstein statistic is not available in closed form, \texttt{waddR} relies on a permutation procedure to assess significance. Specifically, the method uses a two-stage scheme in which group labels are permuted for the non-zero expression values, followed by a GPD-based tail approximation to refine very small permutation $p$-values. A detailed description of the Wasserstein two-stage test and its permutation framework is provided in Appendix~\ref{app:sec:DD_waddR_details}.

Although the GPD refinement in \texttt{waddR} successfully increases the resolution of small permutation $p$-values beyond the coarse empirical grid, it simultaneously introduces a substantial number of zero $p$-values. To demonstrate that \texttt{permApprox} can provide more robust and strictly positive tail probabilities, we revisited the real-data case study presented in the original \texttt{waddR} paper, which compares natural-killer cells from blood and decidua in the scRNA-seq dataset of \citet{vento2018single}. 

For reproducibility, we followed the preprocessing steps provided in the \texttt{waddR} package vignette, which includes a preprocessed and normalized replicate of the same underlying dataset. From these data, we selected the 1000 most prevalent genes and equalized the sample sizes by randomly subsampling from the larger decidua group (3249 cells) to match the number of blood cells. This yielded two equally sized expression matrices, each with dimensions 1000 × 569. A heatmap with gene expressions of the 20 most prevalent genes are provided in Figure~\ref{app:fig:scRNA_heatmap}. 

Because \texttt{waddR} does not export permutation statistics, we modified its source code to return the observed and permuted Wasserstein distances. We then applied \texttt{permApprox} to recompute the permutation $p$-values using two variants: (i) the default constrained workflow described in Section~\ref{sec:methods:default_config} ensuring that the evaluation point lies within the GPD support, and (ii) an unconstrained version omitting the support condition, which isolates the specific contribution of the constraint. Both variants were applied only when the empirical permutation $p$-value was below 0.01, in accordance with the original \texttt{waddR} procedure.

\begin{figure}[ht]
  \centering
  \includegraphics[width=\textwidth]{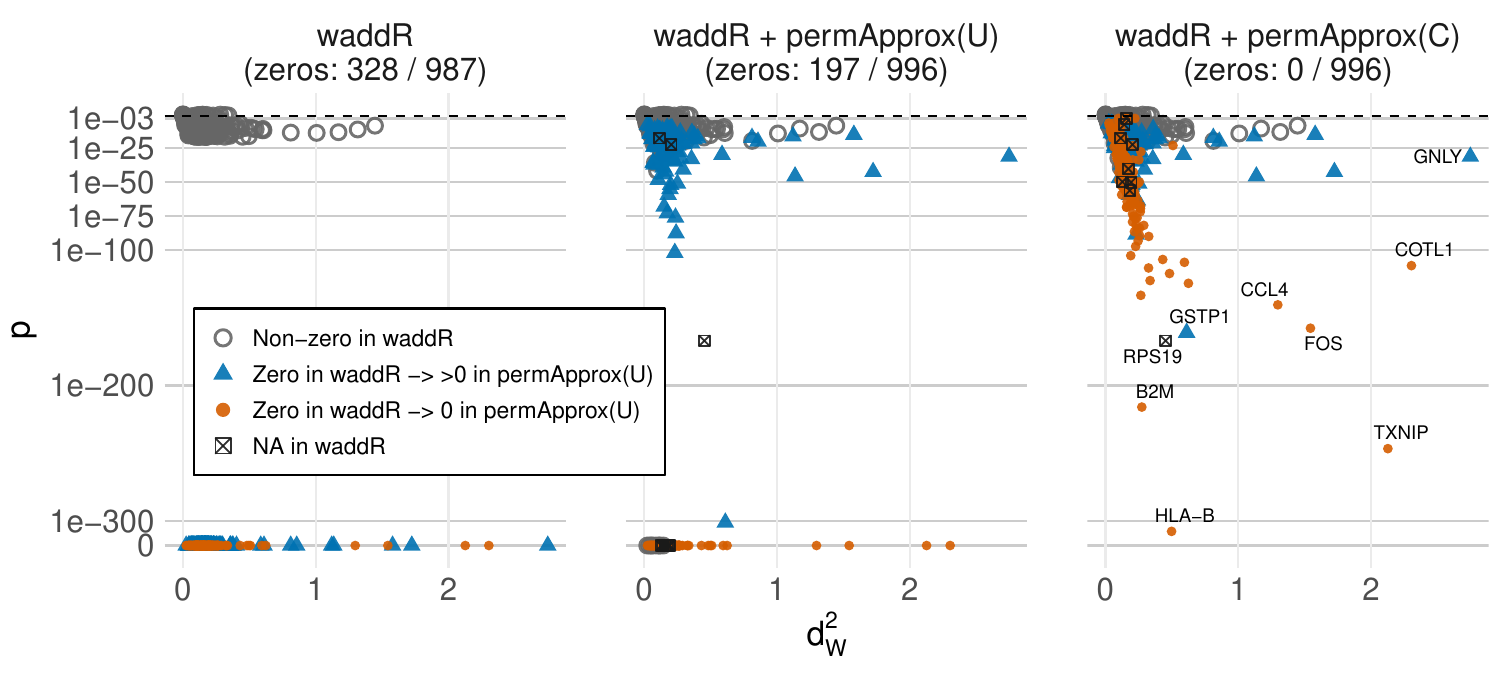}
  \caption{\textbf{Differential distribution analysis in single-cell RNA-seq data.} 
  Comparison of permutation $p$-values from \texttt{waddR} and \texttt{permApprox} for the non-zero expression component of the Wasserstein two-stage test ($B = 1000$).  
  Each point represents one gene, and the color and shape represent the zero-pattern of its \texttt{waddR} and \texttt{permApprox(U)} $p$-values:
  non-zero \texttt{waddR} $p$-values (open circle); genes with $p=0$ under \texttt{waddR} but $p>0$ under \texttt{permApprox(U)} (blue triangle); genes with $p=0$ under both methods (orange circle), and genes with undefined (\texttt{NA}) \texttt{waddR} $p$-values (square with cross). Zero $p$-values are mapped to a small constant floor so they appear at ``0'' on the $y$-axis.
  \textbf{Left:} Standard \texttt{waddR}, showing many zero $p$-values caused by numerical rounding in upper-tail probability computation and bounded tails when the fitted GPD has negative shape.  
  \textbf{Middle:} Unconstrained \texttt{permApprox} (U) removes rounding-induced zeros but still produces zeros when the estimated GPD shape is negative.
  \textbf{Right:} Constrained \texttt{permApprox} (C) enforces that the evaluation point lies within the GPD support, eliminating all zero $p$-values.}
  \label{fig:app_waddr_permapprox}
\end{figure}

Figure~\ref{fig:app_waddr_permapprox} compares the $p$-values from the non-zero expression component of \texttt{waddR} with those obtained from the unconstrained and constrained \texttt{permApprox} approaches. The left panel shows that \texttt{waddR} produces a substantial number of zero $p$-values, arising partly from numerical rounding in the computation of upper-tail probabilities as $\bar{F}_{\mathrm{GPD}}(Y_{\text{obs}}) = 1 - F_{\mathrm{GPD}}(Y_{\text{obs}})$ and partly from bounded GPD support when the GPD shape parameter is negative and the observed statistic lies outside this support. The middle panel demonstrates that unconstrained \texttt{permApprox} effectively resolves the first issue (blue triangles), producing positive tail probabilities whenever the original zeros were caused by rounding. However, unconstrained fitting still inherits zeros from bounded GPD tails (orange points), just like \texttt{waddR}. 

The default \texttt{permApprox} workflow (\texttt{permApprox}(C) indicating constrained fitting, right panel) eliminates all remaining zeros by enforcing the evaluation point to lie strictly within the estimated GPD support (see also Tables~\ref{app:tab:waddr_unadjusted_pvals} and \ref{app:tab:waddr_adjusted_pvals} for the corresponding raw and BH-adjusted $p$-values).
Several genes exhibited extremely large Wasserstein distances so that machine-underflow $p$-values (reported as zero) occurred in the constrained variant. Here, our data-adaptive $\varepsilon$ refinement (Section~\ref{app:sec:methods:epsilon_refinement}) became active, increasing the target factor $\tau$ from its default value of 0.25 to 29.75, thereby ensuring positive and numerically stable tail probabilities for all genes. Consequently, some $p$-values increase, yet the number of genes with significant differences remains nearly unchanged (Table~\ref{app:tab:waddr_sig_counts}).

Beyond eliminating spurious zeros, the refined tail probabilities produced by \texttt{permApprox} also change the relative ranking of genes with extremely small $p$-values.  When inspecting the unconstrained panel, one might conclude that genes with very large Wasserstein distances (e.g.,\ those with $d^2_\text{W} > 1$) represent the strongest distributional differences. However, the constrained tail approximation reveals that several genes with moderate Wasserstein distances, including \textit{HLA-B} and \textit{B2M}, exhibit equally compelling evidence against the null. 
This is consistent with the distributional patterns seen in the histogram–quantile plot of the top-ranking genes (Figure~\ref{app:fig:scRNA_hist_quant}), where these genes with moderate $d^2_\text{W}$ show pronounced and systematic shifts between groups in their non-zero expression distributions. 

Together, these results illustrate that effect size and tail probability are not always aligned in the extreme tail, and that accurate $p$-value approximation is essential for identifying the truly most significant distributional shifts.

%%%%%%%%%%%%%%%%%%%%%%%%%%%%%%%%%%%%%%%%%%
\section{Discussion} \label{sec:discussion}
%%%%%%%%%%%%%%%%%%%%%%%%%%%%%%%%%%%%%%%%%%

Accurate inference in permutation-based testing is often limited by the coarse resolution of empirical permutation $p$-values. Because these values lie on a grid determined by the number of permutations $B$, extremely small $p$-values cannot be resolved unless $B$ is very large, which is often computationally infeasible. Parametric tail modeling based on the Generalized Pareto Distribution (GPD) has been proposed to overcome this limitation by extrapolating beyond the empirical permutation range \citep{knijnenburg2009fewer}. However, existing GPD-based refinements can yield zero $p$-values when the estimated shape parameter is negative and the fitted tail has a finite upper bound, a problem that has remained largely unresolved.

The central contribution of this work is a constrained GPD tail-fitting strategy that guarantees positive tail probabilities at the observed test statistic. Specifically, we propose to fit the GPD under a support constraint, enforced through a data-adaptive $\varepsilon$ rule that balances the trade-off between unrealistically small $p$-values and overly conservative extrapolation. In combination with a robust threshold-detection procedure and a carefully chosen GPD estimator, this approach yields smooth, strictly positive, and numerically stable tail probabilities even when $B$ is modest. In doing so, \texttt{permApprox} maintains the flexibility and distribution-free validity of permutation testing while substantially improving the resolution and reliability of $p$-values in the extreme tail.

The performance of the proposed workflow was evaluated in two controlled simulation settings reflecting common parametric and nonparametric testing scenarios. In the two-sample $t$-test with Gaussian data, where the theoretical null distribution is known, \texttt{permApprox} closely matched the true $p$-values across a wide range of sample sizes and effect sizes. By contrast, the three existing GPD-based refinements considered (i.e., the classical approach of \citet{knijnenburg2009fewer}, its power-transformed variant, and our ``\,+obs'' implementation following \citet{winkler2016faster}) frequently produced zero $p$-values due to negative shape estimates or became markedly conservative as sample size increased. 

In the nonparametric Wilcoxon rank-sum test with exponential data, which induces highly skewed permutation distributions, constrained GPD fitting again yielded accurate and stable tail probabilities. Competing methods were unable to track the true $p$-values even for moderate test statistics, either collapsing to zero or again being overly conservative. In contrast, \texttt{permApprox} remained robust across all configurations considered in this study, producing smooth and strictly positive approximations even for extreme test statistics. These findings highlight that the benefits of our proposed constrained tail modeling approach are not limited to idealized Gaussian settings but may extend to distribution-free tests commonly used in practice.

Beyond controlled simulations, the two biological applications further demonstrated the practical advantages of constrained tail modeling. In the microbiome differential abundance analysis, empirical permutation $p$-values were limited by the coarse resolution inherent to $B=1000$ permutations, obscuring the significance of several strongly differentially abundant genera. \texttt{permApprox} resolved this limitation by producing smoothly varying, strictly positive tail probabilities, enabling meaningful ranking of effects that were indistinguishable under empirical testing. Moreover, the method prevented the zero $p$-value that arose under unconstrained GPD fitting for \emph{Terrisporobacter}, illustrating the necessity of enforcing a valid support in real-world settings where negative shape estimates are common. Comparison with substantially larger permutation budgets ($B=10^6$ and $10^7$) further showed that the \texttt{permApprox} tail estimates closely reproduce the empirical tail trends observed at high resolution for the majority of taxa, at dramatically reduced computational cost.

In the single-cell RNA-seq case study, the benefits of constrained approximation were even more pronounced. The original Wasserstein-based differential distribution test implemented in \texttt{waddR} uses a GPD refinement but produces numerous zero $p$-values due to bounded tails and numerical underflow. By reusing the underlying permutation statistics, \texttt{permApprox} eliminated these zeros while preserving the structure of the permutation null. Importantly, the refined $p$-values revealed additional genes with strong distributional differences that would have been missed with the original implementation. 
These findings underscore that constrained tail modeling offers clear benefits across complex, non-Gaussian data structures, where permutation tests are essential.

Despite its advantages, \texttt{permApprox} has a few practical limitations. First, the approach assumes that the upper tail of the permutation distribution is well approximated by a GPD and that a suitable tail threshold can be identified. In settings where the permutation tail is highly irregular or contains too few exceedances, the chosen goodness-of-fit test may reject the GPD model, in which case tail approximation is inappropriate. We observed such a scenario in a high-throughput chemical genomics application, where the Anderson-Darling test rejected the GPD fit for most tail distributions. Here, \texttt{permApprox}'s implementation of the Gamma approximation of the full permutation distribution provided a better empirical fit \citep{feldl2025statistical}. In cases where no suitable parametric approximation can be identified, the empirical permutation $p$-value remains the appropriate fallback.

Second, although the proposed data-adaptive $\varepsilon$ rule performed robustly across both Gaussian and nonparametric simulation settings, it necessarily influences the balance between overly optimistic and overly conservative tail probabilities. In rare cases with exceptionally extreme test statistics, the constraint may require additional refinement to prevent numerical underflow, at the cost of slightly more conservative $p$-values for other tests. We observed such a situation in the single-cell DD analysis, where the $\varepsilon$ refinement slightly increased the $p$-value of the gene \textit{GSTP1} under the constrained version.
Finally, constrained tail fitting introduces some computational overhead compared to relying solely on empirical permutation $p$-values, though the additional cost is modest in typical analysis settings and negligible compared to the computational burden of increasing the number of permutations.

Taken together, our results suggest that \texttt{permApprox} successfully eliminates the long-standing problem of zero $p$-values in GPD-based permutation $p$-value approximation and provides a broadly applicable workflow for refining permutation tails. While parametric tail modeling already offers an effective remedy for the coarse resolution of empirical $p$-values, existing approaches remain vulnerable to negative shape estimates in combination with support-boundary violations. \texttt{permApprox} addresses these limitations by combining constrained GPD fitting, robust threshold detection, and a data-adaptive $\varepsilon$ rule into a coherent framework that integrates seamlessly with standard permutation pipelines. 

Importantly, \texttt{permApprox} is not tied to any specific test statistic or scientific domain. Any permutation-based method that returns (i) a vector of observed test statistics and (ii) the corresponding matrix of permuted statistics can be directly extended with our workflow. This makes \texttt{permApprox} applicable far beyond the biological examples considered here. Potential use cases include permutation-based inference in association network analysis, where network metrics often lack analytically tractable null distributions \citep{sommer2022randomization}, gene set and pathway enrichment tests that rely on label permutations \citep{mishra2014gene}, genome-wide association studies \citep{john2022efficient}, and permutation-based clustering or classification assessments. 

\texttt{permApprox} is also applicable in permutation-based significance testing in fields beyond life sciences such as economics, social sciences, or environmental studies. 
By improving the reliability of small $p$-values in these contexts, \texttt{permApprox} broadens the scope of accurate and computationally feasible permutation-based inference across a wide range of data-analytic applications.

%%%%%%%%%%%%%%%%%%%%%%%%%%%%%%%%%%%%%%%%%%
% Final information
%%%%%%%%%%%%%%%%%%%%%%%%%%%%%%%%%%%%%%%%%%

\section*{Code availability}

All scripts used to produce the simulation studies and applications presented in this paper are publicly available at \url{https://github.com/stefpeschel/permApprox-manuscript}.
The real-data applications were performed using \texttt{permApprox} (v1.0.2) \citep{peschel2025permapprox}.
The simulation studies relied on a more flexible development version of \texttt{permApprox}, which comprises additional internal options and is included in the same repository.

\section*{Data availability}

The raw DNA sequence data (demultiplexed 16S rRNA amplicon reads) 
from the PASTURE cohort have been deposited in the NCBI Sequence 
Read Archive under accession numbers PRJNA1068358 (2-month data) 
and PRJNA1068358 (12-month data). In this study, we used genus-level count data derived from the 
2-month samples. Access to additional metadata is subject to 
the PASTURE consortium’s data protection regulations and can be requested from the PASTURE consortium directly.

\section*{Competing interests}
No competing interests are declared.

\section*{Funding}

This work was supported by the European Commission [LSHB-CT-2006-018996]; and the European Research Council [ERC-2009-AdG\_20090506\_250268].

\section*{Acknowledgments}

We gratefully acknowledge the PASTURE study team and all participating 
families for their commitment and contribution to the cohort.
We especially thank Martin Depner and Pirkka Kirjavainen for their work 
on the preprocessing and taxonomic processing of the PASTURE 2-month 
stool microbiome data, which were used in this study.

\bibliographystyle{apalike}
\bibliography{references}

%%%%%%%%%%%%%%%%%%%%%%%%%%%%%%%%%%%%%%%%%%
% Appendix
%%%%%%%%%%%%%%%%%%%%%%%%%%%%%%%%%%%%%%%%%%

\clearpage
\appendix

\counterwithin{figure}{section}
\counterwithin{table}{section}
\counterwithin{equation}{section}

\renewcommand{\thesection}{\Alph{section}}
\renewcommand{\thefigure}{\thesection.\arabic{figure}}
\renewcommand{\thetable}{\thesection.\arabic{table}}
\renewcommand{\theequation}{\thesection.\arabic{equation}}

%%%%%%%%%%%%%%%%%%%%%%%%%%%%%%%%%%%%%%%%%%
% Appendix A
%%%%%%%%%%%%%%%%%%%%%%%%%%%%%%%%%%%%%%%%%%s

\section{Extended Methods}
\label{app:sec:methods}

%=========================================
\subsection{Notation}
\label{app:sec:methods:notation}
%=========================================

\begin{table}[H]
\centering
\caption{Notation used throughout the paper.}
\label{app:tab:notation}
\begin{tabularx}{\textwidth}{lX}
\toprule
\textbf{Symbol} & \textbf{Meaning} \\
\midrule
\multicolumn{2}{l}{\textbf{General permutation test notation}} \\[2pt]
$n$ & Per-group sample size \\
$m$ & Number of hypothesis tests (features, genes, taxa) \\
$B$ & Number of permutations \\
$b \in \{1,\dots,B\}$ & Index for permutation replicates \\
$j \in \{1,\dots,m\}$ & Index for hypothesis tests \\
$T_{\text{obs}}^{(j)}$ & Observed test statistic for test $j$ \\
$T_b^{*(j)}$ & Permuted test statistic for test $j$ in permutation $b$ \\
$\mathcal{T}^{(j)} = \{T_b^{*(j)}\}_{b=1}^B$ & Permutation distribution for test $j$ \\
\midrule

\multicolumn{2}{l}{\textbf{$p$-values}} \\[2pt]
$p_{\mathrm{emp}}$ & Empirical permutation $p$-value (Eq.~\ref{eq:empirical_p-value}) \\
$p_{\mathrm{GPD}}$ & Unconstrained GPD-based $p$-value (Eq.~\ref{eq:GPD_p-value})\\
$p_{\mathrm{cGPD}}$ & Constrained GPD-based $p$-value \\
$p$ & Hybrid $p$-value combining empirical and GPD-based $p$-values (Eq.~\ref{eq:hybrid_pvalue}) \\
\midrule

\multicolumn{2}{l}{\textbf{GPD tail modeling}} \\[2pt]
$p_{\mathrm{thr}}$ & GPD tail modeling threshold (on $p_{\mathrm{emp}}$) \\
$u$ & Threshold defining the GPD tail region \\
$k$ & Number of exceedances above $u$\\
$Y_{\text{obs}}$ & Observed excess above $u$ \\
$\mathcal{Y} = \{Y_b^*\}$ & Permutation excesses above $u$ \\
$\sigma, \xi$ & Scale and shape parameters of the GPD \\
$s = -\sigma/\xi$ & Upper support boundary when $\xi < 0$ \\
$\hat{\sigma}_{(U)}, \hat{\xi}_{(U)}$ & Unconstrained GPD parameter estimates \\
$\hat{\sigma}_{(C)}, \hat{\xi}_{(C)}$ & Constrained GPD parameter estimates \\
$\hat{s}_{(U)}, \hat{s}_{(C)}$ & Unconstrained / constrained support boundaries \\
$\bar{F}_{\mathrm{GPD}}(\cdot) = 1 - F_{\mathrm{GPD}}(\cdot)$ & Upper GPD tail probability\\
\midrule

\multicolumn{2}{l}{\textbf{Data-adaptive $\varepsilon$ rule (SLLS)}} \\[2pt]
$\varepsilon$ & Safety margin ensuring GPD support at $Y_{\text{obs}}$ (Eq.~\ref{eq:epsilon_final})
\\
$\kappa_{\mathrm{factor}}$ & Tuning constant controlling curvature in the log-saturation component\\
$\tau$ & Tuning constant controlling plateau height \\
$\rho_{\mathrm{lift}}$ & Weighting parameter in lift component \\
\bottomrule
\end{tabularx}
\end{table}

%=========================================
\subsection{Derivation of the SLLS \texorpdfstring{$\varepsilon$}{epsilon} rule}
\label{app:sec:methods:eps_derivation}
%=========================================

This section summarizes the empirical considerations that guided the construction of the standardized lifted log-saturation (SLLS) rule for selecting the safety margin $\varepsilon$. The presentation follows the sequence of observations made in the exploratory analyses in Section~\ref{app:sec:methods:eps_comparison} and makes explicit how the dependence on sample size, effect size, and test statistic magnitude is incorporated into the final rule.
The complete exploratory workflow is documented in the reproducible analysis scripts accompanying this manuscript available at \url{https://github.com/stefpeschel/permApprox-manuscript}, under \texttt{explorations/perm\_ttest/03\_find\_eps\_rule}.

\paragraph{Joint dependence on test statistic and sample size.} 
Simulation studies showed that the safety margin $\varepsilon$ must increase with the magnitude of the observed test statistic $|T_{\mathrm{obs}}|$ to stabilize constrained GPD tail evaluation. A simple linear rule $\varepsilon = c\,|T_{\mathrm{obs}}|$, however, was inadequate: the factor $c$ varied systematically with sample size $n$, and no single linear scaling performed well across different $n$.

Moreover, the empirical relationship between the required $\varepsilon$ and
$|T_{\mathrm{obs}}|$ was clearly nonlinear, exhibiting pronounced curvature across
increasing effect sizes. An exemplary visualization for $c = 0.1$ is given in Figure~\ref{app:fig:eps_comparison_ttest}\.(b). This indicated that linear rules are too restrictive to capture the behavior of the tail region across a broad range of statistics.

\paragraph{Logarithmic growth and saturation.}
To accommodate this curvature, sublinear growth in $|T_{\mathrm{obs}}|$ was considered. Among several candidates, logarithmic growth provided a particularly good empirical fit. A rule of the form
\[
\varepsilon \;\propto\; \log\!\bigl(1 + c\,|T_{\mathrm{obs}}|\bigr)
\]
increases rapidly for small to moderate statistics while flattening for larger values, yielding substantially more stable tail extrapolation than linear alternatives.

However, without further modification, this rule still allowed $\varepsilon$ to grow without bound for extreme statistics. To prevent unnecessary inflation, the logarithmic growth was therefore normalized to saturate at a finite upper level, leading to a saturation rule of the form
\[
\varepsilon
=
\varepsilon_{\max}
\frac{\log\!\bigl(1 + c\,|T_{\mathrm{obs}}|\bigr)}
     {\log\!\bigl(1 + c\,T_{\max}\bigr)},
\]
where $T_{\max}$ denotes a reference statistic defining the upper end of the relevant range and $\varepsilon_{\max}>0$ is the maximal safety margin on the scale of the test statistic, i.e., the value approached as $|T_{\mathrm{obs}}|$ reaches the upper end of the relevant range. This normalization preserves the nonlinear increase for moderate statistics while ensuring bounded behavior in the extreme regime. A visualization of the rule at this stage is given in Figure~\ref{app:fig:eps_comparison_ttest}\.(c).

\paragraph{Lift for intermediate statistics.}
While the saturated log rule yielded stable behavior for very large
$|T_{\mathrm{obs}}|$, systematic comparisons showed a tendency to produce slightly
too small $p$-values for test statistics in an intermediate range (see Figure~\ref{app:fig:eps_comparison_ttest}\.(c)). This range
corresponds to situations in which tail extrapolation is already required, but
the saturation regime has not yet been reached. To correct this localized
deviation while preserving the asymptotic behavior for extreme statistics, an
additional lift term was incorporated into the safety margin.

Specifically, the lift is defined through the function
\[
\psi(l) = (1-l)^4(1+4l), \qquad l\in[0,1],
\]
which is a compactly supported $C^2$ Wendland kernel. Wendland kernels are widely
used in scattered data approximation and radial basis function methods due to
their smoothness, compact support, and controlled local influence. In the present
context, these properties ensure that the lift selectively increases the safety
margin for moderate values of the standardized statistic, while vanishing
smoothly as $l \to 1$ and leaving the saturation behavior for extreme statistics
unchanged.

\paragraph{Standardization and separation of scale.}
To obtain a formulation that is comparable across tests and distributions, the dependence on $|T_{\mathrm{obs}}|$ is transferred to a standardized scale. Figure~\ref{app:fig:eps_comparison_wilcox} illustrates the motivation for this step: when the $\varepsilon$ rule is applied directly on the raw test statistic scale, its behavior depends strongly on the scale and distribution of the statistic, leading to distorted tail approximations for non-Gaussian tests. Standardization mitigates this effect by expressing the observed statistic relative to its permutation distribution.

Let $Z_{\mathrm{obs}}^{(j)}$ denote the standardized observed statistic for test $j$, and let $Z_{\mathrm{cap}}$ denote an upper reference bound. Defining
\begin{equation}
   l_j = \frac{Z_{\mathrm{obs}}^{(j)}}{Z_{\mathrm{cap}}} \in [0,1] 
\end{equation}
compresses the relevant range of test statistics into a unit interval. The standardized safety margin can then be expressed as a function of $l_j$ alone.

\paragraph{Log-saturation, curvature control, and final rule.}
The standardized saturation component is defined as
\[
\varepsilon^\ast(l)
=
\varepsilon^\ast_{\max}
\frac{\log\!\bigl(1+\kappa l\bigr)}{\log\!\bigl(1+\kappa\bigr)},
\qquad l\in[0,1],
\]
which ensures $\varepsilon^\ast(0)=0$ and
$\varepsilon^\ast(1)=\varepsilon^\ast_{\max}$. The parameter $\kappa>0$ controls
the curvature of the transition, while the plateau height
$\varepsilon^\ast_{\max}$ absorbs the sample-size scaling identified above.

Adding the Wendland lift yields the standardized lifted log-saturation margin
\[
\varepsilon_j^{\ast}
=
\max\!\left\{
\varepsilon^\ast_{\max}
\frac{\log(1 + \kappa\, l_j)}{\log(1+\kappa)}
+
\rho_{\mathrm{lift}}\,\psi(l_j)
,\;
\varepsilon_{\min}
\right\}.
\]

Combining logarithmic saturation, Wendland lift, and rescaling back to the test statistic scale yields the standardized lifted log-saturation (SLLS) rule reported in Section~\ref{sec:methods:epsilon}. Its empirical behavior relative to simpler alternatives is illustrated in Section~\ref{app:sec:methods:eps_comparison}.

%=========================================
\subsection{Comparison of \texorpdfstring{$\varepsilon$}{epsilon} rules}
\label{app:sec:methods:eps_comparison}
%=========================================

\begin{figure}[H]
  \centering
  \includegraphics[width=\linewidth]{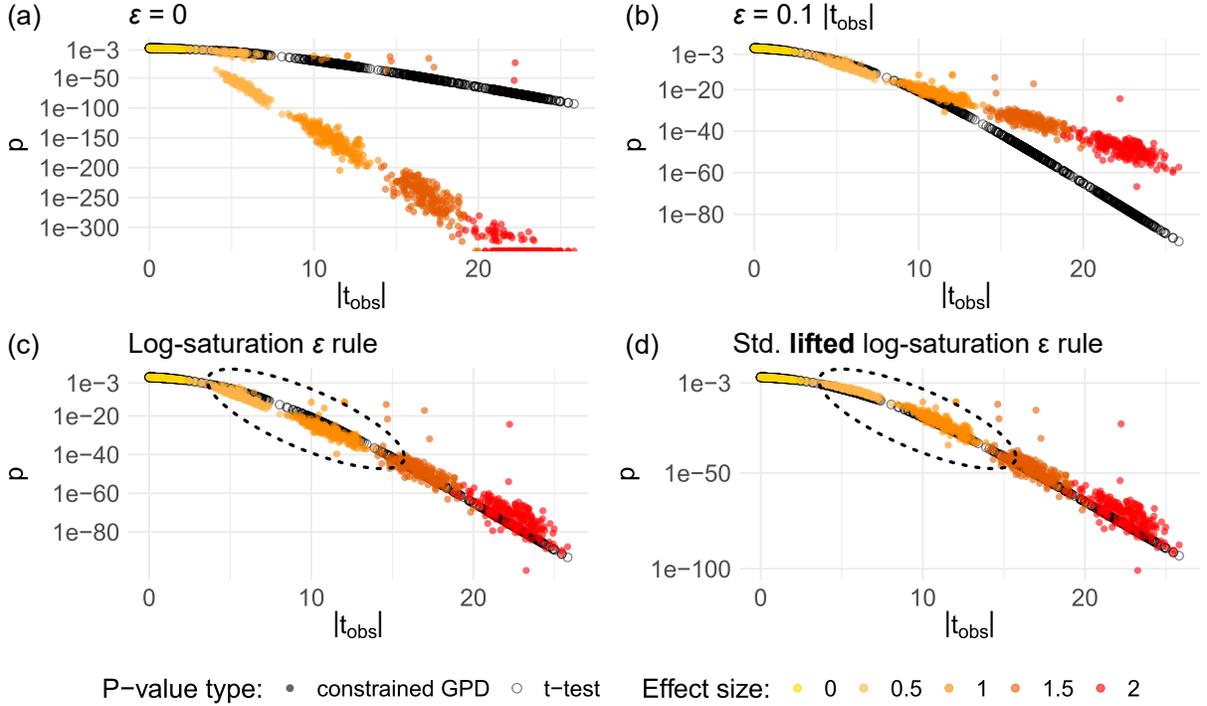}
  \caption{
  \textbf{Effect of $\varepsilon$-rules on constrained permutation $p$-values.}
  Data were simulated under Gaussian two-sample $t$-tests with $n=250$, $\sigma=1$, and mean differences $d \in \{0, 0.5, 1, 1.5, 2\}$, with $200$ tests per effect size ($1000$ in total). 
  Each panel shows $p$ from the constrained GPD approximation (colored points) and $p$ from the two-sided Student’s $t$-test (open circles) versus the observed statistic $t_{\mathrm{obs}}$. Colors indicate the true effect size. 
  Panels correspond to four choices of the support-constraint margin $\varepsilon$: 
  \textbf{(a)} $\varepsilon = 0$ (no margin); 
  \textbf{(b)} $\varepsilon = 0.1\,|t_{\mathrm{obs}}|$ (linear-in-$t$ baseline);
  \textbf{(c)} $\varepsilon = A\log\!\bigl(1 + B|t_{\mathrm{obs}}|\bigr)$ (log-saturation rule); and 
  \textbf{(d)} the standardized lifted log-saturation (SLLS) rule, applied on $Z$-scores and mapped back to the $T$-scale (with $k_{\text{factor}} = 1000$ and $\tau = 0.25$). The dashed circles indicate the area in which the \textit{lift} mainly takes effect. The $y$-axis is shown on a $\log_{10}$ scale with tick labels in the original $p$-value scale.
}
  \label{app:fig:eps_comparison_ttest}
\end{figure}

To illustrate the behavior of constrained GPD $p$-value approximations under different $\varepsilon$-rules, we simulated Gaussian two-sample data with equal variance ($\sigma=1$), increasing mean differences (effect sizes $d \in \{0, 0.5, 1, 1.5, 2\}$), and varying sample sizes. 
For each effect size, $200$ independent tests were generated, resulting in $1000$ tests per setting. 
Ground-truth $p$-values were obtained from the classical Student’s $t$-test, and permutation distributions with $B = 1000$ replicates were used for the constrained GPD-based approximations. 
Figure~\ref{app:fig:eps_comparison_ttest} compares the resulting $p$-values across four $\varepsilon$-rules for $n=250$ (per group). 
With $\varepsilon = 0$, the smallest $p$-values are heavily underestimated, especially for strong effects. 
The linear $\varepsilon$-rule ($\varepsilon = 0.1,|t_{\mathrm{obs}}|$) improves stability but fails to reproduce the nonlinear curvature of the $t$-test $p$-values.
The log-saturation rule yields generally well-calibrated results, though the approximated $p$-values tend to fall slightly below the $t$-test values for test statistics in the intermediate range ($t_{\mathrm{obs}} \approx 5$-$12$). 
In contrast, the standardized lifted log-saturation (SLLS) rule exhibits the most balanced behavior.
Its additional lift component effectively raises the approximated $p$-values for intermediate test statistics slightly above the corresponding \textit{t}-test values, while maintaining agreement across all effect sizes.

To emphasize the importance of standardization in the $\varepsilon$-rule, we repeated the analysis using the nonparametric Wilcoxon rank-sum test (with Mann-Whitney $U$ statistic). 
We simulated two groups of exponentially distributed observations means $1$ and $1 + d$, corresponding to effect sizes $d \in \{0, 0.5, 1, 1.5, 2\}$, and generated $200$ independent tests per effect size ($1000$ in total) for varying sample sizes. 
The ground-truth $p$-values were obtained from the classical Wilcoxon test, while permutation distributions with $B = 1000$ replicates were used for the GPD-based approximations. 
We compared the lifted log-saturation (LLS) rule, which operates directly on the raw $U$-statistics, to the standardized lifted log-saturation (SLLS) rule, which applies the same transformation on standardized $Z$-scores and then maps the resulting $\varepsilon$ values back to the original $U$-scale.

\begin{figure}[H]
  \centering
  \includegraphics[width=\linewidth]{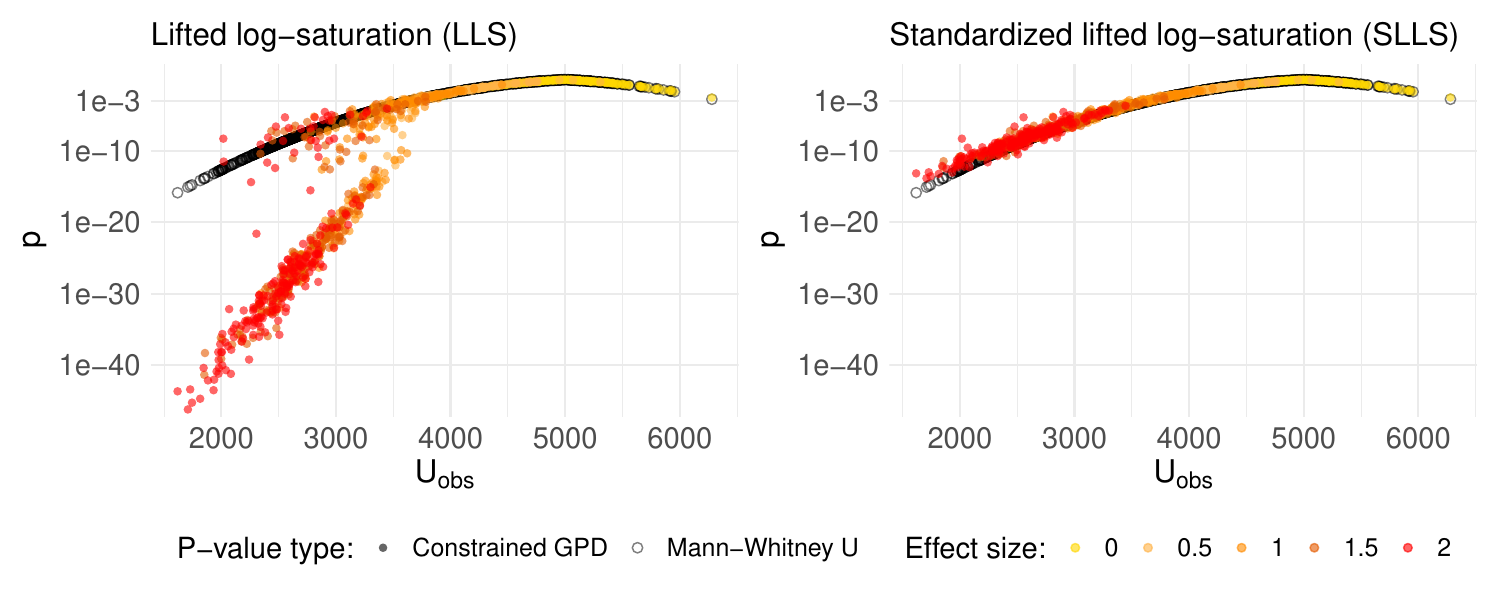}
  \caption{
  \textbf{Effect of standardization on constrained permutation $p$-values for the Wilcoxon test.}
  Data were generated from exponential distributions with means $1$ and $1 + d$ ($d \in \{0, 0.5, 1, 1.5, 2\}$), using equal group sizes ($n=100$) and $200$ tests per effect size ($1000$ in total). 
  Each panel shows the constrained GPD $p$-values (colored points) versus the observed Mann-Whitney $U$ statistic, compared to the ``ground-truth'' $p$-values from the Wilcoxon test (open circles). Colors indicate the true effect size. 
  The left panel uses a non-standardized $\varepsilon$-rule, based on the $U$-statistics itself, while the right panel uses our proposed standardized version (SLLS), which operates on $Z$-scores.
  }
  \label{app:fig:eps_comparison_wilcox}
\end{figure}

For the Wilcoxon test, standardization becomes crucial (Figure~\ref{app:fig:eps_comparison_wilcox}). 
While it has little impact in the Gaussian $t$-test setting, the absence of standardization leads to strong deviations from the Wilcoxon $p$-values for nonzero effect sizes. In contrast, the standardized version yields $p$-values that closely follow the Wilcoxon results across all effect sizes, confirming that this step is essential for the $\varepsilon$-rule to remain robust across test statistics of different scales and distributions.

%=========================================
\subsection{Data-adaptive \texorpdfstring{$\varepsilon$}{epsilon} refinement}
\label{app:sec:methods:epsilon_refinement}
%=========================================

Even with the SLLS $\varepsilon$-rule, machine-underflow $p$-values (reported as zero) may still occur for negative shape estimates ($\hat\xi < 0$), either because the test statistic is extremely large or because the fitted upper support is very close to the constraint, causing the tail probability to fall below machine precision. We therefore employ a \textit{data-adaptive refinement} for the target factor $\tau$, one of the SLLS tuning parameters, to identify the minimum $\tau$ value that eliminates underflow zeros while preserving moderate cases.

Let $\varepsilon(t_{\text{obs}};\tau)$ denote the SLLS epsilon produced at target factor $\tau$ and define the index set of underflowing tests at $\tau$ as

\begin{equation}
 \mathcal Z(\tau)=\{j:\ \hat p_j(\tau)\leq \text{.Machine.double.xmin}\}.
\end{equation}

We seek the minimal $\tau^\star=\inf \{\tau:\ |\mathcal Z(\tau)|=0\}$.
Assuming monotonicity of the “no-underflow” event with increasing $\tau$ (empirically observed), we use a two-phase search:

\begin{itemize}
    \item \textbf{Phase 1 (expansion / bracketing):} Starting at $\tau_0$ (default 0.25), we step additively by $s$ and inflate the step geometrically $s\leftarrow s\cdot\rho$ until $|\mathcal Z(\tau)|=0$ is achieved, producing a bracket $[\tau_{\text{lo}},\tau_{\text{hi}}]$ with underflow at $\tau_{\text{lo}}$ and none at $\tau_{\text{hi}}$. To reduce cost, at each try we fit the GPD only for tests currently in $\mathcal Z(\tau)$.

    \item \textbf{Phase 2 (bisection / refinement):} Bisection of $[\tau_{\text{lo}},\tau_{\text{hi}}]$ until the width falls below a numerical tolerance, returning $\tau^\star\approx\tau_{\text{hi}}$. During bisection we again refit only the currently underflowing subset. Finally, we perform one full refit for all selected tests at $\tau^\star$ and report the resulting $p$-values.
\end{itemize}

\subsubsection*{Defaults and stopping}
Expansion step $s_{\text{init}}=10$, growth $\rho=\frac{1+\sqrt5}{2}\approx1.618$, at most 20 expansion steps; bisection tolerance 0.1 with at most 30 iterations. The procedure logs $\tau$ and $|\mathcal Z(\tau)|$ per iteration for transparency. If expansion hits its cap with $|\mathcal Z(\tau)|>0$, we return the best $\tau$ found and flag this in the messages (users can increase \texttt{step\_init}, \texttt{grow}, or the iteration caps).

\subsubsection*{Computational notes}
The search is subset-aware: only tests with underflow are refit during exploration, yielding substantial speedups when zeros are rare. The final results are obtained from a single full refit at $\tau^\star$, so intermediate exploration does not affect the reported estimates. Progress bars (optional) mirror those used in threshold detection.

This refinement aims to eliminate, up to machine precision, all reported zeros due to underflow while keeping $\tau$ as small as possible, thereby preserving power and avoiding overly conservative inflation of $\varepsilon$ in routine cases. If the expansion phase hits its iteration cap with remaining underflowing tests, the best available $\tau$ is returned and this is flagged in the output.

%%%%%%%%%%%%%%%%%%%%%%%%%%%%%%%%%%%%%%%%%%
% Appendix B
%%%%%%%%%%%%%%%%%%%%%%%%%%%%%%%%%%%%%%%%%%
\newpage
\section{Simulation studies to find a default \texttt{permApprox} configuration}
\label{app:sec:default}

The simulation studies in this section serve to determine an optimal configuration for \texttt{permApprox}, with the goal of estimating small $p$-values for a limited number of permutations as accurately as possible.

%============================================================
\subsection{Study on the number of starting exceedances}
\label{app:sec:default:exceed0}
%============================================================

This analysis investigates how the number of starting exceedances used in the GPD tail modeling, denoted by $k_0$, affects the stability and accuracy of permutation-based $p$-value approximation.

The simulated data are identical to those described in Section~\ref{sec:sim:ttest}, consisting of two-sample $t$-test statistics based on Gaussian data with $n = 100$ observations per group and an effect size of $d = 1$.  
Permutation-based test statistics were computed for varying numbers of permutations ($B \in \{500, 1000, 5000, 10000\}$), and the resulting permutation distributions were used to assess the influence of $k_0$ on GPD fitting and $p$-value approximation.

Five strategies were compared: a fixed number of 100 and 250 exceedances, and relative fractions of 10\%, 25\%, and 50\% of all permutations.  
All fits were obtained using \texttt{permApprox} with the proposed constrained GPD fitting approach (with LME as estimator and FTR for threshold detection).  

\begin{figure}[H]
    \centering
    \includegraphics[width=\textwidth]{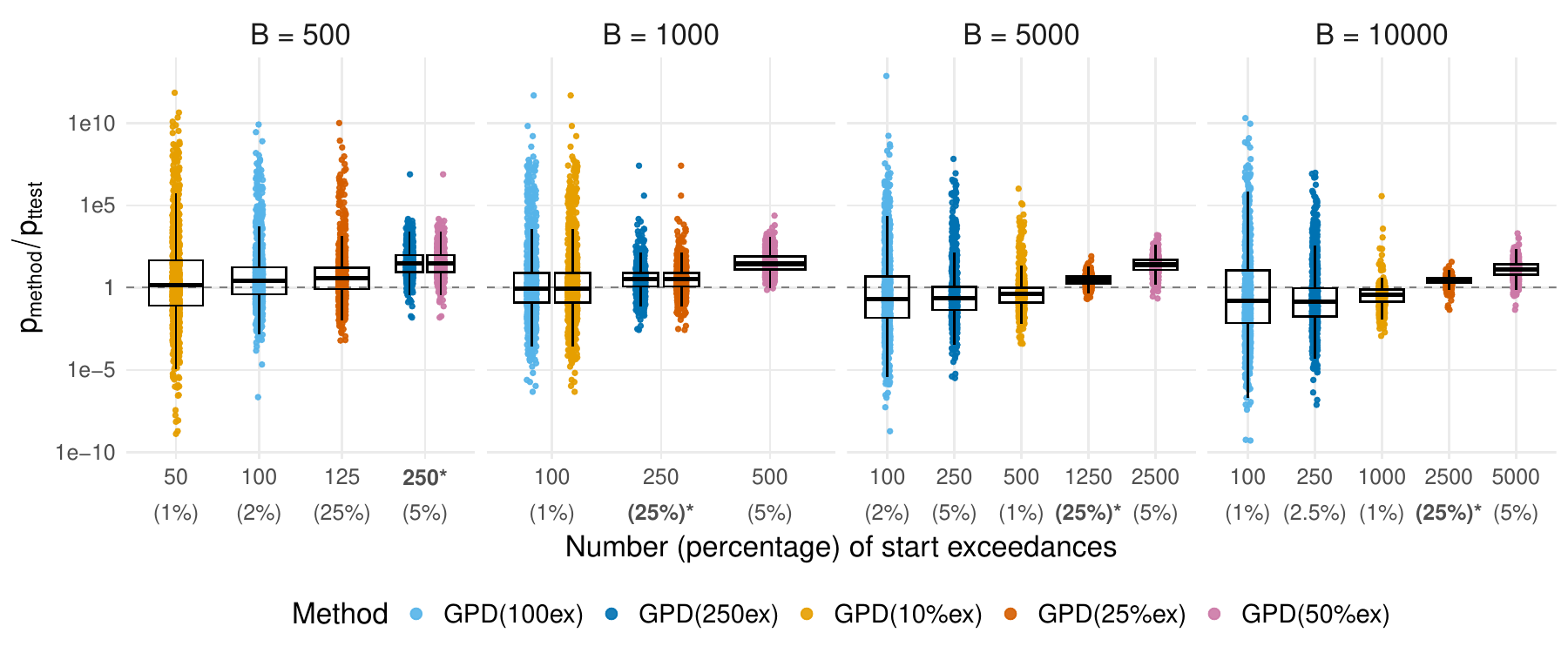}
    \caption{Ratio of approximated permutation $p$-values to the parametric $t$-test reference ($p_{\text{method}} / p_{t\text{-test}}$) for different numbers of permutations ($B$). Each facet corresponds to one $B$, and boxplots summarize the distribution of ratios across 1000 simulation replicates. On the $x$-axis, the numbers denote the starting exceedances $k_0$, while the percentages refer to fractions of $B$, used to initiate the FTR-based threshold detection. The starred methods denote the default choice in \texttt{permApprox} ($k_0 = \mathrm{max}(0.25B, 250)$).}
    \label{app:fig:nexceed}
\end{figure}

Figure~\ref{app:fig:nexceed} displays the ratios of approximated permutation $p$-values to the parametric $t$-test reference ($p_{\text{method}} / p_{t\text{-test}}$) across different $B$.  
For $B = 500$, using $k_0 = 250$ (corresponding to 50\% of $B$) resulted in $p$-values slightly above those of the $t$-test but with substantially lower variation than smaller $k_0$ values.  
For $B = 1000$, both $k_0 = 250$ and $k_0 = 500$ yielded acceptable variance, while the median of $k_0 = 250$ was closest to the $t$-test reference.  
For larger $B$, the variation decreased up to $k_0 = 25\%$ of $B$ and increased thereafter, with the median remaining closest to the $t$-test values at 25\%.  
We chose $k_0 = \mathrm{max}(0.25B, 250)$ as the default starting point in \texttt{permApprox} because it is a good tradeoff between bias and stability across all numbers of permutations.

%============================================================
\subsection{Comparison of threshold detection methods}
\label{app:sec:default:thresh_methods}
%============================================================

This analysis compares the performance of different threshold detection methods. 
We used the same simulation setup as described in Section~\ref{sec:sim:ttest}, applying the constrained GPD fitting procedure (with LME as estimator and the SLLS rule for the support constraint) to approximate permutation-based $p$-values for a range of sample sizes (\(n \in \{50, 100, 500, 1000\}\)). 
The number of permutations was fixed to $B = 1000$ and the threshold search was initiated with $k_0 = 0.25\,B = 250$.

All threshold methods described in Section~\ref{sec:methods:threshold} were evaluated. 
For the FTR-based methods (FTR, robFTR), a step size of 10 was used for scanning candidate thresholds. 
For the remaining methods (FwdStop, GOF-CP, Shape-Var), the search was performed over all possible thresholds above $k_0$ (step size 1).

\begin{figure}[H]
    \centering
    \includegraphics[width=\textwidth]{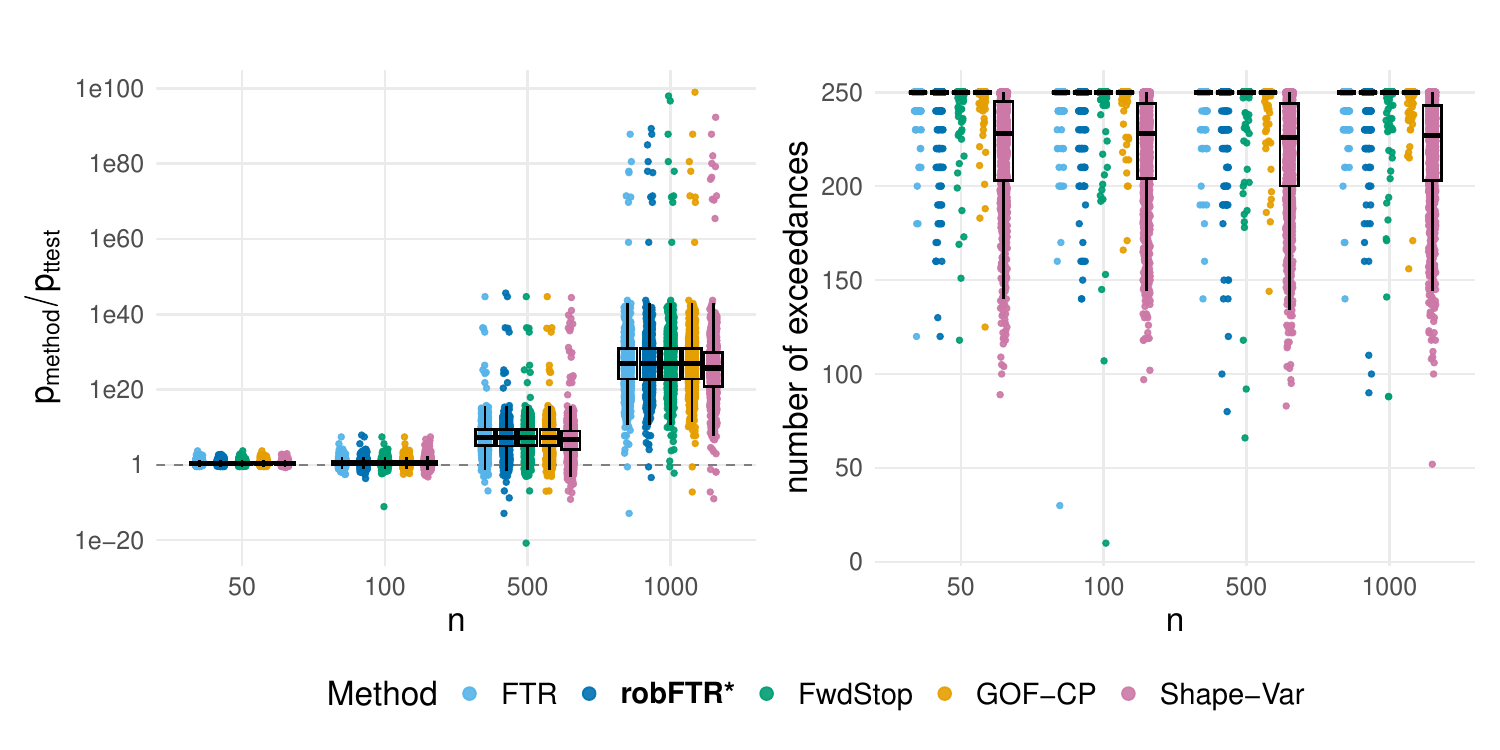}
    \caption{Comparison of threshold detection methods. Left: Ratios of approximated permutation $p$-values to the parametric $t$-test reference ($p_{\text{method}} / p_{t\text{-test}}$). Right: Number of exceedances selected by each method. 
    Each panel shows jittered boxplots across simulation replicates, grouped by sample size ($n$) on the $x$-axis and colored by threshold detection method. The $p$-value approximations were obtained using constrained LME fitting with the SLLS $\varepsilon$-rule. All methods started with $k_0 = 250$ exceedances. $k_0$ was iteratively increased by 10 for FTR and robFTR, and by 1 for the other methods. The starred method (robFTR*) is the default method in \texttt{permApprox}.}
    \label{app:fig:threshold_methods}
\end{figure}

Figure~\ref{app:fig:threshold_methods} summarizes the results. 
The left panel shows the ratio between the approximated permutation $p$-values and the parametric $t$-test reference. 
All methods yield nearly identical $p$-values, with only the Shape-Var method producing values slightly closer to the reference. This method also selects thresholds with much higher variability, as seen in the right panel, which shows the number of exceedances selected by each method. 
For the other methods, the chosen number of exceedances is either exactly or close to the initial number of exceedances, which is 250.

Due to their comparable accuracy, we base the default choice on computational considerations. 
The FTR methods are substantially faster because they (i) perform reliably even with a coarser threshold grid (step size 10), and (ii) stop the search early once the Anderson–Darling (AD) test is not rejected.
Among them, we select robFTR as the default in \texttt{permApprox}, as it is more robust to outliers in the AD $p$-value distribution, while maintaining the speed advantage of early stopping.

%=========================================
\subsection{Comparison of GPD parameter estimation methods}
\label{app:sec:default:estimators}
%=========================================

In this section, we evaluate the GPD parameter estimation methods explained in Section~\ref{sec:methods:gpd_estimators} in terms of accuracy, robustness, and runtime. The evaluated methods are the Method of Moments (MOM), the one- and two-parameter maximum likelihood estimators (MLE1D and MLE2D), the likelihood moment estimator (LME), the Zhang-Stephens estimator (ZSE), the two-step nonlinear least squares estimator (NLS2), and the weighted nonlinear log-least squares estimator (WNLLSM).
These estimators form the foundation of our approach, as all subsequent steps in the \texttt{permApprox} pipeline rely on reliable estimation of the GPD scale ($\sigma$) and shape ($\xi$) parameters.

Our goal in this section is to identify a default estimator that performs consistently well across different conditions and can be efficiently adapted to the support constraint introduced in Section~\ref{sec:methods:constraint}. 
We therefore examine the behavior of the methods under both unconstrained and constrained estimation settings. 

%-----------------------------------------
\subsubsection{Tolerance selection for GPD optimization methods}
\label{app:sec:default:estimators:tol}
%-----------------------------------------

All optimization-based GPD estimators in \texttt{permApprox} depend on a numerical tolerance parameter that determines the convergence criterion of the underlying optimization routine.
Because each method minimizes a different objective function (e.g., log-likelihood or least-squares loss), the same tolerance value does not correspond to the same numerical or statistical accuracy across estimators.
To ensure comparability and numerical stability across the optimization-based GPD parameter estimation methods, we systematically investigated the influence of the optimization tolerance (\texttt{tol}) on the fitted densities at a given evaluation point above the true boundary. For each method, we fitted the GPD to simulated data with shape parameter $\xi = -0.2$ and scale parameter $\sigma = 1$, resulting in a true boundary at 
$s = -\sigma / \xi = 5$. 
The GPD was fitted under a boundary constraint at the evaluation point $X_{\mathrm{eval}} =7$, and the corresponding probability density at this point, 
$f_{\mathrm{GPD}}(X_{\mathrm{eval}})$, was evaluated across a range of tolerances ($10^{-5}$–$10^{-15}$). 

\begin{figure}[H]
\centering
\includegraphics[width=\linewidth]{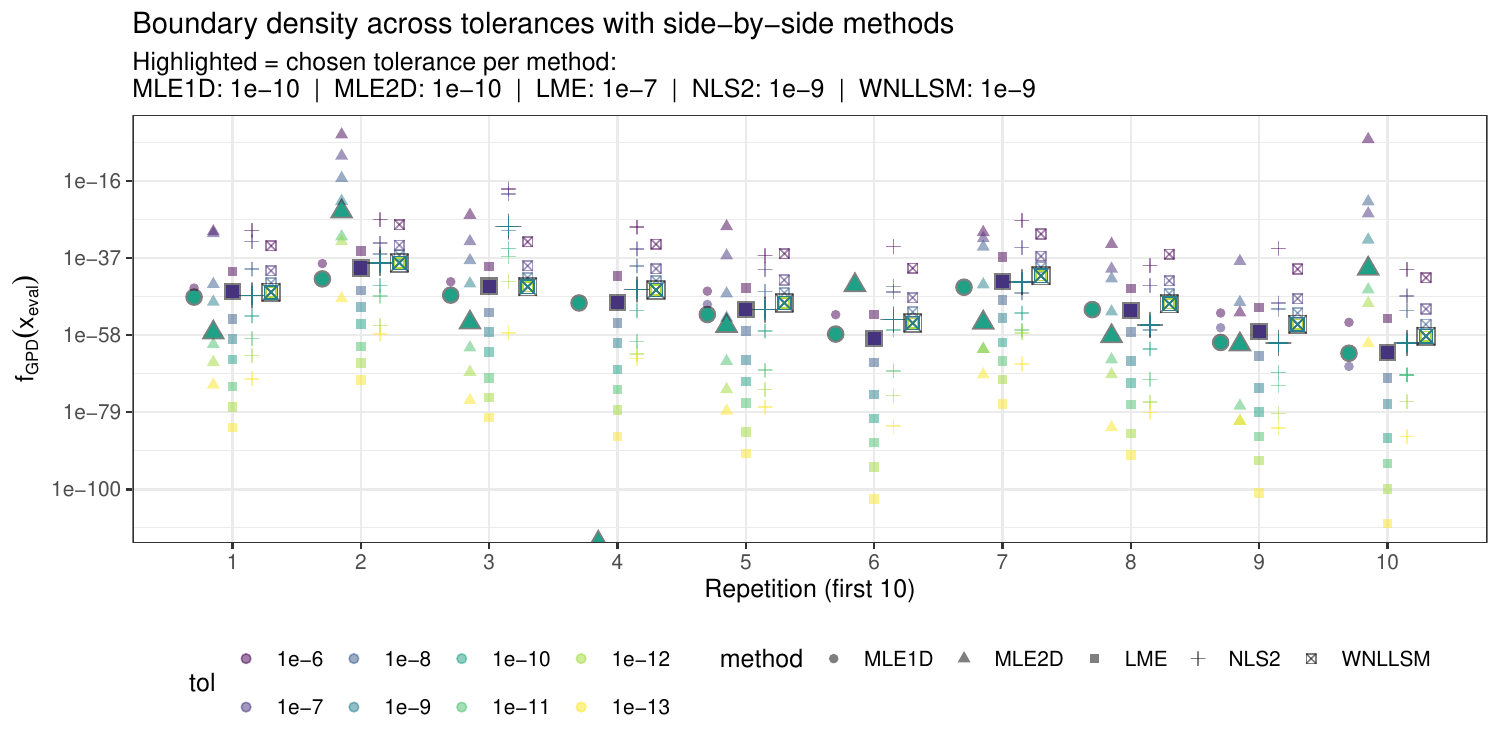}
\caption{GPD densities at an evaluation point ($X_{\mathrm{eval}}$ = 7) above the true boundary ($-\sigma/\xi=5$) across repetitions and optimization tolerances ($10^{-6}$–$10^{-13}$). GPD parameters were fitted with constraint $-\sigma/\xi>X_{\mathrm{eval}}$. Each point represents one simulation repetition, colored by tolerance and shaped by method. Highlighted points mark the chosen tolerance for each method.}
\label{app:fig:tol_calibration}
\end{figure}

Figure~\ref{app:fig:tol_calibration} illustrates that some methods produce a wide range of density values depending on the optimization tolerance, i.e., the density can become extremely small for MLE1D, LME, and NLS2, while MLE1D and WNLLSM becomes stable for certain tolerance values and won't get any smaller. Such variation would considerably affect the tail probability and thus the resulting permutation $p$-value approximation within the \texttt{permApprox} workflow.

To decide on an optimal tolerance value for each method, we first identified the most numerically stable estimator across tolerances. The one-dimensional maximum likelihood estimator (MLE1D) showed the smallest variation in $f_{\mathrm{GPD}}(X_{\mathrm{eval}})$ and was therefore used as a \emph{ground truth reference}. For all other methods, we computed the mean absolute error (MAE) of the log-transformed densities at $X_{\mathrm{eval}}$ relative to the MLE1D reference across repetitions and tolerance values. Instead of selecting the tolerance with the absolute minimum MAE, we applied a plateau criterion: among all tolerances yielding an MAE within 0.005 of the minimum value (on the log scale), we selected the largest tolerance value. This strategy favors numerically stable yet computationally efficient settings, ensuring that further tightening of \texttt{tol} would not meaningfully change the estimates.

The selected tolerance values for each method are summarized in Table~\ref{app:tab:tol_settings}, and were used consistently in all subsequent simulation studies and applications.

\begin{table}[H]
\centering
\caption{Chosen optimization tolerances for each GPD parameter estimation method. The ZSE method is not optimization-based and therefore does not require a tolerance parameter.}
\label{app:tab:tol_settings}
\begin{tabular}{ll}
\toprule
\textbf{Method} & \textbf{Selected tolerance} \\
\midrule
LME      & 1e-07  \\
MLE1D    & 1e-10 \\
MLE2D    & 1e-10 \\
NLS2     & 1e-09  \\
WNLLSM   & 1e-09  \\
ZSE      & --- (no optimization) \\
\bottomrule
\end{tabular}
\end{table}

%-----------------------------------------
\subsubsection{Approximation accuracy}
\label{app:sec:default:estimators:unconstrained}
%-----------------------------------------

We next evaluate each estimator’s ability to recover the true GPD parameters from synthetic exceedances generated without any boundary constraint. 
Samples were drawn as
\begin{equation}
Y_i \sim \mathrm{GPD}(\sigma_0 = 1,\, \xi_0),
\quad
\xi_0 \in \{-0.6, -0.4, -0.2, 0, 0.2, 0.4, 0.6\},
\end{equation}
with sample sizes
\begin{equation}
n \in \{25, 50, 75, 100, 250, 500, 750, 1000\},
\end{equation}
and 500 independent replicates per configuration. 
For each replicate, parameter estimates $(\hat{\sigma}, \hat{\xi})$ were obtained using \texttt{fit\_gpd()} from the \texttt{permApprox} package. 
Estimation accuracy was quantified by the empirical root mean squared error (RMSE) across simulation replicates:
\begin{equation}
\mathrm{RMSE}(\xi) = 
\sqrt{\frac{1}{R}\sum_{r=1}^{R}(\hat{\xi}_r - \xi_0)^2},
\qquad
\mathrm{RMSE}(\sigma) = 
\sqrt{\frac{1}{R}\sum_{r=1}^{R}(\hat{\sigma}_r - \sigma_0)^2},
\end{equation}
where $R = 500$ is the number of replicates. 
Lower RMSE values indicate higher estimation accuracy.

\begin{figure}[H]
  \centering
  \includegraphics[width=\textwidth]{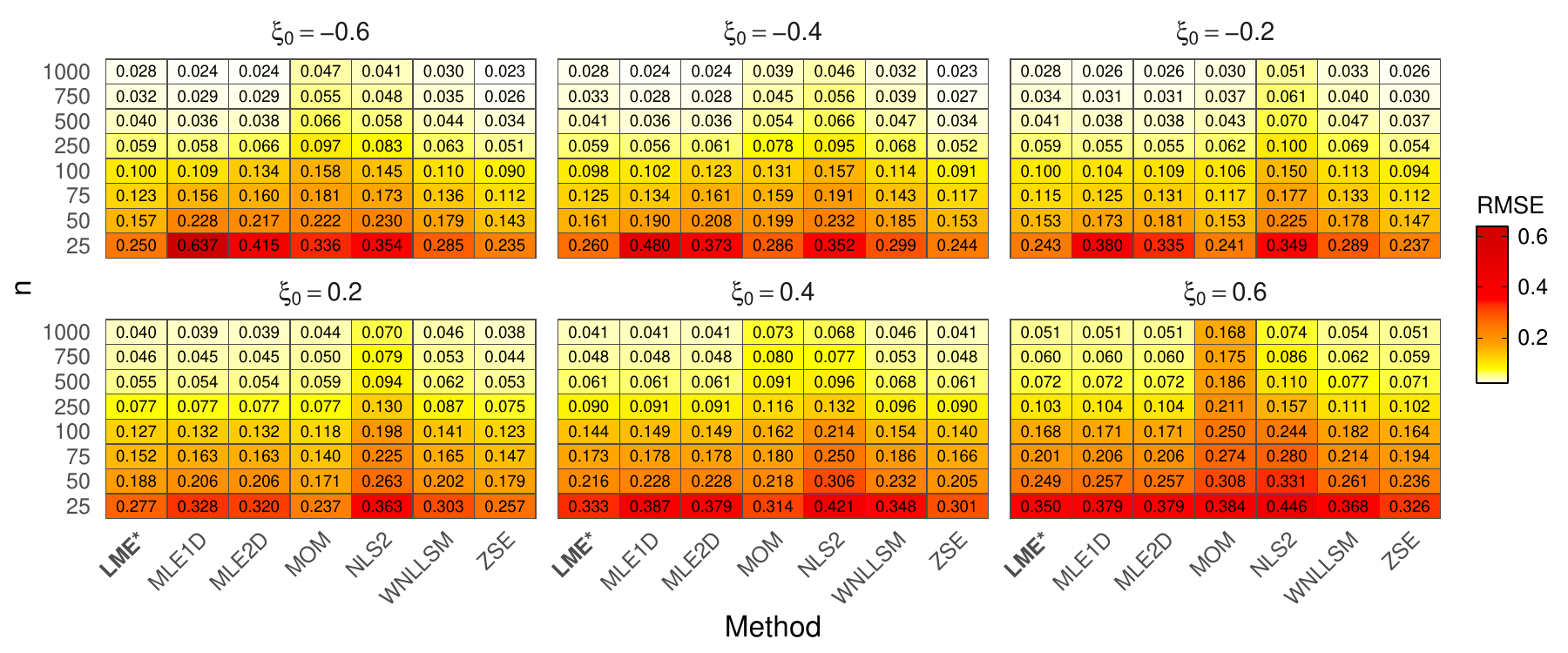}
  \caption{Root mean squared error (RMSE) of the estimated GPD shape parameter $\hat{\xi}$ compared to the true value $\xi_0$ across sample sizes $n$, estimation methods, and true shape values $\xi_0\in \{-0.6, -0.4, -0.2, 0.2, 0.4, 0.6\}$. 
  The color scale represents absolute RMSE values, ranging from the minimum to the maximum observed across all scenarios. Lighter colors correspond to lower RMSE and thus higher estimation accuracy. The starred method (LME*) is the default estimator in \texttt{permApprox}.}
  \label{app:fig:sim_unconstrained_rmse_shape}
\end{figure}

\begin{figure}[H]
  \centering
  \includegraphics[width=\textwidth]{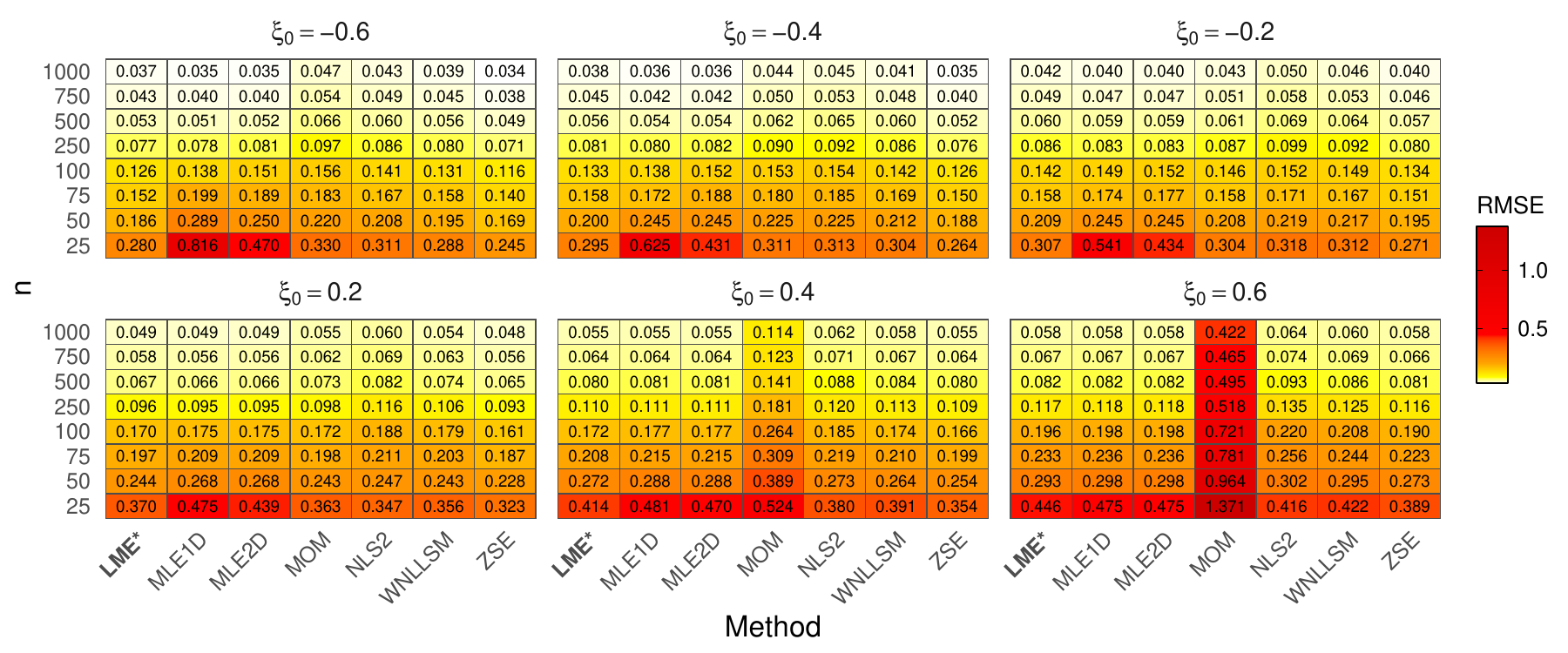}
  \caption{Root mean squared error (RMSE) of the estimated GPD scale parameter $\hat{\sigma}$ compared to the true value $\sigma_0 = 1$ across sample sizes $n$, estimation methods, and true shape values $\xi_0\in \{-0.6, -0.4, -0.2, 0.2, 0.4, 0.6\}$. 
  The color scale represents absolute RMSE values, ranging from the minimum to the maximum observed across all scenarios. Lighter colors correspond to lower RMSE and thus higher estimation accuracy. The starred method (LME*) is the default estimator in \texttt{permApprox}.}
  \label{app:fig:sim_unconstrained_rmse_scale}
\end{figure}

\begin{figure}[H]
  \centering
  \includegraphics[width=\textwidth]{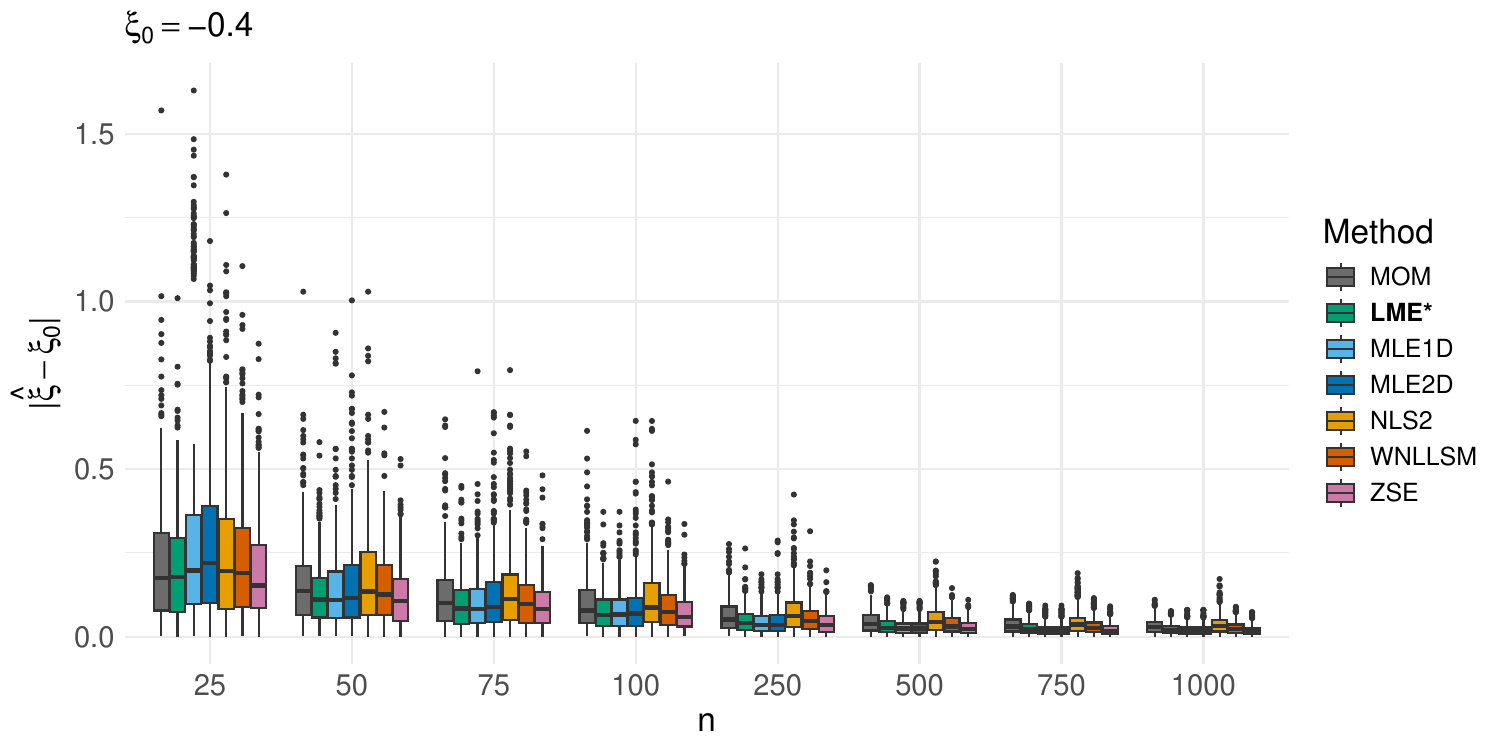}
  \caption{Absolute estimation errors of the GPD shape parameter ($|\hat{\xi} - \xi_0|$) for $\xi_0 = -0.4$ across sample sizes.
  Boxplots summarize the distribution of absolute errors across 500 replicates per setting, with colors indicating the estimation methods. The starred method (LME*) is the default estimator in \texttt{permApprox}.}
  \label{app:fig:sim_unconstrained_error_shape}
\end{figure}

Figures~\ref{app:fig:sim_unconstrained_rmse_shape} and~\ref{app:fig:sim_unconstrained_rmse_scale} summarize the results for the shape and scale parameters, respectively, across all estimators, sample sizes, and true shape values. 
In both panels, the color scale represents the absolute RMSE values, ranging from the minimum to the maximum observed across all scenarios, thereby allowing a direct comparison of estimation accuracy between methods and parameter settings. The distribution of absolute estimation errors for $\xi_0 = -0.4$ across sample sizes is shown in Figure~\ref{app:fig:sim_unconstrained_error_shape}.

As expected, absolute errors and RMSE values generally decrease with increasing sample size for all estimators, reflecting the asymptotic consistency of the methods. 
Across all shape parameters, the Zhang–Stephens estimator (ZSE) achieved the lowest or near-lowest RMSE within each sample size, occasionally surpassed by the method of moments (MOM) for shape values close to zero.
The likelihood moment estimator (LME) exhibited comparably high accuracy and closely followed the performance of ZSE across the entire range of $\xi_0$ and $n$.
Both maximum-likelihood estimators (MLE1D and MLE2D) converged to LME’s performance for large $n \ge 250$ but were less accurate for small samples ($n \le 50$). 
MOM performed best when $\xi_0 \sim 0$ but deteriorated quickly for nonzero $\xi_0$, particularly for positives shapes. 
By contrast, NLS2 consistently produced the largest RMSE values for both parameters. 
Overall, ZSE and LME emerged as the most reliable estimators for unconstrained GPD fitting in terms of both shape and scale recovery.

%-----------------------------------------
\subsubsection{Robustness under support constraints}
\label{app:sec:default:estimators:constrained}
%-----------------------------------------

To examine robustness under constrained tail fitting, each estimator is evaluated in a setting where the fitted GPD must satisfy a support constraint at an evaluation point above the true boundary. 
Data are generated from a bounded GPD with $\sigma_0 = 1$ and $\xi_0 = -0.2$, implying a true boundary at $s_0 = -\sigma_0 / \xi_0 = 5$. 
For each sample size $n \in \{25, 50, 75, 100, 250, 500, 750, 1000\}$, the evaluation point is set to
\begin{equation}
x_{\mathrm{eval}} = s_0 + \varepsilon \quad\text{with}\quad \varepsilon \in \{1, 2, 3\}
\;\;(\Rightarrow\; x_{\mathrm{eval}} \in \{6, 7, 8\}),
\end{equation}
and each estimator is fitted under the constraint $\hat{s} = -\hat{\sigma} / \hat{\xi} > x_{\mathrm{eval}}$. 

In this setting, direct comparison of parameter estimates to their true values is not meaningful, as the imposed support constraint introduces a deliberate model misspecification ($s_0 < x_{\mathrm{eval}} < \hat{s}$). 
A suitable default estimator for \texttt{permApprox} should produce densities that are not only stable across samples but also sufficiently small to yield accurate approximations of small \(p\)-values. 
We therefore assess the stability of the fitted density at the evaluation point, \(f_{\widehat{\mathrm{GPD}}}(x_{\mathrm{eval}})\), which reflects the numerical robustness of the fit and directly impacts the resulting tail probabilities. 

\begin{figure}[H]
  \centering
  \includegraphics[width=\textwidth]{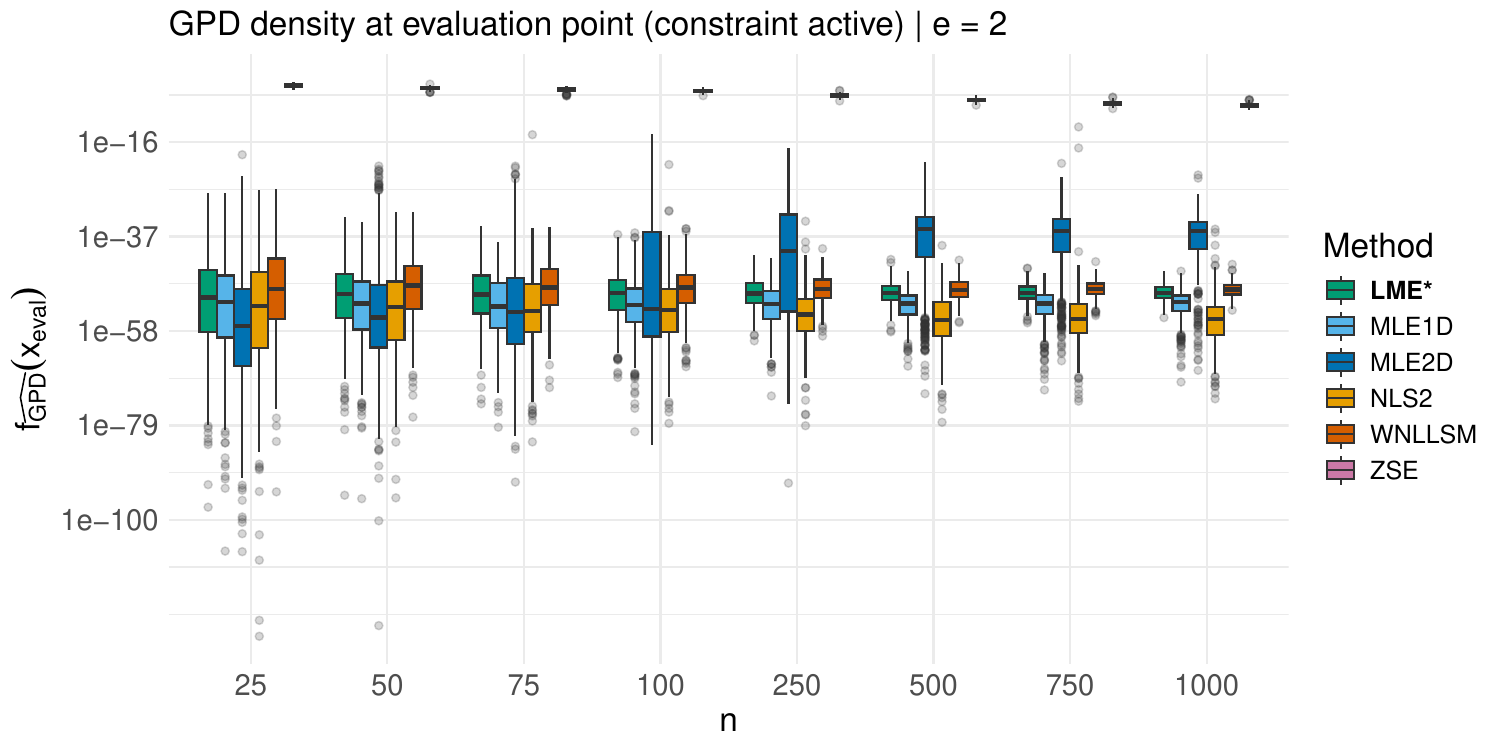}
  \caption{
    Constrained fits at \(x_{\mathrm{eval}} = 7\) (\(\varepsilon = 2\)): grouped boxplots of the fitted density values \(f_{\widehat{\mathrm{GPD}}}(x_{\mathrm{eval}})\) across sample sizes \(n\) and estimation methods. 
    Only repetitions where the constraint \(\hat{s} = -\hat{\sigma}/\hat{\xi} > x_{\mathrm{eval}}\) was active are shown. 
    The y-axis is on a log10 scale, while tick labels show the original $p$-values. The starred method (LME*) is the default in \texttt{permApprox}.
  }
  \label{app:fig:sim_constrained_dgpd_eps2}
\end{figure}

Figure~\ref{app:fig:sim_constrained_dgpd_eps2} presents the results for \(\varepsilon = 2\) (corresponding to \(x_{\mathrm{eval}} = 7\)), which are representative of all evaluated settings, as the cases \(\varepsilon \in \{1, 3\}\) showed qualitatively similar behavior with differences only in magnitude. 
The grouped boxplots display the fitted densities \(f_{\widehat{\mathrm{GPD}}}(x_{\mathrm{eval}})\) across sample sizes and estimation methods.

Although ZSE yields by far the lowest variability in \(f_{\widehat{\mathrm{GPD}}}(x_{\mathrm{eval}})\), its resulting density values are extremely large across all settings. 
Since such inflated values translate into overly conservative \(p\)-value approximations, ZSE is not a suitable candidate for the default configuration in \texttt{permApprox}.  
MLE2D shows substantial variability in both spread and magnitude across sample sizes, indicating instability under the support constraint.

\begin{figure}[H]
  \centering
  \includegraphics[width=\textwidth]{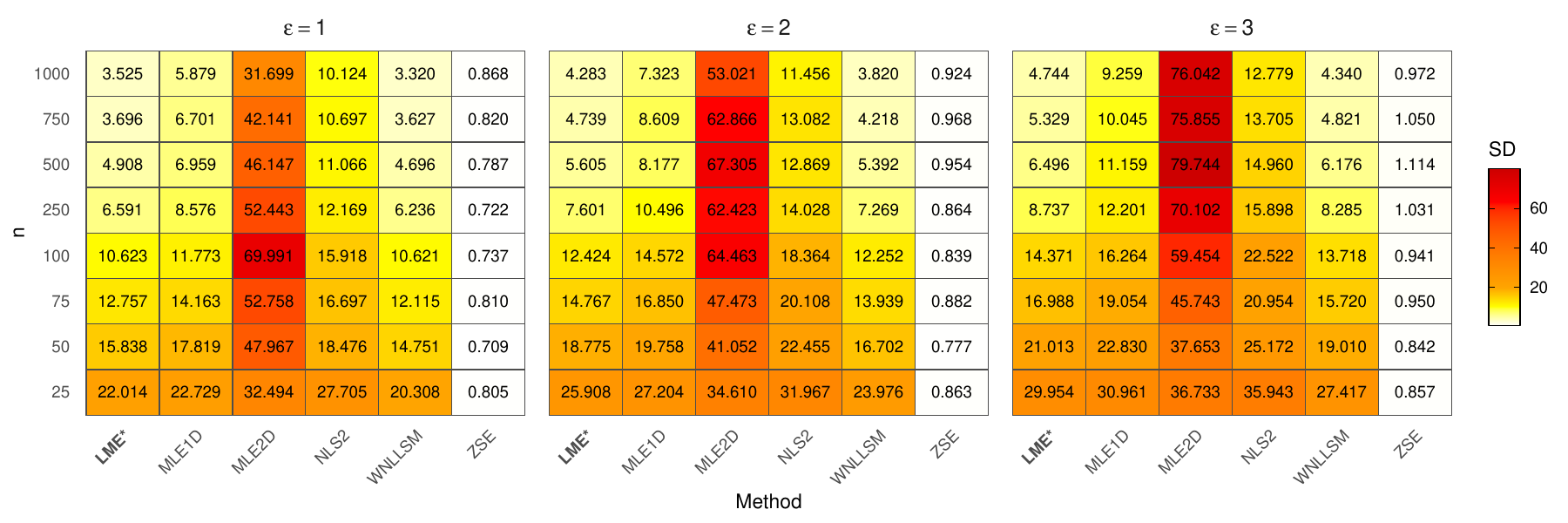}
  \caption{
  Standard deviation of the fitted density values \(f_{\widehat{\mathrm{GPD}}}(x_{\mathrm{eval}})\) across 500 replicates per setting, shown for $\varepsilon \in \{1, 2, 3\}\;\;(\Rightarrow\; x_{\mathrm{eval}} \in \{6, 7, 8\})$. 
  Color intensity reflects standard deviation (SD) of the density: $\sqrt(\mathrm{Var}(f_{\widehat{\mathrm{GPD}}}(x_{\mathrm{eval}})))$, with darker shades indicating higher variability. The starred method (LME*) is the default in \texttt{permApprox}
  }
  \label{app:fig:sim_constrained_dgpd_sd}
\end{figure}

Among the remaining methods, variability is low and consistent. 
Within each sample size, the standard deviations follow the consistent order:  
$\mathrm{SD}_n(\text{WNLLSM}) < \mathrm{SD}_n(\text{LME}) < \mathrm{SD}_n(\text{MLE1D}) < \mathrm{SD}_n(\text{NLS2}),$ though the differences are small.  
These standard deviations are visualized for all three $\varepsilon$ values in Figure~\ref{app:fig:sim_constrained_dgpd_sd}.

%-----------------------------------------
\subsubsection{Runtime comparison of GPD fitting methods}
\label{app:sec:default:estimators:runtime}
%-----------------------------------------

\begin{figure}[H]
  \centering
  \includegraphics[width=\linewidth]{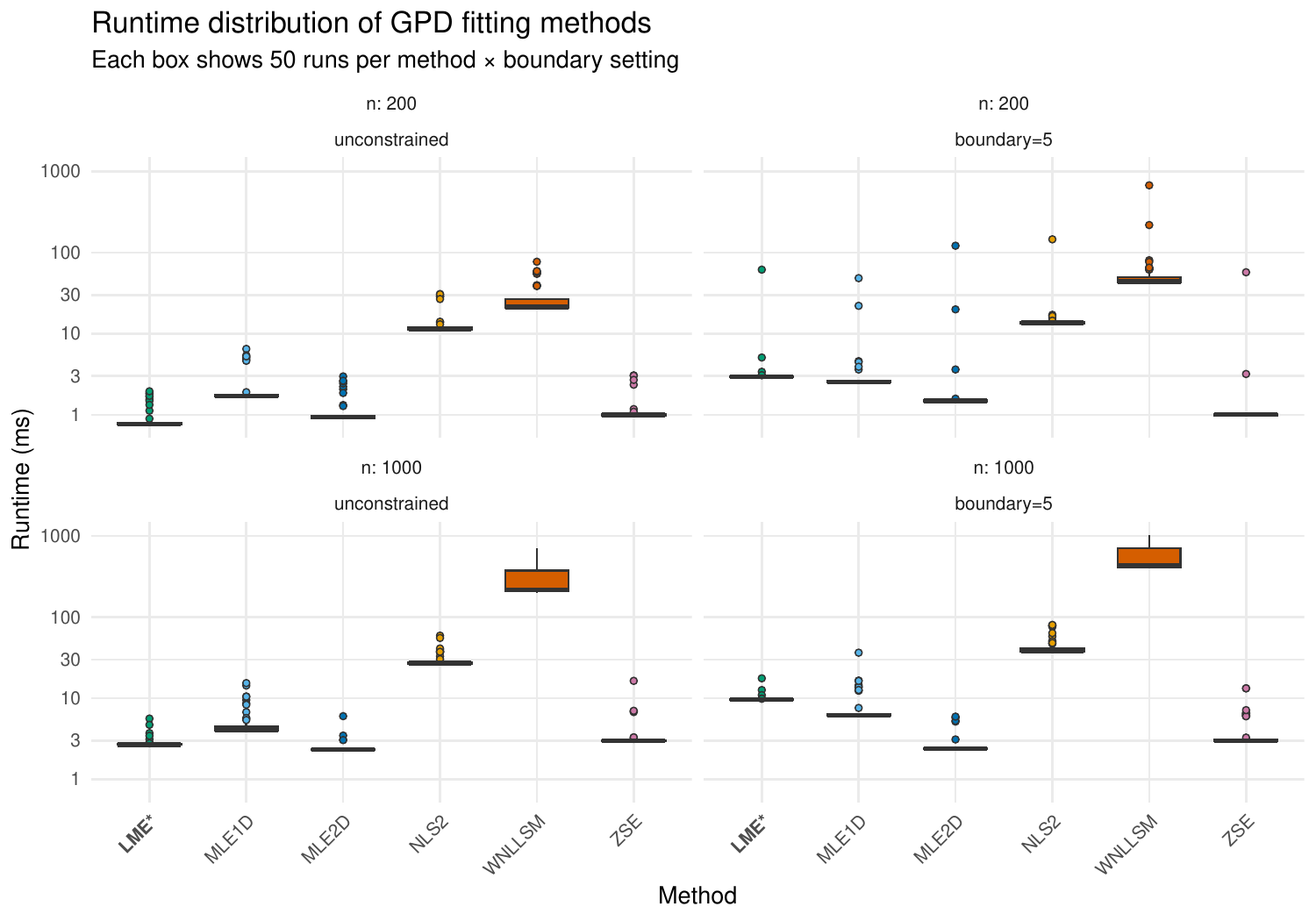}
  \caption{Runtime distribution of GPD fitting methods across sample sizes ($n = 200, 1000$) and boundary settings (with vs.\ without boundary constraint). Each box represents 50 runs per method and scenario. The starred method (LME*) is the default estimator in \texttt{permApprox}.}
  \label{app:fig:runtime_gpd_methods}
\end{figure}

To assess the computational efficiency of the GPD parameter estimation approaches, we benchmarked all methods under different sample sizes ($n = 200$ and $n = 1000$) and boundary settings (with and without constraint). Each combination was evaluated 50 times per method, and runtimes were recorded in milliseconds (Figure~\ref{app:fig:runtime_gpd_methods}). Overall, the runtime of all methods increased with sample size. NLS2 and WNLLSM were by far the slowest methods across all scenarios, reflecting their iterative nature and two-step optimization procedures. The runtime of the remaining methods was comparable, with one fastest method depending on the setting: LME was fastest for small $n$ and unconstrained fitting, ZSE was fastest for small $n$ in the constrained case, and MLE2D was generally the fastest for large sample sizes.

%-----------------------------------------
\subsubsection{Decision for a default estimation method}
\label{app:sec:default:estimators:decision}
%-----------------------------------------

Although, in principle, all evaluated estimators are compatible with the \texttt{permApprox} framework, our choice of a default value is based on the following criteria: balance between estimation accuracy, robustness under support constraints, and computational efficiency.

The likelihood moment estimator (LME) offers the most favorable combination of these criteria. 
In the unconstrained setting, it performs nearly as well as ZSE, which showed the lowest RMSE overall. 
However, under the support constraint, ZSE produces highly inflated density values at the evaluation point. This would result in overly conservative \(p\)-value approximations, making it unsuitable as a default.

Among the remaining methods, MLE2D shows considerable variability in the constrained setting and unstable behavior across sample sizes. 
WNLLSM achieves slightly lower variability than LME but is between 15 and 80 times slower in our runtime benchmarks, which limits its practical use in large-scale applications. 
MLE1D performs comparably to LME but tends to be slightly more variable. 
NLS2 is consistently the least accurate and one of the slowest methods.

Taken together, LME provides a reliable and computationally efficient solution that performs well across scenarios. 
It is therefore chosen as the default estimator in \texttt{permApprox}, while the other methods remain accessible for comparative or specialized use.

%%%%%%%%%%%%%%%%%%%%%%%%%%%%%%%%%%%%%%%%%%
% Appendix C
%%%%%%%%%%%%%%%%%%%%%%%%%%%%%%%%%%%%%%%%%%
\newpage
\section{Material to supplement the simulation studies}
\label{app:sec:simulation_supplement}

% t-test: shape estimates
\begin{figure}[H]
  \centering
  \includegraphics[width=\textwidth]{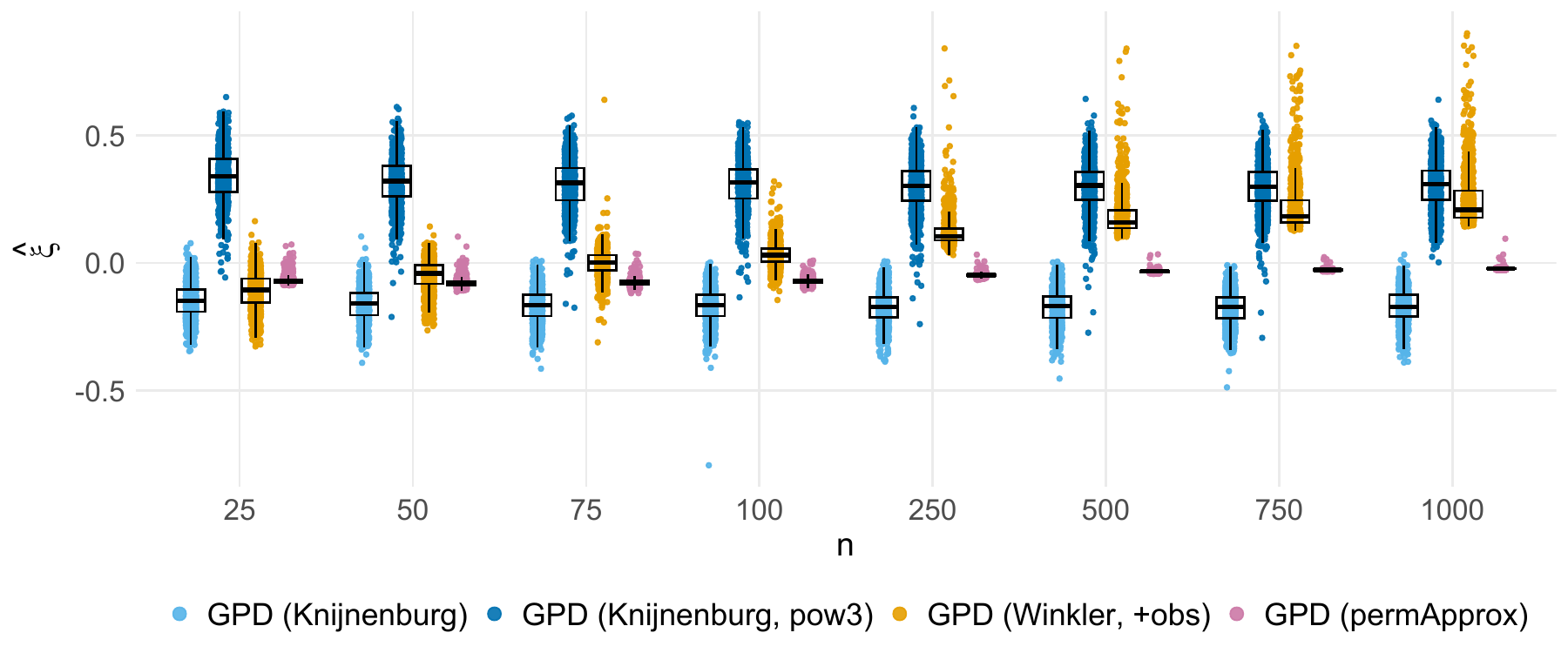}
  \caption{\textbf{Two-sample t-test with Gaussian data:} Estimated GPD shape values across sample sizes for the compared GPD-based methods. Effect size and number of permutations are fixed to $d=1$ and $B=1000$, respectively. Points indicate individual replicates (1000 per setting).}
  \label{app:fig:ttest_shapes_by_n}
\end{figure}

% t-test: p-values by n
\begin{figure}[ht]
  \centering
  \includegraphics[width=\textwidth]{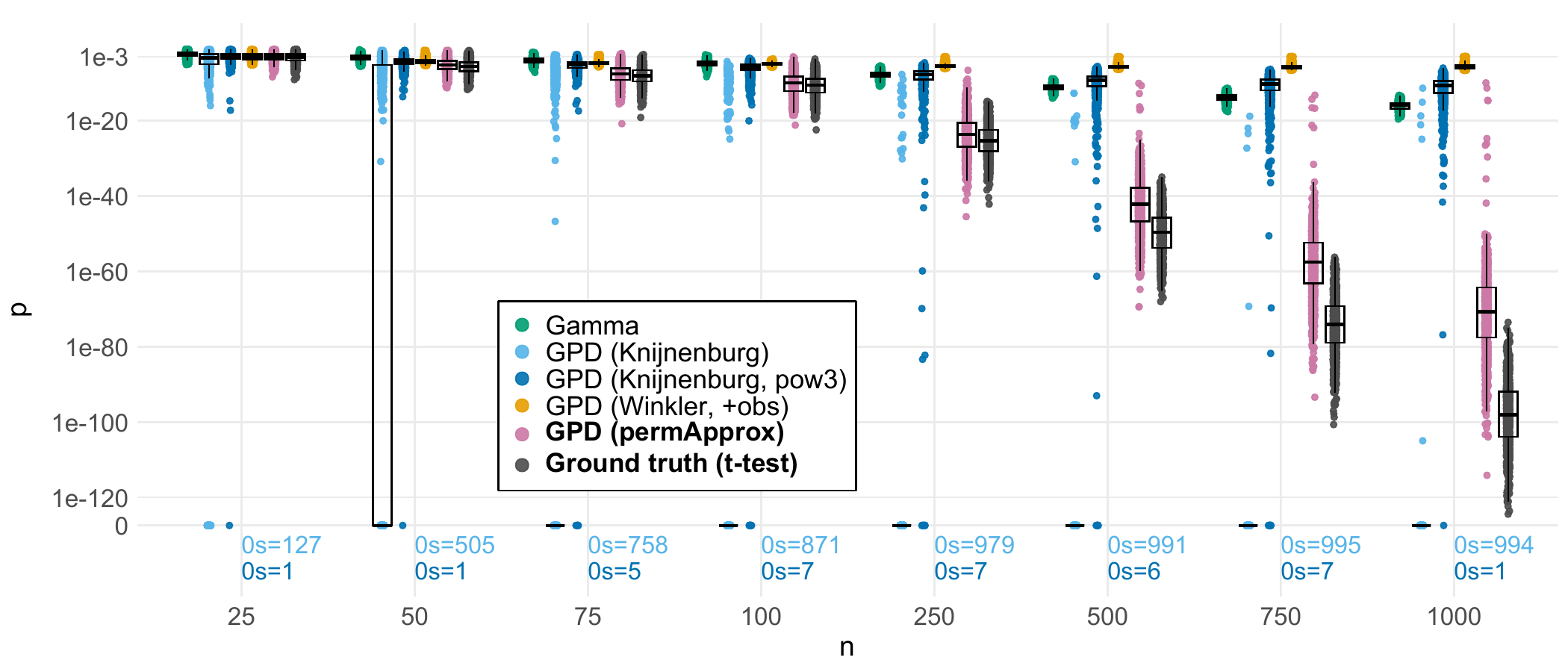}
  \caption{\textbf{Two-sample t-test with Gaussian data:} $p$-values produced with the different approximation methods, in comparison to Student's \textit{t}-test $p$-values across varying sample sizes.
  Effect size and number of permutations are fixed to $d=1$ and $B=1000$, respectively. Points indicate individual replicates (1000 per setting). 
  Counts of exact zeros are shown below each group as ``\texttt{0s}=x'' in the corresponding color.
  Zero $p$-values are mapped to a small constant floor so they appear at ``0'' on the $y$-axis. The y-axis is on a log10 scale, while tick labels show the original $p$-values.}
  \label{app:fig:ttest_pvals_by_n}
\end{figure}

% t-test: p-values by d and B
\begin{figure}[H]
  \centering
  \includegraphics[width=\textwidth]{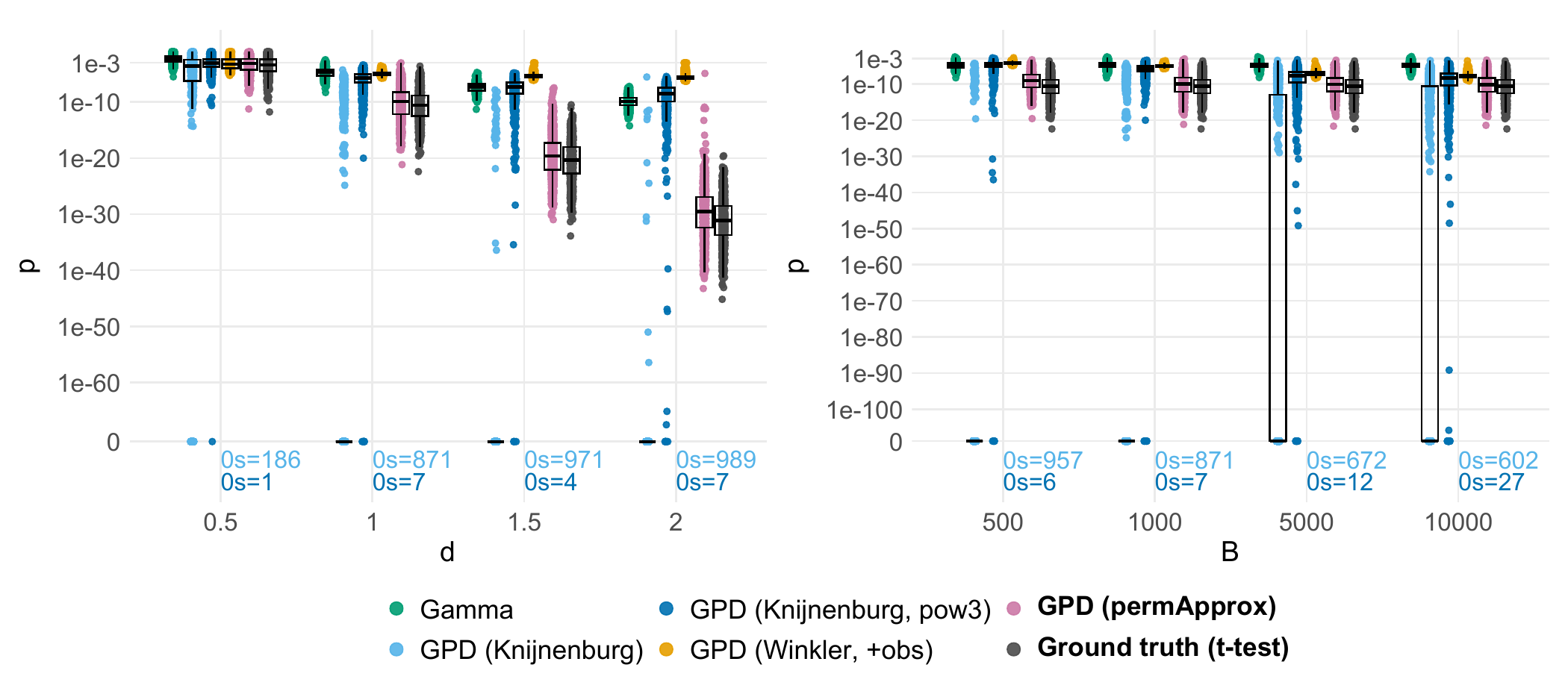}
  \caption{\textbf{Two-sample t-test with Gaussian data:} Approximated permutation $p$-values, stratified by effect size $d$ (left panel) and number of permutations $B$ (right panel). Fixed values (if not stratified): $n=100$, $d=1$ and $B=1000$. Points indicate individual replicates (1000 per setting). 
  Counts of exact zeros are shown below each group as ``\texttt{0s}=x'' in the corresponding color.
  For plotting only, zero $p$-values are mapped to a small constant floor so they appear at ``0'' on the $y$-axis. The y-axis is on a log10 scale, while tick labels show the original $p$-values.}
  \label{app:fig:ttest_pvals_by_d_and_B}
\end{figure}

% t-test: ratios by d and B
\begin{figure}[H]
  \centering
  \includegraphics[width=\textwidth]{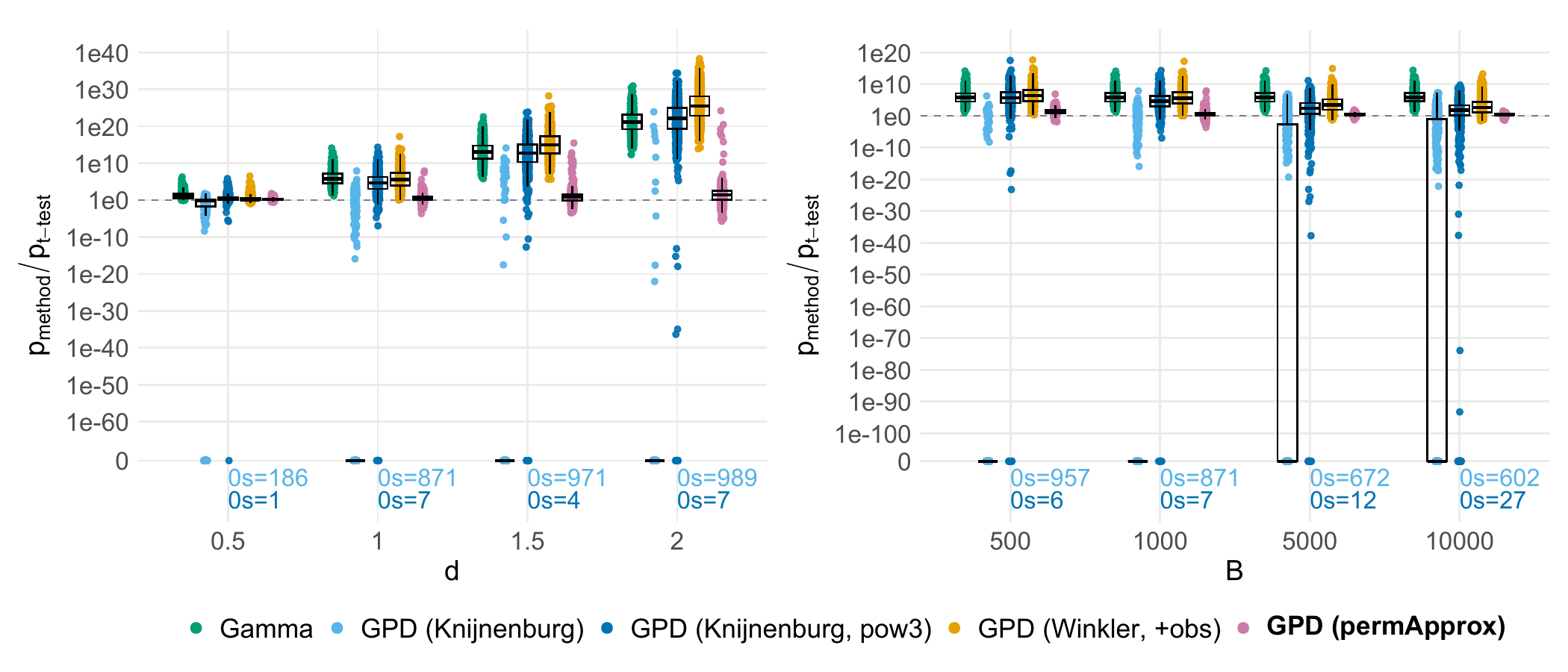}
  \caption{\textbf{Two-sample t-test with Gaussian data:} Ratios of approximated to ground-truth $p$-values: $p_{\text{method}} / p_{t\text{-test}}$, stratified by effect size $d$ (left panel) and number of permutations $B$ (right panel). Fixed values (if not stratified): $n=100$, $d=1$ and $B=1000$. Points indicate individual replicates (1000 per setting). 
  Counts of exact zeros are shown below each group as ``\texttt{0s}=x'' in the corresponding color.
  For plotting only, zero $p$-values are mapped to a small constant floor so they appear at ``0'' on the $y$-axis ratios.}
  \label{app:fig:ttest_ratios_by_d_and_B}
\end{figure}

% Wilcoxon: p-values by n
\begin{figure}[H]
  \centering
  \includegraphics[width=\textwidth]{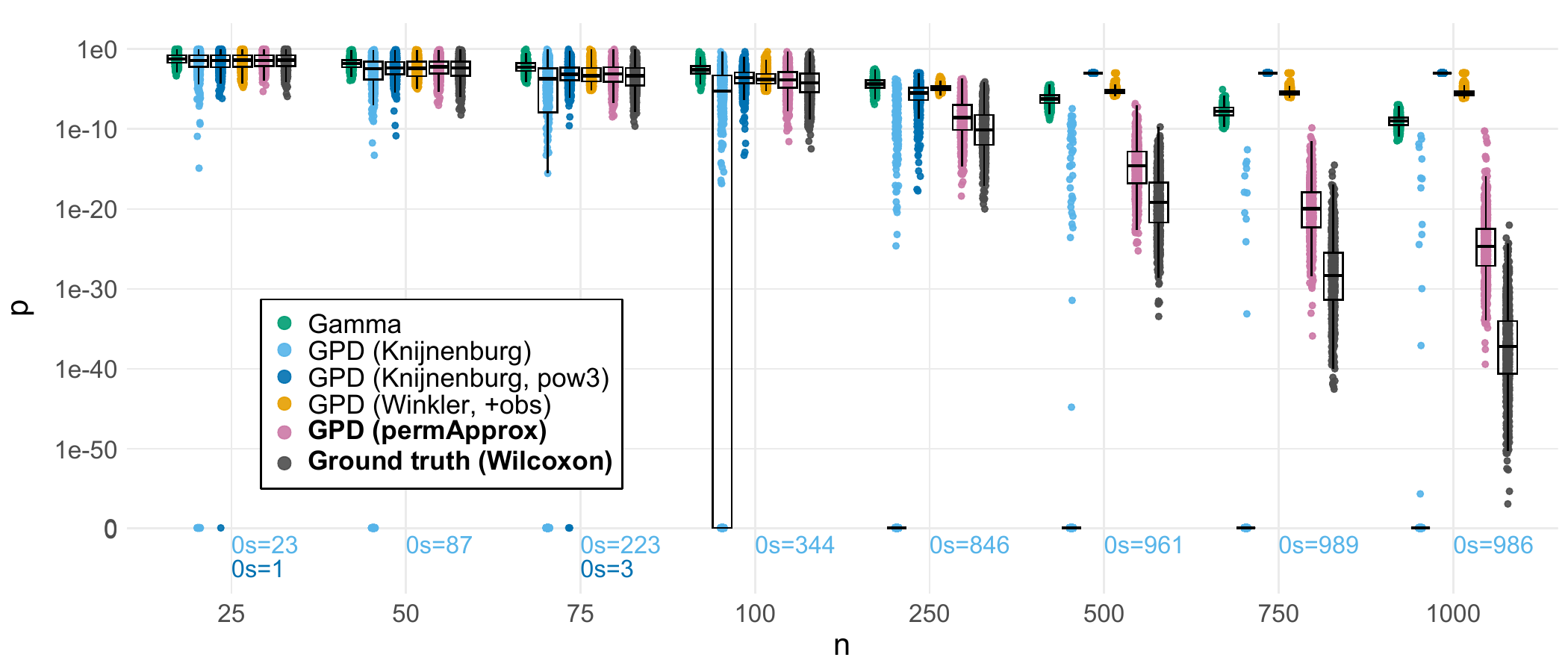}
    \caption{\textbf{Wilcoxon rank-sum test with exponential data:} Approximated permutation $p$-values, stratified by sample size $n$. Fixed values: $d=1$ and $B=1000$. Points indicate individual replicates (1000 per setting). 
    Counts of exact zeros are shown below each group as ``\texttt{0s}=x'' in the corresponding color. 
    For plotting only, zero $p$-values are mapped to a small constant floor so they appear at ``0'' on the $y$-axis. The y-axis is on a log10 scale, while tick labels show the original $p$-values.}
  \label{app:fig:wilcox_pvals_by_n}
\end{figure}

% Wilcoxon: p-values by d and B
\begin{figure}[H]
  \centering
  \includegraphics[width=\textwidth]{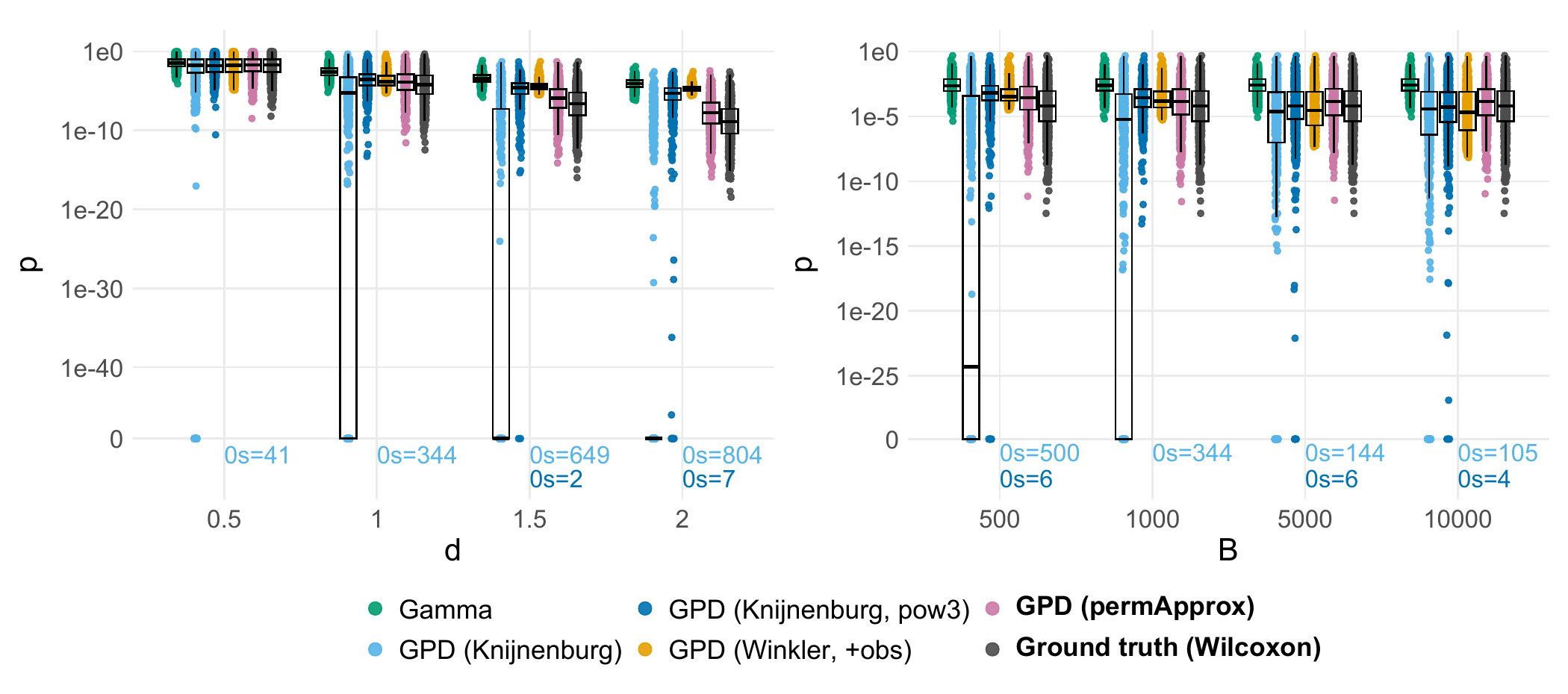}
  \caption{\textbf{Wilcoxon rank-sum test with exponential data:} Approximated permutation $p$-values, stratified by effect size $d$ (left panel) and number of permutations $B$ (right panel). Fixed values (if not stratified): $n=100$, $d=1$ and $B=1000$. Points indicate individual replicates (1000 per setting). 
  Counts of exact zeros are shown below each group as ``\texttt{0s}=x'' in the corresponding color.
  For plotting only, zero $p$-values are mapped to a small constant floor so they appear at ``0'' on the $y$-axis. The y-axis is on a log10 scale, while tick labels show the original $p$-values.}
  \label{app:fig:wilcox_pvals_by_d_and_B}
\end{figure}

% Wilcoxon: ratios by d and B
\begin{figure}[H]
  \centering
  \includegraphics[width=\textwidth]{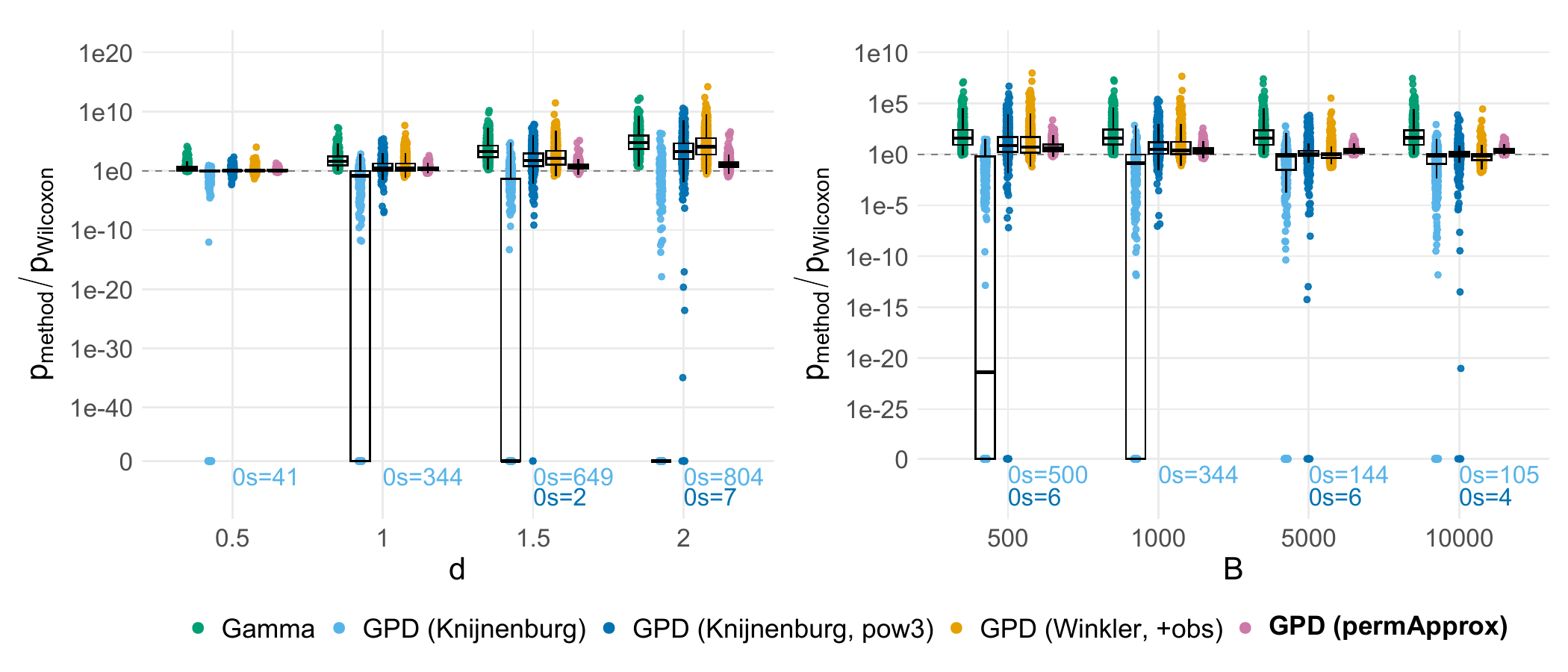}
  \caption{\textbf{Wilcoxon rank-sum test with exponential data:} Ratios of approximated to ground-truth $p$-values: $p_{\text{Method}} / p_{\text{Wilcoxon}}$, stratified by effect size $d$ (left panel) and number of permutations $B$ (right panel). Fixed values (if not stratified): $n=100$, $d=1$ and $B=1000$. Points indicate individual replicates (1000 per setting). 
  Counts of exact zeros are shown below each group as ``\texttt{0s}=x'' in the corresponding color.
  For plotting only, zero $p$-values are mapped to a small constant floor so they appear at ``0'' on the $y$-axis. The y-axis is on a log10 scale, while tick labels show the original ratios.}
  \label{app:fig:wilcox_ratios_by_d_and_B}
\end{figure}

%%%%%%%%%%%%%%%%%%%%%%%%%%%%%%%%%%%%%%%%%%
% Appendix D
%%%%%%%%%%%%%%%%%%%%%%%%%%%%%%%%%%%%%%%%%%
\newpage
\section{Material to supplement the application sections}
\label{app:sec:application_supplement}

%============================================================
\subsection{Differential abundance analysis - The dacomp framework}
\label{app:sec:DA_microbiome_dacomp}
%============================================================

We performed differential abundance (DA) testing with the \texttt{dacomp} \texttt{R} package \citep{brill2021dacomp}, which implements the framework of \citet{brill2022testing}. A central idea of \texttt{dacomp} is that, in compositional microbiome data, differential abundance of an individual taxon cannot be inferred from its raw counts alone, because the sequencing depth is arbitrary and the counts of all taxa are constrained to sum to this depth.  Instead, \texttt{dacomp} tests each taxon relative to a set of \emph{reference taxa} whose absolute abundances are assumed to not change with the phenotype. 

The reference set $\mathcal{B}$ is constructed by identifying taxa whose log-ratios relative to all other taxa vary minimally across samples. For each taxon $j$, dacomp computes
\begin{equation}
\label{eq:dacomp_log-ratios}
\log_{10}\!\left(\frac{X_{ij}+1}{X_{ik}+1}\right),
\qquad k\neq j,
\end{equation}
and summarizes their sample variability by a median score $S_j$. Taxa with $S_j \le S_{\mathrm{crit}}$ form the reference set, where $S_{\mathrm{crit}}$ is chosen to ensure that the total reference counts per sample remain within a practical range.

For each taxon $j \notin \mathcal{B}$, dacomp performs a rarefaction-based test. Let $R_i = \sum_{\ell\in\mathcal{B}} X_{i\ell}$ denote the total reference abundance in sample $i$ and define the rarefaction depth $\lambda_j = \min_i R_i$. For each sample $i$, a rarefied count
$Z_{ij} \sim \mathrm{Hypergeometric}\!\left(\lambda_j,\; X_{ij},\; R_i\right)$
is drawn by sampling $\lambda_j$ reads without replacement from the pool consisting of taxon $j$ and the reference taxa.

The test statistic for taxon $j$ is the Wilcoxon rank-sum statistic applied to the rarefied counts,
\begin{equation}
\label{eq:dacomp_test_statistic}
T_j = \sum_{i : Y_i = 1} \operatorname{rank}(Z_{ij}),
\end{equation}
and its permutation distribution $\{T_j^{*(b)}\}_{b=1}^B$ is obtained by shuffling group labels. Both $T_j$ and the permuted statistics are exported by \texttt{dacomp} and can therefore be used directly as input to \texttt{permApprox}.

%============================================================
\subsection{Differential abundance analysis - Additional results}
\label{app:sec:DA_microbiome_results}
%============================================================

% alpha diversity
\begin{figure}[H]
  \centering
  \includegraphics[width=0.9\textwidth]{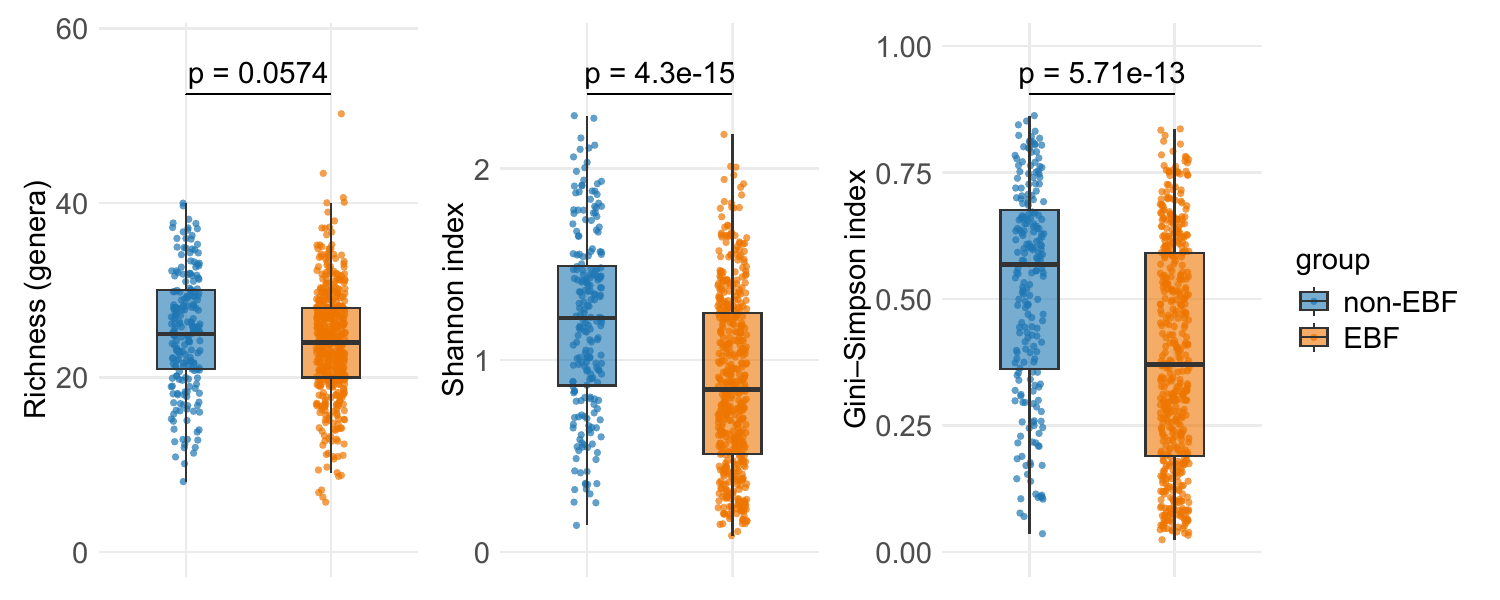}
  \caption{\textbf{Microbiome study:} Alpha diversity in exclusively breastfed (EBF, $n=485$) and non-EBF infants ($n=212$) at age 2 months.
  Richness (number of genera), Shannon diversity, and Gini–Simpson diversity were compared using Wilcoxon rank-sum tests ($p$-values above panels). Across all three measures, EBF infants exhibited lower diversity than non-EBF infants, consistent with recent findings in early-life microbiome studies.
}
  \label{app:fig:alpha_diversity}
\end{figure}

%----------------------------------
% Runtime comparison
\begin{table}[ht]
\centering
\caption{\textbf{Microbiome study:} Runtime comparison for differential abundance testing with \texttt{dacomp} and \texttt{permApprox} under varying permutation budgets. Reported are wall-clock runtimes for increasing numbers of permutations $B$, measured in seconds, minutes, and hours. Experiments were conducted on a Linux-based compute node equipped with an Intel Xeon E7-4850 v4 CPU (2.10\,GHz) and 170\,GB RAM.}
\label{app:tab:dacomp_runtime}
\begin{tabular}{lrrrr}
\toprule
Method & $B$ & Time (s) & Time (min) & Time (h) \\
\midrule
\texttt{dacomp} only                          & $10^{3}$  &    1.84   &   0.03   & 0.001 \\
\texttt{dacomp} only                          & $10^{4}$  &   19.06   &   0.32   & 0.005 \\
\texttt{dacomp} only                          & $10^{5}$  &  169.21   &   2.82   & 0.047 \\
\texttt{dacomp} only                          & $10^{6}$  & 1663.48   &  27.72   & 0.462 \\
\texttt{dacomp} only                          & $10^{7}$  & 16984.33  & 283.07   & 4.718 \\
\midrule
\texttt{dacomp} + \texttt{permApprox} (SLLS, $k_0=250$)   & $10^{3}$  &    3.69   &   0.06   & 0.001 \\
\texttt{dacomp} + \texttt{permApprox} (SLLS, $k_0=250$)   & $10^{4}$  &   20.51   &   0.34   & 0.006 \\
\texttt{dacomp} + \texttt{permApprox} (SLLS, $k_0=250$)   & $10^{5}$  &  171.43   &   2.86   & 0.048 \\
\midrule
\texttt{dacomp} + \texttt{permApprox} (SLLS, $k_0=0.25B$) & $10^{3}$  &    3.46   &   0.06   & 0.001 \\
\texttt{dacomp} + \texttt{permApprox} (SLLS, $k_0=0.25B$) & $10^{4}$  &   25.67   &   0.43   & 0.007 \\
\texttt{dacomp} + \texttt{permApprox} (SLLS, $k_0=0.25B$) & $10^{5}$  &  407.11   &   6.79   & 0.113 \\
\bottomrule
\end{tabular}
\end{table}

%----------------------------------
% Runtime plot
\begin{figure}[H]
  \centering
  \includegraphics[width=\textwidth]{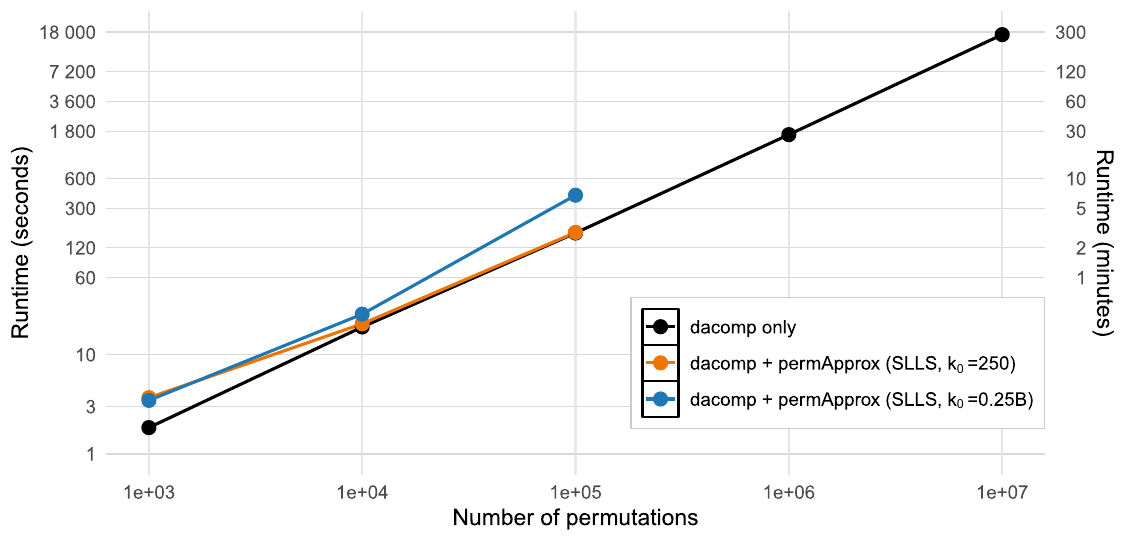}
  \caption{\textbf{Microbiome study:}
  Visualization of the runtime results reported in Table~\ref{app:tab:dacomp_runtime}. Shown is the runtime as a function of the number of permutations $B$ for genus-level differential abundance analysis in the PASTURE cohort using \texttt{dacomp} alone (black) and \texttt{dacomp} combined with \texttt{permApprox} using the SLLS rule with two threshold settings: a fixed number of starting exceedances ($k_0=250$; orange) and a relative number of starting exceedances ($k_0=0.25\,B$; blue). Runtimes are displayed on a logarithmic scale in seconds (left axis) with a secondary axis in minutes (right).}
  \label{app:fig:dacomp_runtime}
\end{figure}

%----------------------------------
% dacomp prevalence score
\begin{figure}[H]
  \centering
  \includegraphics[width=0.75\textwidth]{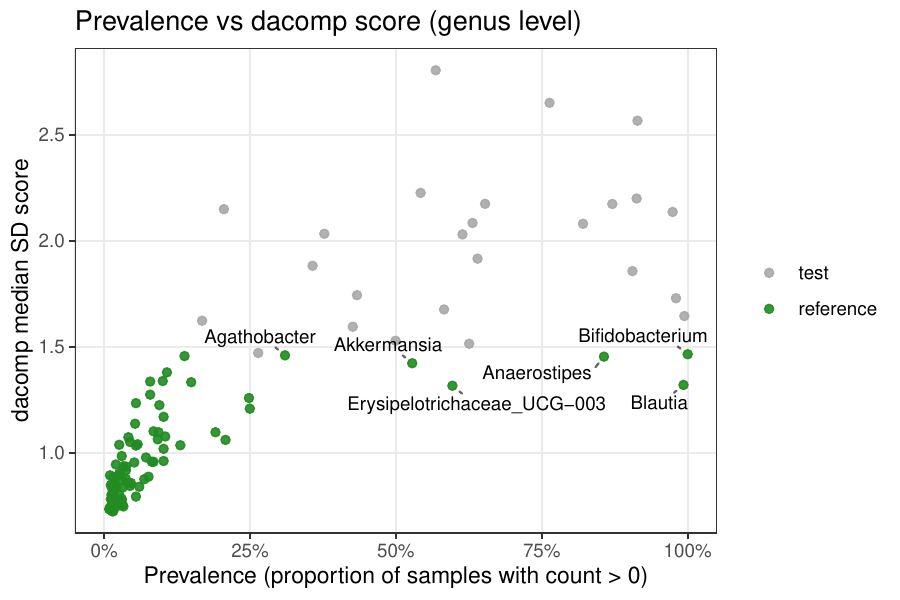}
  \caption{
  \textbf{Microbiome study:} Prevalence versus \texttt{dacomp} median SD score $S_j$ (see Equation~\ref{eq:dacomp_log-ratios}) on genus level.
  Each point represents one genus.  
  Green points indicate taxa selected as reference by \texttt{dacomp}, while gray points represent genera to be tested.}
  \label{app:fig:dacomp_prev_vs_score}
\end{figure}

%----------------------------------
% Phylum composition
\begin{figure}[H]
  \centering
    \begin{subfigure}[t]{0.48\textwidth}
    \centering
    \includegraphics[width=\linewidth]{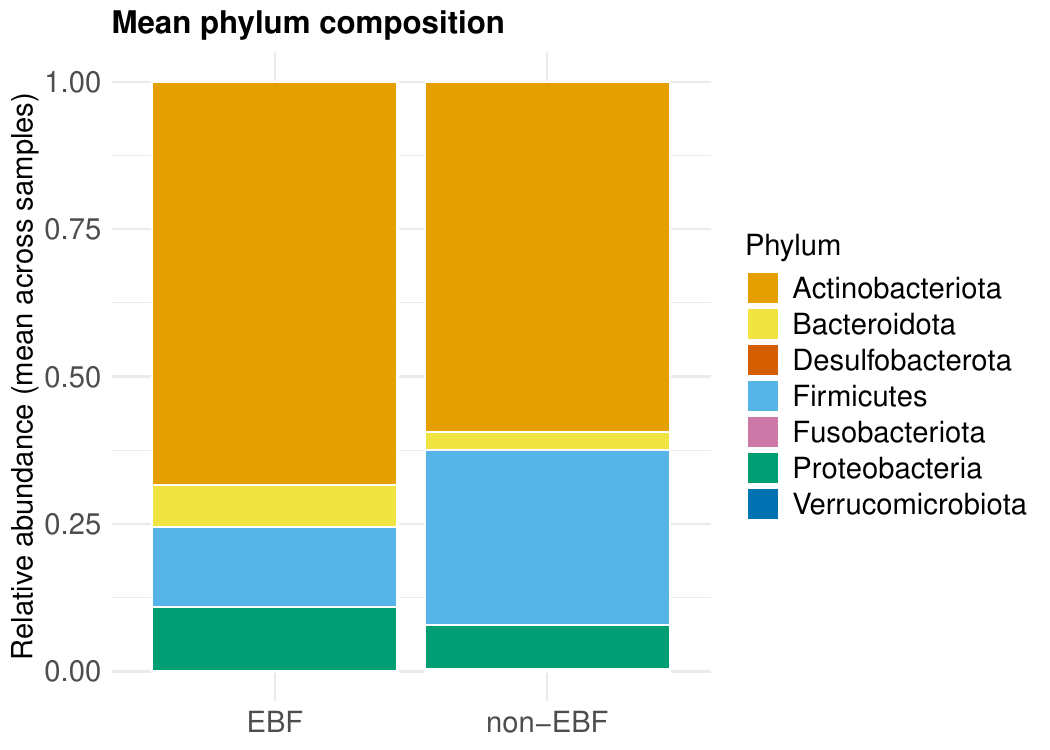}
    \subcaption{Mean phylum composition by feeding group.}
    \label{fig:mean_comp_phylum}
  \end{subfigure}
  \hfill
  \begin{subfigure}[t]{0.48\textwidth}
    \centering
    \includegraphics[width=\linewidth]{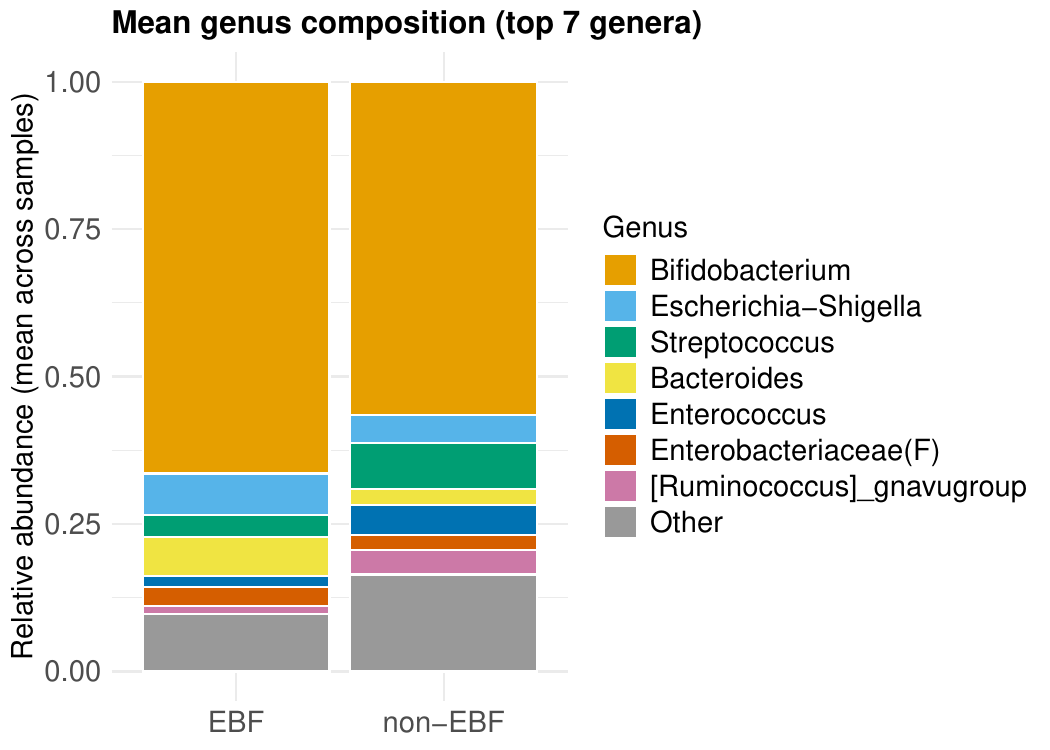}
    \subcaption{Mean genus composition by feeding group (top 7 genera).}
    \label{fig:mean_comp_genus}
  \end{subfigure}
  \caption{\textbf{Microbiome study:} Mean relative gut microbiome composition in exclusively breastfed (EBF) and non-EBF infants at age 2 months, shown at (a) phylum level and (b) genus level.}
  \label{app:fig:mean_composition_genus_phylum}
\end{figure}

%----------------------------------
% Abundance boxplots
\begin{figure}[H]
  \centering
  \includegraphics[width=\textwidth]{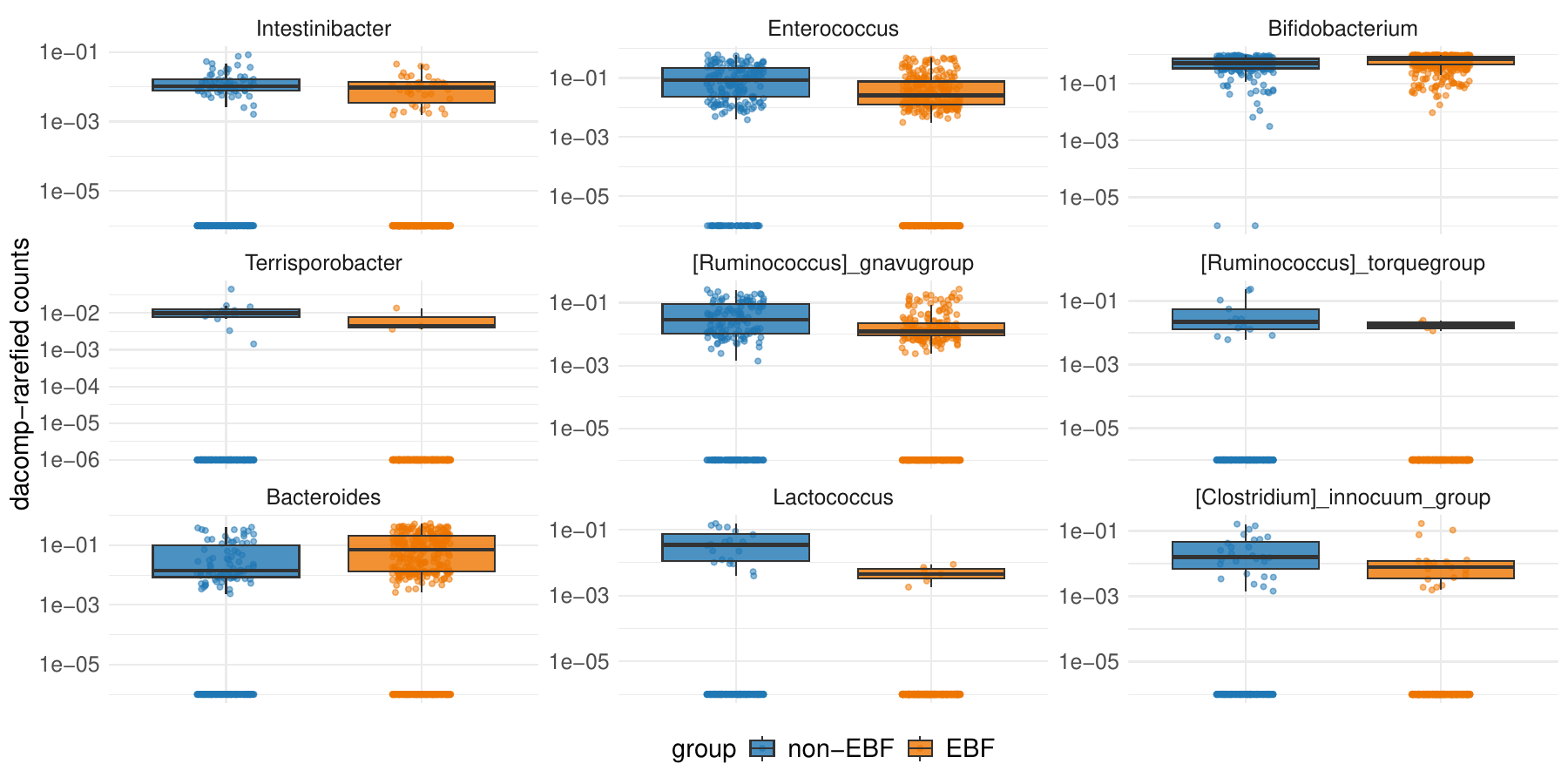}
  \caption{
  \textbf{Microbiome study:} \texttt{dacomp}-rarefied counts for the nine genera discussed in the main section. Each panel shows the distribution of \texttt{dacomp}-normalized counts for one genus, grouped by feeding mode (exclusively breastfed (EBF) for at least two months vs no exclusive breast feeding).}
  \label{app:fig:dacomp_rarefied_counts}
\end{figure}

%----------------------------------
% Permutation histograms
\begin{figure}[H]
  \centering
  \includegraphics[width=\textwidth]{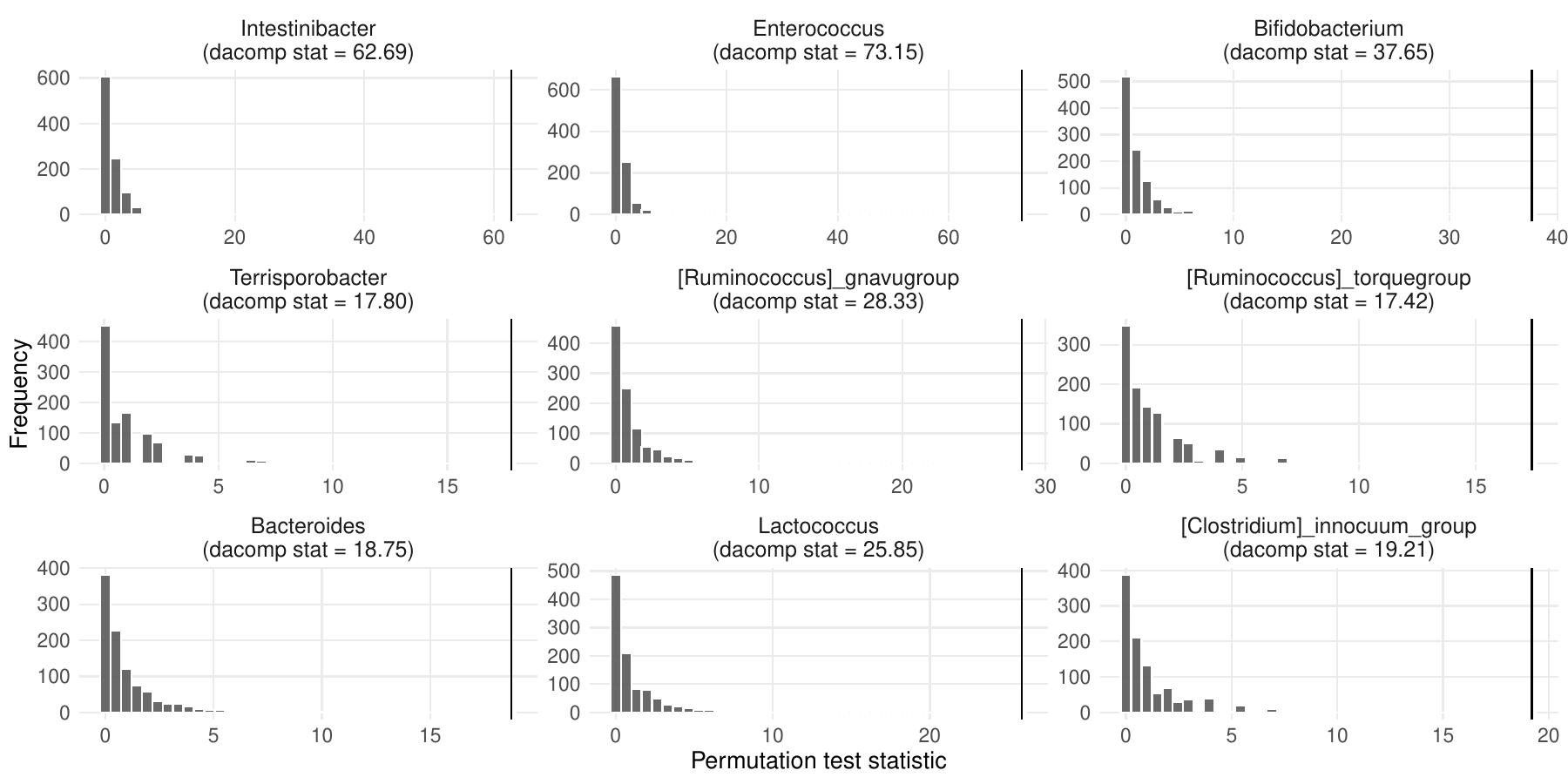}
  \caption{
  \textbf{Microbiome study:} Histograms of the \texttt{dacomp} test statistics for the nine genera discussed in the main section. Each panel shows the distribution of permutation test statistics for 1000 permutations, together with the observed test statistic for that genus.}
  \label{app:fig:dacomp_perm_histograms}
\end{figure}

%----------------------------------
% Unadjusted p-values
\begin{table}[H]
\centering
\caption{\textbf{Microbiome study:}
Unadjusted $p$-values for genera that are significant in at least one of the reported methods at level $\alpha = 0.01$.
For each genus, the table reports the \texttt{dacomp} test statistic $T_{\text{dacomp}}$ and the corresponding $p$-values obtained with \texttt{dacomp} for different permutation budgets $B$, and with the unconstrained (U) and constrained (C) \texttt{permApprox} approach for $B = 1000$.
Asterisks indicate $p$-values below $\alpha = 0.01$.}
\label{app:tab:dacomp_unadjusted_pvals}
\centering
\fontsize{8}{10}\selectfont
\begin{tabular}[t]{lrlllll}
\toprule
Genus &
\multicolumn{1}{c}{$T_\text{dacomp}$} &
\multicolumn{1}{c}{dacomp} &
\multicolumn{1}{c}{dacomp} &
\multicolumn{1}{c}{dacomp} &
\multicolumn{1}{c}{permApprox(U)} &
\multicolumn{1}{c}{permApprox(C)} \\
~ & ~ &
\multicolumn{1}{c}{($B=1000$)} &
\multicolumn{1}{c}{($B=10^6$)} &
\multicolumn{1}{c}{($B=10^7$)} &
\multicolumn{1}{c}{($B=1000$)} &
\multicolumn{1}{c}{($B=1000$)} \\
\midrule
\cellcolor{gray!10}{Enterococcus} & \cellcolor{gray!10}{67.42} & \cellcolor{gray!10}{9.99e-04*} & \cellcolor{gray!10}{1.00e-06*} & \cellcolor{gray!10}{1.00e-07*} & \cellcolor{gray!10}{4.25e-13*} & \cellcolor{gray!10}{4.25e-13*}\\
Intestinibacter & 45.20 & 9.99e-04* & 1.00e-06* & 1.00e-07* & 1.97e-18* & 1.97e-18*\\
\cellcolor{gray!10}{Bifidobacterium} & \cellcolor{gray!10}{45.12} & \cellcolor{gray!10}{9.99e-04*} & \cellcolor{gray!10}{1.00e-06*} & \cellcolor{gray!10}{1.00e-07*} & \cellcolor{gray!10}{2.18e-11*} & \cellcolor{gray!10}{2.18e-11*}\\
{}[Ruminococcus]\_gnavugroup & 24.16 & 9.99e-04* & 1.00e-06* & 4.00e-07* & 1.06e-07* & 1.06e-07*\\
\cellcolor{gray!10}{{}[Clostridium]\_innocuum\_group} & \cellcolor{gray!10}{26.41} & \cellcolor{gray!10}{9.99e-04*} & \cellcolor{gray!10}{1.00e-06*} & \cellcolor{gray!10}{8.00e-07*} & \cellcolor{gray!10}{1.07e-03*} & \cellcolor{gray!10}{1.07e-03*}\\
\addlinespace
Bacteroides & 20.71 & 9.99e-04* & 1.37e-04* & 4.50e-06* & 2.93e-05* & 2.93e-05*\\
\cellcolor{gray!10}{Terrisporobacter} & \cellcolor{gray!10}{22.09} & \cellcolor{gray!10}{9.99e-04*} & \cellcolor{gray!10}{1.04e-04*} & \cellcolor{gray!10}{1.02e-05*} & \cellcolor{gray!10}{0.00e+00*} & \cellcolor{gray!10}{2.52e-09*}\\
Lactococcus & 17.77 & 9.99e-04* & 4.00e-06* & 1.55e-05* & 4.08e-05* & 4.08e-05*\\
\cellcolor{gray!10}{Streptococcus} & \cellcolor{gray!10}{13.75} & \cellcolor{gray!10}{9.99e-04*} & \cellcolor{gray!10}{3.99e-04*} & \cellcolor{gray!10}{2.02e-04*} & \cellcolor{gray!10}{2.07e-04*} & \cellcolor{gray!10}{2.07e-04*}\\
{}[Ruminococcus]\_torquegroup & 13.39 & 9.99e-04* & 7.78e-03* & 2.31e-04* & 3.35e-06* & 3.35e-06*\\
\addlinespace
\cellcolor{gray!10}{Granulicatella} & \cellcolor{gray!10}{10.24} & \cellcolor{gray!10}{7.99e-03*} & \cellcolor{gray!10}{4.10e-05*} & \cellcolor{gray!10}{1.28e-03*} & \cellcolor{gray!10}{7.99e-03*} & \cellcolor{gray!10}{7.99e-03*}\\
Clostridium\_sensu\_strict1 & 9.55 & 9.99e-04* & 2.54e-03* & 1.97e-03* & 4.97e-05* & 4.97e-05*\\
\cellcolor{gray!10}{Sellimonas} & \cellcolor{gray!10}{7.64} & \cellcolor{gray!10}{1.80e-02} & \cellcolor{gray!10}{9.89e-02} & \cellcolor{gray!10}{2.63e-03*} & \cellcolor{gray!10}{2.25e-02} & \cellcolor{gray!10}{2.25e-02}\\
Blautia & 8.89 & 9.99e-04* & 2.60e-04* & 2.84e-03* & 5.94e-05* & 5.94e-05*\\
\cellcolor{gray!10}{Haemophilus} & \cellcolor{gray!10}{8.47} & \cellcolor{gray!10}{2.90e-02} & \cellcolor{gray!10}{7.70e-04*} & \cellcolor{gray!10}{3.17e-03*} & \cellcolor{gray!10}{2.80e-02} & \cellcolor{gray!10}{2.80e-02}\\
\addlinespace
uncultured18 & 9.64 & 2.10e-02 & 5.69e-02 & 3.22e-03* & 2.86e-02 & 2.86e-02\\
\cellcolor{gray!10}{Tyzzerella} & \cellcolor{gray!10}{8.22} & \cellcolor{gray!10}{4.00e-03*} & \cellcolor{gray!10}{2.11e-03*} & \cellcolor{gray!10}{3.72e-03*} & \cellcolor{gray!10}{2.02e-03*} & \cellcolor{gray!10}{2.02e-03*}\\
Leuconostoc & 9.19 & 2.70e-02 & 9.21e-02 & 8.33e-03* & 2.70e-02 & 2.70e-02\\
\cellcolor{gray!10}{Coprobacillus} & \cellcolor{gray!10}{9.19} & \cellcolor{gray!10}{2.90e-02} & \cellcolor{gray!10}{8.42e-03*} & \cellcolor{gray!10}{8.38e-03*} & \cellcolor{gray!10}{2.90e-02} & \cellcolor{gray!10}{2.90e-02}\\
Collinsella & 5.69 & 9.99e-04* & 3.39e-04* & 1.70e-02 & 3.30e-04* & 3.30e-04*\\
\addlinespace
\cellcolor{gray!10}{Gordonibacter} & \cellcolor{gray!10}{6.88} & \cellcolor{gray!10}{4.37e-01} & \cellcolor{gray!10}{2.60e-03*} & \cellcolor{gray!10}{2.79e-02} & \cellcolor{gray!10}{4.37e-01} & \cellcolor{gray!10}{4.37e-01}\\
Clostridioides & 3.49 & 9.99e-03* & 1.98e-03* & 3.90e-02 & 5.59e-03* & 5.59e-03*\\
\cellcolor{gray!10}{22\_Lachnospiraceae(F)} & \cellcolor{gray!10}{3.81} & \cellcolor{gray!10}{3.19e-01} & \cellcolor{gray!10}{7.43e-03*} & \cellcolor{gray!10}{6.17e-02} & \cellcolor{gray!10}{3.19e-01} & \cellcolor{gray!10}{3.19e-01}\\
Holdemanella & 2.78 & 5.30e-01 & 3.34e-03* & 6.55e-02 & 5.30e-01 & 5.30e-01\\
\cellcolor{gray!10}{Paeniclostridium} & \cellcolor{gray!10}{4.58} & \cellcolor{gray!10}{8.99e-03*} & \cellcolor{gray!10}{8.38e-03*} & \cellcolor{gray!10}{9.21e-02} & \cellcolor{gray!10}{8.99e-03*} & \cellcolor{gray!10}{8.99e-03*}\\
\addlinespace
Epulopiscium & 2.08 & 2.60e-02 & 3.97e-02 & 2.62e-01 & 6.62e-03* & 6.62e-03*\\
\cellcolor{gray!10}{Veillonella} & \cellcolor{gray!10}{1.18} & \cellcolor{gray!10}{3.70e-02} & \cellcolor{gray!10}{3.73e-03*} & \cellcolor{gray!10}{2.78e-01} & \cellcolor{gray!10}{3.94e-02} & \cellcolor{gray!10}{3.94e-02}\\
\bottomrule
\end{tabular}
\end{table}

% Significance agreement table
\begin{table}[H]
\centering
\caption{\textbf{Microbiome study:}
Summary of agreement between \texttt{dacomp} with $B=10^7$ permutations and the constrained \texttt{permApprox} approach (\texttt{permApprox(C)}, $B=1000$) based on raw, unadjusted $p$-values at significance level $\alpha = 0.01$.
Results are shown for all 118 genus-level hypothesis tests.}
\label{app:tab:dacomp_sig_agreement_raw}
\begin{tabular}{lr}
\toprule
\textbf{Metric} & \textbf{Value} \\
\midrule
Total number of tests & 118 \\
Significant genera identified by \texttt{dacomp} ($B=10^7$) & 19 \\
Significant genera identified by \texttt{permApprox(C)} ($B=1000$) & 18 \\
Shared significant genera & 14 \\
Sensitivity of \texttt{permApprox(C)} relative to \texttt{dacomp} & 73.7\% \\
Precision of \texttt{permApprox(C)} relative to \texttt{dacomp} & 77.8\% \\
Overall agreement rate & 92.4\% \\
\bottomrule
\end{tabular}
\end{table}

%----------------------------------
% Adjusted p-values

\begin{table}[H]
\centering
\caption{Same as Table~\ref{app:tab:dacomp_unadjusted_pvals} but with BH-adjusted $p$-values.}
\label{app:tab:dacomp_adjusted_pvals}
\centering
\fontsize{8}{10}\selectfont
\begin{tabular}[t]{lrlllll}
\toprule
Genus &
\multicolumn{1}{c}{$T_\text{dacomp}$} &
\multicolumn{1}{c}{dacomp} &
\multicolumn{1}{c}{dacomp} &
\multicolumn{1}{c}{dacomp} &
\multicolumn{1}{c}{permApprox(U)} &
\multicolumn{1}{c}{permApprox(C)} \\
~ & ~ &
\multicolumn{1}{c}{($B=1000$)} &
\multicolumn{1}{c}{($B=10^6$)} &
\multicolumn{1}{c}{($B=10^7$)} &
\multicolumn{1}{c}{($B=1000$)} &
\multicolumn{1}{c}{($B=1000$)} \\
\midrule
\cellcolor{gray!10}{Enterococcus} & \cellcolor{gray!10}{67.42} & \cellcolor{gray!10}{5.30e-03*} & \cellcolor{gray!10}{9.60e-06*} & \cellcolor{gray!10}{1.47e-06*} & \cellcolor{gray!10}{1.51e-11*} & \cellcolor{gray!10}{2.27e-11*}\\
Intestinibacter & 45.20 & 5.30e-03* & 9.60e-06* & 1.47e-06* & 1.05e-16* & 2.10e-16*\\
\cellcolor{gray!10}{Bifidobacterium} & \cellcolor{gray!10}{45.12} & \cellcolor{gray!10}{5.30e-03*} & \cellcolor{gray!10}{9.60e-06*} & \cellcolor{gray!10}{1.47e-06*} & \cellcolor{gray!10}{5.83e-10*} & \cellcolor{gray!10}{7.77e-10*}\\
{}[Ruminococcus]\_gnavugroup & 24.16 & 5.30e-03* & 9.60e-06* & 4.62e-06* & 2.27e-06* & 2.27e-06*\\
\cellcolor{gray!10}{{}[Clostridium]\_innocuum\_group} & \cellcolor{gray!10}{26.41} & \cellcolor{gray!10}{5.30e-03*} & \cellcolor{gray!10}{9.60e-06*} & \cellcolor{gray!10}{7.56e-06*} & \cellcolor{gray!10}{8.77e-03*} & \cellcolor{gray!10}{8.77e-03*}\\
\addlinespace
Bacteroides & 20.71 & 5.30e-03* & 8.44e-04* & 3.69e-05* & 4.48e-04* & 4.48e-04*\\
\cellcolor{gray!10}{Terrisporobacter} & \cellcolor{gray!10}{22.09} & \cellcolor{gray!10}{5.30e-03*} & \cellcolor{gray!10}{7.26e-04*} & \cellcolor{gray!10}{7.32e-05*} & \cellcolor{gray!10}{0.00e+00*} & \cellcolor{gray!10}{6.74e-08*}\\
Lactococcus & 17.77 & 5.30e-03* & 3.32e-05* & 9.58e-05* & 5.46e-04* & 5.46e-04*\\
\cellcolor{gray!10}{Streptococcus} & \cellcolor{gray!10}{13.75} & \cellcolor{gray!10}{5.30e-03*} & \cellcolor{gray!10}{1.93e-03*} & \cellcolor{gray!10}{1.23e-03*} & \cellcolor{gray!10}{2.01e-03*} & \cellcolor{gray!10}{2.01e-03*}\\
{}[Ruminococcus]\_torquegroup & 13.39 & 5.30e-03* & 2.64e-02 & 1.36e-03* & 5.97e-05* & 5.97e-05*\\
\addlinespace
\cellcolor{gray!10}{Granulicatella} & \cellcolor{gray!10}{10.24} & \cellcolor{gray!10}{3.96e-02} & \cellcolor{gray!10}{3.16e-04*} & \cellcolor{gray!10}{7.29e-03*} & \cellcolor{gray!10}{5.03e-02} & \cellcolor{gray!10}{5.03e-02}\\
Clostridium\_sensu\_strict1 & 9.55 & 5.30e-03* & 1.08e-02 & 1.01e-02 & 5.91e-04* & 5.91e-04*\\
\cellcolor{gray!10}{Haemophilus} & \cellcolor{gray!10}{8.47} & \cellcolor{gray!10}{8.41e-02} & \cellcolor{gray!10}{3.72e-03*} & \cellcolor{gray!10}{1.43e-02} & \cellcolor{gray!10}{1.07e-01} & \cellcolor{gray!10}{1.07e-01}\\
Blautia & 8.89 & 5.30e-03* & 1.55e-03* & 1.43e-02 & 6.36e-04* & 6.36e-04*\\
\cellcolor{gray!10}{Tyzzerella} & \cellcolor{gray!10}{8.22} & \cellcolor{gray!10}{1.97e-02} & \cellcolor{gray!10}{8.85e-03*} & \cellcolor{gray!10}{1.55e-02} & \cellcolor{gray!10}{1.54e-02} & \cellcolor{gray!10}{1.54e-02}\\
\addlinespace
Collinsella & 5.69 & 5.30e-03* & 1.82e-03* & 6.18e-02 & 2.94e-03* & 2.94e-03*\\
\cellcolor{gray!10}{Clostridioides} & \cellcolor{gray!10}{3.49} & \cellcolor{gray!10}{4.57e-02} & \cellcolor{gray!10}{8.85e-03*} & \cellcolor{gray!10}{1.12e-01} & \cellcolor{gray!10}{3.98e-02} & \cellcolor{gray!10}{3.98e-02}\\
\bottomrule
\end{tabular}
\end{table}

% Significance agreement table
\begin{table}[H]
\centering
\caption{\textbf{Microbiome study:}
Summary of agreement between \texttt{dacomp} with $B=10^7$ permutations and the constrained \texttt{permApprox} approach (\texttt{permApprox(C)}, $B=1000$) based on Benjamini--Hochberg–adjusted $p$-values at significance level $\alpha = 0.01$.
Results are shown for all 118 genus-level hypothesis tests.}
\label{app:tab:dacomp_sig_agreement_adj}
\begin{tabular}{lr}
\toprule
\textbf{Metric} & \textbf{Value} \\
\midrule
Total number of tests & 118 \\
Significant genera identified by \texttt{dacomp} ($B=10^7$) & 11 \\
Significant genera identified by \texttt{permApprox(C)} ($B=1000$) & 13 \\
Shared significant genera & 10 \\
Sensitivity of \texttt{permApprox(C)} relative to \texttt{dacomp} & 90.9\% \\
Precision of \texttt{permApprox(C)} relative to \texttt{dacomp} & 76.9\% \\
Overall agreement rate & 96.6\% \\
\bottomrule
\end{tabular}
\end{table}

%============================================================
\subsection{Differential distribution analysis - The waddR approach}
\label{app:sec:DD_waddR_details}
%============================================================

Differential distribution (DD) analysis was performed with the \texttt{waddR} package \citep{schefzik2021fast}. The framework proposed by \citet{schefzik2021fast} can be summarized as follows:
For a given gene, let $\hat{F}_A$ and $\hat{F}_B$ denote the empirical distribution functions of the expression values in groups $A$ and $B$. 
Testing for distributional differences between the two groups can be formulated as
\begin{equation}
  H_0: F_A = F_B 
  \qquad\text{versus}\qquad 
  H_1: F_A \neq F_B .
\end{equation}

To quantify distributional differences, \texttt{waddR} uses the squared 2-Wasserstein distance
\begin{equation}
\label{eq:Wasserstein_distance}
  d_W^2(\hat{F}_A, \hat{F}_B)
    = \int_0^1 \bigl( \hat{F}_A^{-1}(u) - \hat{F}_B^{-1}(u) \bigr)^2 \, du,
\end{equation}
where $u\in[0,1]$ denotes the quantile level. \eqref{eq:Wasserstein_distance} is non-negative and equals zero if and only if the two distributions are identical. Using this metric, the hypothesis can equivalently be expressed as the one-sided test
\begin{equation}
  H_0: d_W^2 = 0 \qquad\text{versus}\qquad H_1: d_W^2 > 0.
\end{equation}

Under $H_0$, the non-zero expression values are exchangeable across groups, and the null distribution of $d_W^2$ is obtained by repeatedly permuting the group labels. The resulting permutation distribution is then used to compute an empirical $p$-value.

To improve the resolution of very small $p$-values, \texttt{waddR} applies a semi-parametric tail approximation: if the number of permutation statistics exceeding $T_{\mathrm{obs}}$ is smaller than 10, a generalized Pareto distribution is fitted to the upper tail of the permutation distribution and used to refine the tail probability. Since the principle of GPD-based tail approximation is already described in Section~\ref{sec:methods:gpd_tail_approximation}, we refer the reader to that subsection for details. 

It is important to note that \texttt{waddR} employs a two-stage procedure for differential distribution testing. The first stage addresses the excess zeros via a ``differential proportions of zero expression'' test, which models the probability of zero expression in groups $A$ and $B$ using a Bayesian logistic regression model. This yields a $p$-value for detecting differences in the zero-expression proportions. The second stage applies the semi-parametric Wasserstein permutation test to the non-zero expression values and is therefore the only component that relies on permutation statistics and potential tail modeling. Hence, our comparison focuses exclusively on this non-zero component, and we follow the same rule as \texttt{waddR}: whenever the number of exceedances is below~10, we apply \texttt{permApprox} in place of the original semi-parametric refinement to ensure a fair and methodologically aligned comparison.

%============================================================
\subsection{Differential distribution analysis - Additional results}
\label{app:sec:DD_results}
%============================================================

\begin{figure}[H]
\centering
\includegraphics[width=\textwidth]{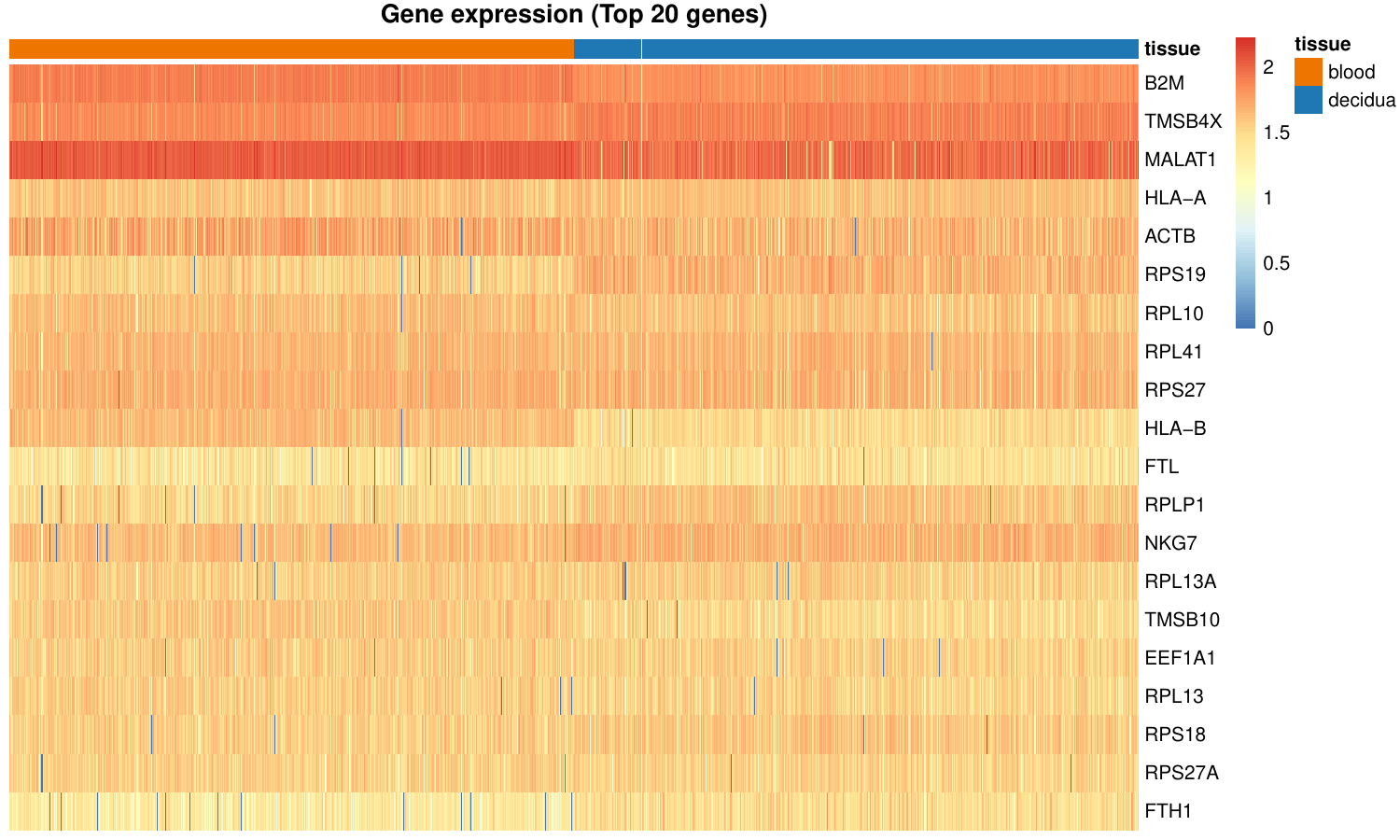}
\caption{\textbf{scRNA-seq study:} Heatmap of log-transformed expression values for the 20 genes with the highest overall prevalence across blood and decidua samples. Each column represents a single cell and each row a gene. Cells are annotated by tissue of origin, and rows and columns are shown without clustering to highlight the global differences in expression levels between the two tissues.
}
\label{app:fig:scRNA_heatmap}
\end{figure}

%----------------------------------
\begin{figure}[H]
\centering
\includegraphics[width=\textwidth]{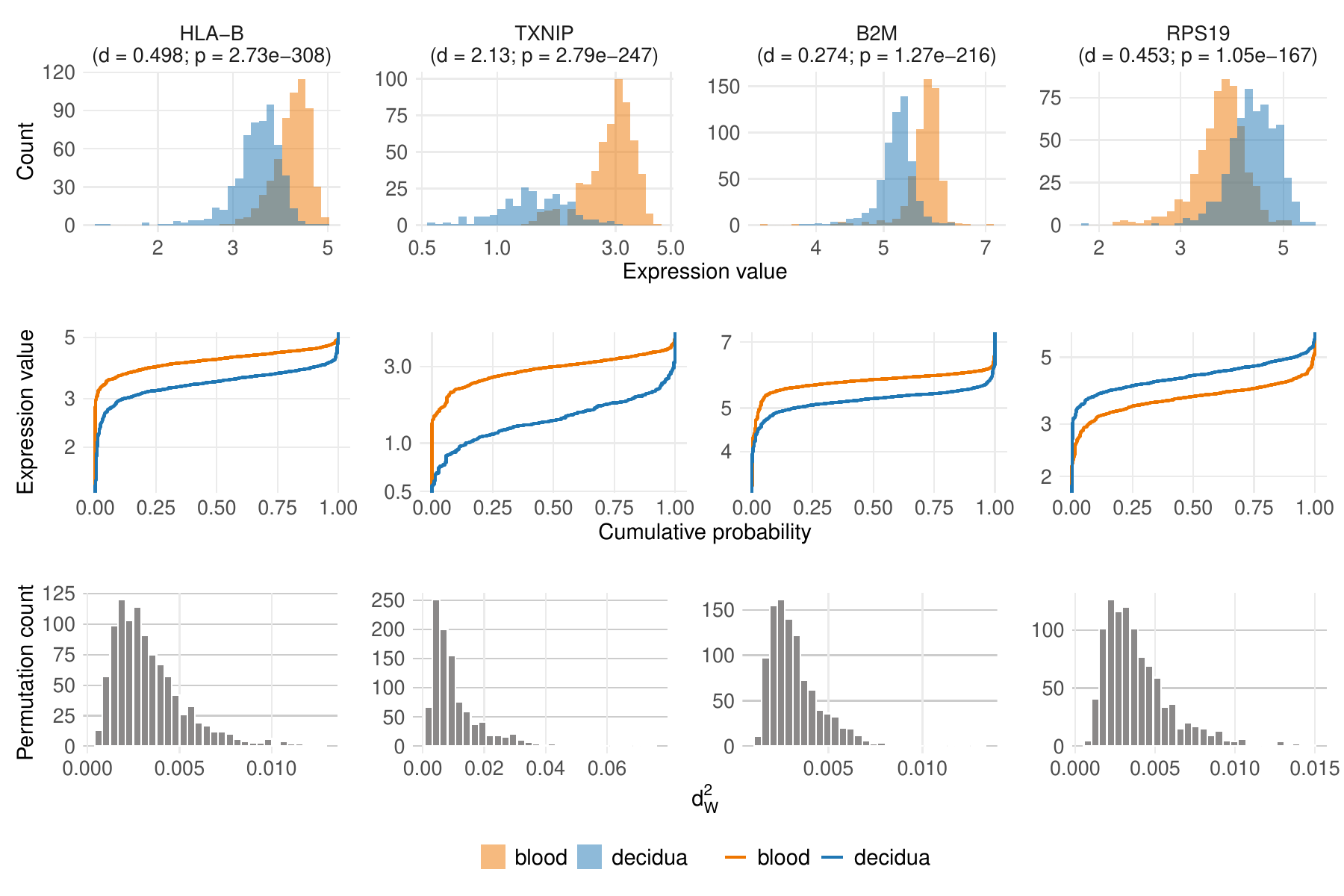}
\caption{\textbf{scRNA-seq study:} 
Histograms (top row), empirical quantile functions (middle row), and permutation distributions (bottom row) of the non-zero expression counts for the four genes with smallest \texttt{permApprox} (constrained) $p$-values ($B = 1000$).
Strip titles report the TS Wasserstein statistic $d$ and the constrained \texttt{permApprox} $p$-value. 
The top and middle panels illustrate the observed distributional differences between blood and decidua that contribute to the 2-Wasserstein distance. 
The bottom panels show the corresponding permutation distributions of $d_W^2$. For all four genes, the observed test statistic lies far beyond the most extreme permutation test statistic.}
\label{app:fig:scRNA_hist_quant}
\end{figure}

%----------------------------------
\begin{table}[H]
\centering
\caption{\textbf{scRNA-seq study:} $p$-values for the top 20 genes, ordered by the constrained \texttt{permApprox} (C) $p$-values (bold).
For each gene, the table reports the 2-Wasserstein distance $d_{W}^2$ and the $p$-values obtained from three approaches: the original \texttt{waddR} method, and \texttt{permApprox(U)} and \texttt{permApprox(C)}, which recompute the tail $p$-values from the same waddR permutation test statistics using unconstrained or constrained GPD fitting, respectively.
$\hat{\xi}_{\text{waddR}}$ denotes the shape parameter estimate from the waddR tail-refinement step, and $\hat{\xi}_{(U)}$ and $\hat{\xi}_{(C)}$ are the corresponding shape estimates from the unconstrained and constrained \texttt{permApprox} fits.}
\label{app:tab:waddr_unadjusted_pvals}
\centering
\fontsize{8}{10}\selectfont
\begin{tabular}[t]{lrlllrrr}
\toprule
Gene & $d_{W}^2$ & $p_{\text{waddR}}$ & $p_{\text{permApprox(U)}}$ & \textbf{$p_{\text{permApprox(C)}}$} & $\hat{\xi}_{\text{waddR}}$ & $\hat{\xi}_{\text{(U)}}$ & $\hat{\xi}_{\text{(C)}}$\\
\midrule
\cellcolor{gray!10}{HLA-B} & \cellcolor{gray!10}{0.498} & \cellcolor{gray!10}{0.00e+00} & \cellcolor{gray!10}{0.00e+00} & \cellcolor{gray!10}{2.73e-308} & \cellcolor{gray!10}{0.003} & \cellcolor{gray!10}{-0.018} & \cellcolor{gray!10}{-0.003}\\
TXNIP & 2.127 & 0.00e+00 & 0.00e+00 & 2.79e-247 & 0.022 & -0.029 & -0.004\\
\cellcolor{gray!10}{B2M} & \cellcolor{gray!10}{0.274} & \cellcolor{gray!10}{0.00e+00} & \cellcolor{gray!10}{0.00e+00} & \cellcolor{gray!10}{1.27e-216} & \cellcolor{gray!10}{0.007} & \cellcolor{gray!10}{-0.072} & \cellcolor{gray!10}{-0.004}\\
RPS19 & 0.453 & NA & 1.05e-167 & 1.05e-167 & 0.005 & -0.003 & -0.003\\
\cellcolor{gray!10}{GSTP1} & \cellcolor{gray!10}{0.612} & \cellcolor{gray!10}{0.00e+00} & \cellcolor{gray!10}{1.07e-301} & \cellcolor{gray!10}{9.53e-162} & \cellcolor{gray!10}{0.025} & \cellcolor{gray!10}{-0.006} & \cellcolor{gray!10}{-0.005}\\
\addlinespace
FOS & 1.544 & 0.00e+00 & 0.00e+00 & 1.85e-158 & 0.043 & -0.092 & -0.006\\
\cellcolor{gray!10}{CCL4} & \cellcolor{gray!10}{1.298} & \cellcolor{gray!10}{0.00e+00} & \cellcolor{gray!10}{0.00e+00} & \cellcolor{gray!10}{3.11e-141} & \cellcolor{gray!10}{0.028} & \cellcolor{gray!10}{-0.169} & \cellcolor{gray!10}{-0.006}\\
CD99 & 0.267 & 0.00e+00 & 0.00e+00 & 3.22e-134 & 0.003 & -0.016 & -0.006\\
\cellcolor{gray!10}{KLRF1} & \cellcolor{gray!10}{0.625} & \cellcolor{gray!10}{0.00e+00} & \cellcolor{gray!10}{0.00e+00} & \cellcolor{gray!10}{2.33e-125} & \cellcolor{gray!10}{0.010} & \cellcolor{gray!10}{-0.021} & \cellcolor{gray!10}{-0.006}\\
APMAP & 0.336 & 0.00e+00 & 0.00e+00 & 2.51e-123 & 0.006 & -0.052 & -0.006\\
\addlinespace
\cellcolor{gray!10}{AES} & \cellcolor{gray!10}{0.483} & \cellcolor{gray!10}{0.00e+00} & \cellcolor{gray!10}{0.00e+00} & \cellcolor{gray!10}{3.89e-118} & \cellcolor{gray!10}{0.006} & \cellcolor{gray!10}{-0.027} & \cellcolor{gray!10}{-0.007}\\
GMFG & 0.325 & 0.00e+00 & 0.00e+00 & 4.89e-114 & 0.003 & -0.025 & -0.007\\
\cellcolor{gray!10}{COTL1} & \cellcolor{gray!10}{2.304} & \cellcolor{gray!10}{0.00e+00} & \cellcolor{gray!10}{0.00e+00} & \cellcolor{gray!10}{2.64e-112} & \cellcolor{gray!10}{0.042} & \cellcolor{gray!10}{-0.088} & \cellcolor{gray!10}{-0.007}\\
ZFP36L2 & 0.595 & 0.00e+00 & 0.00e+00 & 7.06e-110 & 0.030 & -0.100 & -0.007\\
\cellcolor{gray!10}{TRAF3IP3} & \cellcolor{gray!10}{0.432} & \cellcolor{gray!10}{0.00e+00} & \cellcolor{gray!10}{0.00e+00} & \cellcolor{gray!10}{8.88e-108} & \cellcolor{gray!10}{0.006} & \cellcolor{gray!10}{-0.057} & \cellcolor{gray!10}{-0.007}\\
\addlinespace
TMSB10 & 0.191 & 0.00e+00 & 0.00e+00 & 6.05e-105 & 0.001 & -0.056 & -0.007\\
\cellcolor{gray!10}{CD63} & \cellcolor{gray!10}{0.228} & \cellcolor{gray!10}{0.00e+00} & \cellcolor{gray!10}{0.00e+00} & \cellcolor{gray!10}{4.00e-98} & \cellcolor{gray!10}{0.016} & \cellcolor{gray!10}{-0.056} & \cellcolor{gray!10}{-0.007}\\
CDC42 & 0.246 & 0.00e+00 & 0.00e+00 & 4.35e-94 & 0.003 & -0.013 & -0.008\\
\cellcolor{gray!10}{TPST2} & \cellcolor{gray!10}{0.326} & \cellcolor{gray!10}{0.00e+00} & \cellcolor{gray!10}{0.00e+00} & \cellcolor{gray!10}{8.87e-91} & \cellcolor{gray!10}{0.006} & \cellcolor{gray!10}{-0.015} & \cellcolor{gray!10}{-0.008}\\
CTSW & 0.258 & 0.00e+00 & 0.00e+00 & 3.41e-90 & 0.004 & -0.023 & -0.008\\
\bottomrule
\end{tabular}
\end{table}

%----------------------
% Adjusted p-values
\begin{table}[H]
\centering
\caption{Same as Table~\ref{app:tab:waddr_unadjusted_pvals} but with BH-adjusted $p$-values.}
\label{app:tab:waddr_adjusted_pvals}
\centering
\fontsize{8}{10}\selectfont
\begin{tabular}[t]{lrlllrrr}
\toprule
Gene & $d_{W}^2$ & $p_{\text{waddR}}$ & $p_{\text{permApprox(U)}}$ & \textbf{$p_{\text{permApprox(C)}}$} & $\hat{\xi}_{\text{waddR}}$ & $\hat{\xi}_{\text{(U)}}$ & $\hat{\xi}_{\text{(C)}}$\\
\midrule
\cellcolor{gray!10}{HLA-B} & \cellcolor{gray!10}{0.498} & \cellcolor{gray!10}{0.00e+00} & \cellcolor{gray!10}{0.00e+00} & \cellcolor{gray!10}{2.72e-305} & \cellcolor{gray!10}{0.003} & \cellcolor{gray!10}{-0.018} & \cellcolor{gray!10}{-0.003}\\
TXNIP & 2.127 & 0.00e+00 & 0.00e+00 & 1.39e-244 & 0.022 & -0.029 & -0.004\\
\cellcolor{gray!10}{B2M} & \cellcolor{gray!10}{0.274} & \cellcolor{gray!10}{0.00e+00} & \cellcolor{gray!10}{0.00e+00} & \cellcolor{gray!10}{4.22e-214} & \cellcolor{gray!10}{0.007} & \cellcolor{gray!10}{-0.072} & \cellcolor{gray!10}{-0.004}\\
RPS19 & 0.453 & NA & 5.27e-167 & 2.62e-165 & 0.005 & -0.003 & -0.003\\
\cellcolor{gray!10}{GSTP1} & \cellcolor{gray!10}{0.612} & \cellcolor{gray!10}{0.00e+00} & \cellcolor{gray!10}{5.39e-301} & \cellcolor{gray!10}{1.90e-159} & \cellcolor{gray!10}{0.025} & \cellcolor{gray!10}{-0.006} & \cellcolor{gray!10}{-0.005}\\
\addlinespace
FOS & 1.544 & 0.00e+00 & 0.00e+00 & 3.07e-156 & 0.043 & -0.092 & -0.006\\
\cellcolor{gray!10}{CCL4} & \cellcolor{gray!10}{1.298} & \cellcolor{gray!10}{0.00e+00} & \cellcolor{gray!10}{0.00e+00} & \cellcolor{gray!10}{4.43e-139} & \cellcolor{gray!10}{0.028} & \cellcolor{gray!10}{-0.169} & \cellcolor{gray!10}{-0.006}\\
CD99 & 0.267 & 0.00e+00 & 0.00e+00 & 4.01e-132 & 0.003 & -0.016 & -0.006\\
\cellcolor{gray!10}{KLRF1} & \cellcolor{gray!10}{0.625} & \cellcolor{gray!10}{0.00e+00} & \cellcolor{gray!10}{0.00e+00} & \cellcolor{gray!10}{2.58e-123} & \cellcolor{gray!10}{0.010} & \cellcolor{gray!10}{-0.021} & \cellcolor{gray!10}{-0.006}\\
APMAP & 0.336 & 0.00e+00 & 0.00e+00 & 2.50e-121 & 0.006 & -0.052 & -0.006\\
\addlinespace
\cellcolor{gray!10}{AES} & \cellcolor{gray!10}{0.483} & \cellcolor{gray!10}{0.00e+00} & \cellcolor{gray!10}{0.00e+00} & \cellcolor{gray!10}{3.52e-116} & \cellcolor{gray!10}{0.006} & \cellcolor{gray!10}{-0.027} & \cellcolor{gray!10}{-0.007}\\
GMFG & 0.325 & 0.00e+00 & 0.00e+00 & 4.06e-112 & 0.003 & -0.025 & -0.007\\
\cellcolor{gray!10}{COTL1} & \cellcolor{gray!10}{2.304} & \cellcolor{gray!10}{0.00e+00} & \cellcolor{gray!10}{0.00e+00} & \cellcolor{gray!10}{2.02e-110} & \cellcolor{gray!10}{0.042} & \cellcolor{gray!10}{-0.088} & \cellcolor{gray!10}{-0.007}\\
ZFP36L2 & 0.595 & 0.00e+00 & 0.00e+00 & 5.03e-108 & 0.030 & -0.100 & -0.007\\
\cellcolor{gray!10}{TRAF3IP3} & \cellcolor{gray!10}{0.432} & \cellcolor{gray!10}{0.00e+00} & \cellcolor{gray!10}{0.00e+00} & \cellcolor{gray!10}{5.90e-106} & \cellcolor{gray!10}{0.006} & \cellcolor{gray!10}{-0.057} & \cellcolor{gray!10}{-0.007}\\
\addlinespace
TMSB10 & 0.191 & 0.00e+00 & 0.00e+00 & 3.77e-103 & 0.001 & -0.056 & -0.007\\
\cellcolor{gray!10}{CD63} & \cellcolor{gray!10}{0.228} & \cellcolor{gray!10}{0.00e+00} & \cellcolor{gray!10}{0.00e+00} & \cellcolor{gray!10}{2.34e-96} & \cellcolor{gray!10}{0.016} & \cellcolor{gray!10}{-0.056} & \cellcolor{gray!10}{-0.007}\\
CDC42 & 0.246 & 0.00e+00 & 0.00e+00 & 2.41e-92 & 0.003 & -0.013 & -0.008\\
\cellcolor{gray!10}{TPST2} & \cellcolor{gray!10}{0.326} & \cellcolor{gray!10}{0.00e+00} & \cellcolor{gray!10}{0.00e+00} & \cellcolor{gray!10}{4.65e-89} & \cellcolor{gray!10}{0.006} & \cellcolor{gray!10}{-0.015} & \cellcolor{gray!10}{-0.008}\\
CTSW & 0.258 & 0.00e+00 & 0.00e+00 & 1.70e-88 & 0.004 & -0.023 & -0.008\\
\bottomrule
\end{tabular}
\end{table}

%----------------------------------
\begin{table}[H]
\centering
\caption{\textbf{scRNA-seq study:} Summary statistics for the three approaches compared in the application.
\texttt{waddR} refers to the original method, whereas \texttt{permApprox(U)} and \texttt{permApprox(C)} re-use the same waddR permutation test statistics but recompute the $p$-values via unconstrained or constrained GPD fitting, respectively. The column \texttt{n\_tests} reports the number of genes with a non-NA test statistic (out of 1,000). \texttt{n\_zero} denotes the number of zero $p$-values, and the remaining columns list the number of $p$-values below the indicated significance thresholds $\alpha$.
}
\label{app:tab:waddr_sig_counts}
\begin{tabular}{lrrrrr}
\toprule
\textbf{Method} & \textbf{n\_tests} & \textbf{n\_zero} & $\boldsymbol{\alpha = 0.05}$ & $\boldsymbol{\alpha = 0.01}$ & $\boldsymbol{\alpha = 0.001}$ \\
\midrule
waddR                 & 987 & 328 & 959 & 940 & 915 \\
waddR + permApprox(U) & 996 & 197 & 967 & 949 & 923 \\
waddR + permApprox(C) & 996 & 0   & 967 & 949 & 922 \\
\bottomrule
\end{tabular}
\end{table}

\end{document}